\documentclass[%
 reprint,onecolumn,
 amsmath,amssymb,
 aps,
]{revtex4-2}

\usepackage{graphicx}
\usepackage{dcolumn}
\usepackage{bm}
\usepackage{xcolor}
\usepackage{subfig}
\usepackage{amsmath}
\usepackage{mathrsfs}
\usepackage{rotating}

\begin{document}


\title{Surface and internal gravity waves on a viscous liquid layer: initial-value problems}


\author{Ramana Patibandla}
\affiliation{%
 Dept. Applied Mechanics, IIT Madras, India
}%
\author{Saswata Basak}%
 \affiliation{%
 Dept. Chemical Engg., IIT Bombay, India
}%
\author{Anubhab Roy}%
 \email{anubhab@iitm.ac.in}
 \affiliation{%
 Dept. Applied Mechanics, IIT Madras, India
}%
\author{Ratul Dasgupta}%
\email{dasgupta.ratul@iitb.ac.in}
 \affiliation{%
 Dept. Chemical Engg., IIT Bombay, India
}%

\maketitle

\date{\today}

We study a class of initial value problems (IVPs) involving perturbations on a density stratified, quiescent, viscous liquid layer with a free-surface. The  geometry is a two-dimensional, rectangular configuration taking into account surface-tension and gravity. Linearised predictions are obtained by solving the IVP analytically for free-surface and vortical initial perturbations. The viscous spectrum comprises a discrete spectrum of propagating surface and internal modes and damped vorticity modes. Continuous spectra may also exist depending on whether the domain is bounded or unbounded. It is shown that independent of stratification, the vorticity modes in the spectrum play an important role. For an  unstratified pool of a highly viscous liquid, the vorticity modes are found to make a large contribution to the temporal evolution of the transient vorticity layer, at the free-surface. It is demonstrated that the temporal evolution of localised vortical perturbations sufficiently deep inside the pool, cannot be captured by surface modes (capillarity-gravity modes) and require contributions from the vorticity modes. The free-surface signature of a deep, localised, vortical perturbation is however found to be negligible. On a stratified pool of viscous liquid, such vortical initial perturbations are shown to have significant projections on vorticity as well as internal gravity modes. In the infinite depth limit  of an unstratified pool, the contribution from the continuous spectrum vorticity modes is analysed vis-a-vis that from their finite depth, discrete spectrum counterparts. Several analytical predictions are compared against direct numerical simulations (DNS) obtaining excellent agreement. Our results extend classical ones due to \cite{lamb1932,yih1960gravity,prosperetti1976viscous,prosperetti1982small} and receive support from nonlinear simulations.

\section{Introduction}
Solution to initial value problems (IVPs) for surface waves on a quiescent or sheared liquid layer have remained an active topic of research since the classical analysis by Cauchy \cite{cauchy1827theorie} and Poisson \cite{poisson1818memoire}, see \cite{ellingsen2014initial,shardt2016oscillatory,shen2017marangoni,shen2018capillary,tyvand2021nonlinear, tyvand2021nonlinear2}. For a unstratified, quiescent layer modelled as being infinitely deep, this was first done in the linearised, inviscid-irrotational framework (also known as the pebble in the pond problem, see \cite{lamb1932, debnath1994nonlinear, craik2004origins}). For a given initial perturbation, the interface shape and perturbation velocity potential at any later time may be expressed as inverse Fourier (Hankel) integrals over initial conditions weighted by a harmonic time dependence, the frequency of which comes from the dispersion relation \cite{debnath1994nonlinear}. While such integrals can now be solved numerically quite routinely (see fig. 1a and 1b generated using Mathematica \cite{wolfram2013wolfram}), their physical content becomes clear from the application of method of stationary phase (see sec. 2.1.3 in \cite{johnson1997modern}) leading to the notion of a local Fourier mode, whose envelope propagates at the group velocity.

Inclusion of viscosity into the analysis using a monochromatic Fourier mode was first reported by Basset \cite{basset1888treatise} and subsequently by Lamb \cite{lamb1932} and Harrison \cite{harrison1908influence} who obtained the viscous dispersion relation for surface waves on a quiescient, unstratified pool of infinite and finite depth. Lamb \cite{lamb1932} also commented on the continuous spectrum of eigenmodes for an infinitely deep pool of viscous liquid. An explicit solution to the linearised, viscous initial-value problem for a localised surface deformation on an infinitely deep, cylindrical pool was reported by Miles \cite{miles1968cauchy}. Using Hankel transform (in radial direction) and Laplace transform (time), Miles \cite{miles1968cauchy} obtained the solution to the viscous (linearised) Cauchy-Poisson problem comprising a free-surface displacement and a surface impulse. In rectangular geometry, the initial value problem for pure surface deformation (single Fourier mode), as well as a combination of surface impulse and surface deformation, was obtained by Prosperetti and co-workers in a series of studies \cite{prosperetti1976viscous, prosperetti1981motion,cortelezzi1981small,prosperetti1982small}. It was demonstrated that the time evolution of a standing Fourier mode is governed by a damped harmonic oscillator equation with a memory term. Such memory terms arise in various oscillation problems \cite{prosperetti1980normal,berger1988initial} and owe their physical origin to vorticity modes excited due to projections from initial conditions \cite{prosperetti1982small}. In more recent work, the viscous initial-value problem for a pure free-surface deformation (the restoring force being either capillary-gravity or capillarity alone) with zero vorticity initially, has been studied in different base state geometries \cite{farsoiya2017axisymmetric, farsoiya2020azimuthal} leading to similar equations with memory terms. Of interest also, is the persistence of this term, when oscillatory forcing is included leading to parametric instabilities. In the viscous case, viscous standing waves under oscillatory forcing have been shown to be governed by damped Mathieu equation with a memory term, in various configurations \citep{beyer1995faraday,cerda1997faraday,cerda1998faraday,patankar2022dynamic}.
	\begin{figure}
		\centering
			\subfloat[$t=0$ s]{\includegraphics[scale=0.2]{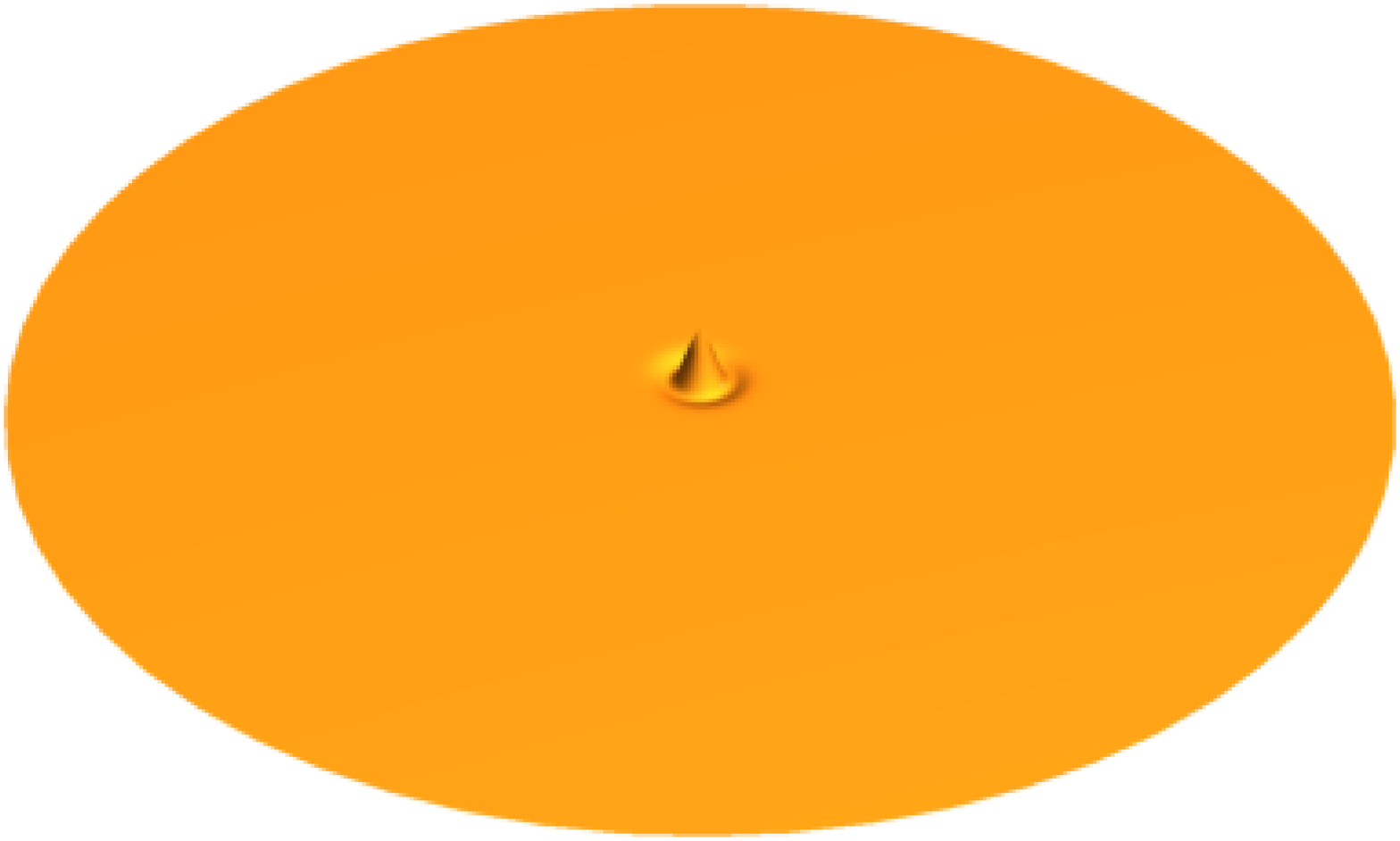}\label{fig1a}}\hspace{8mm}
		\subfloat[$t=15$ s]{\includegraphics[scale=0.2]{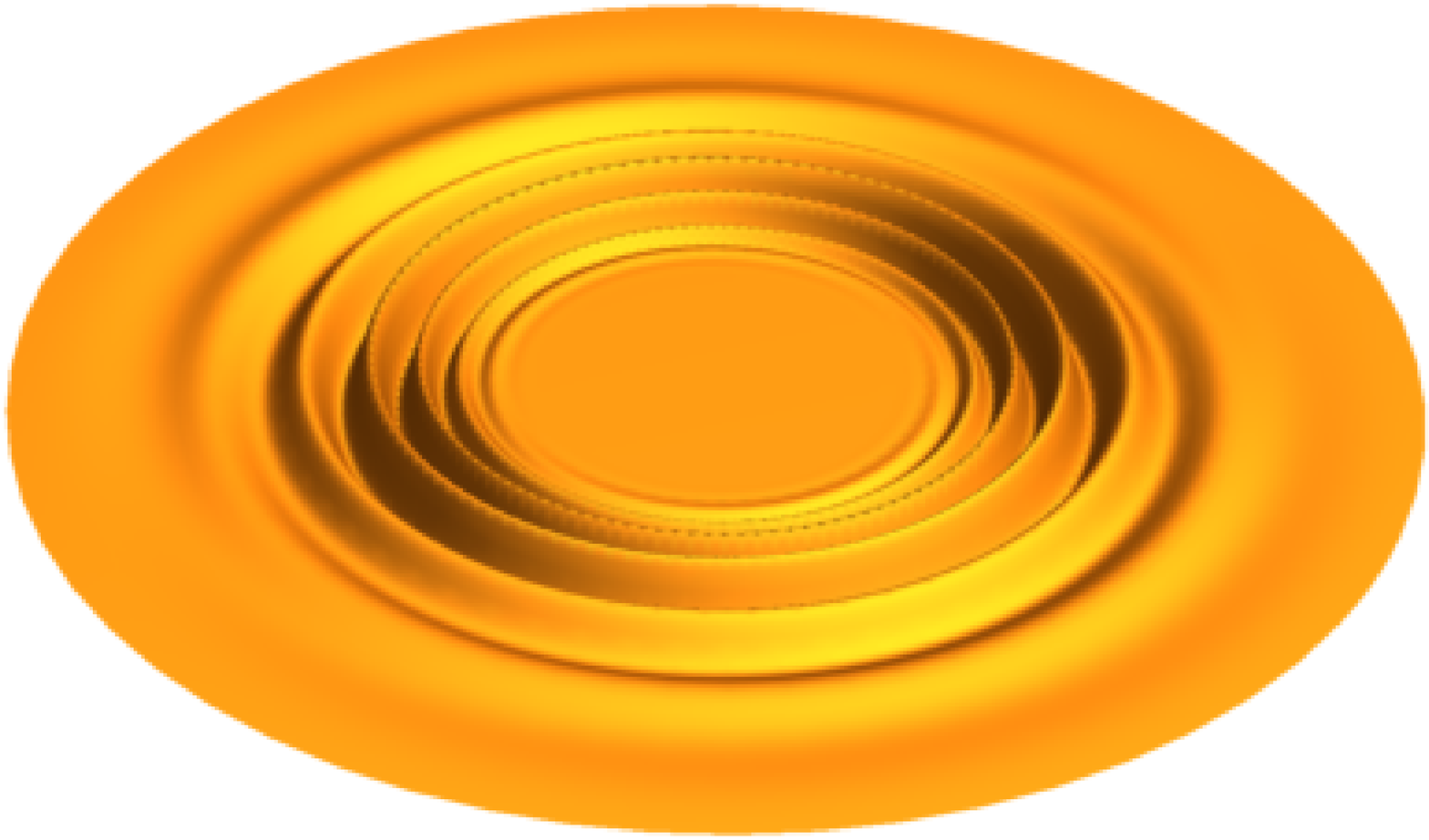}\label{fig1b}}
		\caption{Three dimensional rendering of axisymmetric surface gravity waves from the inviscid, Cauchy-Poisson solution a) The initial surface perturbation \citep{miles1968cauchy} is $\eta(r,0) = 40\left(1-r^2\right)\exp\left(-r^2\right)$ on a quiescent pool of infinite pth. Panel b) At any time $t>0$, the interface shape is represented by \cite{debnath1994nonlinear} $\eta(r,t) = \int_{0}^{\infty}dk\;k J_0(kr)\mathbb{H}(\eta(r,0))\cos(\sqrt{gk}t)$ where $\mathbb{H}(f(r))$ is the $0$th order Hankel transform of $f(r)$.}
		\label{fig1}
	\end{figure}
 \footnotesize
 \begin{sidewaystable} 
 \captionsetup{type=table} 
    \caption{A representative summary of literature}
\begin{tabular}{|p{0.5cm}|p{2.5cm}|p{5.5cm}|p{2.5cm}|p{2cm}|p{8cm}|}
\hline
No. & Type & Reference & Type & Visc./Invisc & Remarks\\ 
 \hline
1 & & Wu and Mei, 1967 \cite{wu1967two}  & IVP & 2D, Inviscid & Point source starting at $t=0$. A free surface is also considered.\\ 
2 & & Thomas and Stevenson, 1972 \citep{thomas1972similarity} & Sim. sol. & 2D, Viscous & Oscillating localized disturbance (sim. sol.). Experimentally verified\\ 
3 & & Debnath and Guha 1989 \citep{debnath1989cauchy} & IVP & 2D, Inviscid & Axisymmetric Cauchy-Poisson problem\\
4 & Stratified & Rollins and Debnath 1992 \citep{rollins1992cauchy} & IVP & 2D, Inviscid & Axisymmetric, rotating, Cauchy-Poisson problem\\
5 & & Voisin 1991 \citep{voisin1991internal} & IVP & 3D, Inviscid & Monochromatic and impulsive point sources. Oscillations of a sphere.\\ 
6 & & Simakov 1993 \citep{simakov1993initial} & IVP & 3D, Inviscid & Impulsive point source. Simple extended source.\\ 
7 & & Voisin 1994 \citep{voisin1994internal} & IVP & 3D , Inviscid & Moving point source: oscillating and non-oscillating\\ 
8 & & Lighthill 1996 \citep{lighthill1996internal} & IVP & 3D, Inviscid & Initial localized disturbance. Effect of viscosity.\\ 
9 & & Ghosh et. al 2000 \citep{ghosh2000waves} & IVP & 2D, Inviscid & Finite depth, inertial surface\\
10 & & Gurski et. al 2002 \citep{gurski2004slow}  & Normal mode & 3D, Viscous & Proved that a countably infinte spectrum of non-oscillatory modes exist\\
\hline
11 & & Cauchy (1827) \citep{cauchy1827theorie} \& Poisson (1818) \citep{poisson1818memoire} & IVP & 2D, Inviscid & A spectrum of modes on a non-quiescent free-surface as initial data.\\
12 & & Basset (1888) \citep{basset1888treatise} & Normal Mode & 2D, Inviscid & Finite depth\\
13 & & Lamb (1932)\citep{lamb1932} & Normal Mode & 2D, Viscous & Continuous spectrum for infinite depth\\
14 & Unstratified & Harrison (1908)  \citep{harrison1908influence}, Lamb (1932) & Normal Mode & 2D, Viscous & Bottom layer is considered as infinite depth.\\
15 & & Prosperetti (1976) \citep{prosperetti1976viscous} & IVP & 2D, Viscous & Derive an integro-differential amplitude equation for the free-surface.\\
16 & & Prosperetti (1981) \citep{prosperetti1981motion} & IVP & 3D, Viscous & Two-fluid; Derive an integro-differential amplitude equation for the interface.\\
17 & & Prosperetti \& Cortelezzi (1982) \citep{prosperetti1982small} &IVP  &3D, Viscous & Finite Depth; Derive an integro-differential amplitude equation for the free-surface for a Dirac-delta vortical disturbance.\\
18 & & Antuono \& Colagrossi (2013)  \citep{antuono2013damping} & Normal Mode & 2D, Viscous & Approximate formula for damping rates of viscous gravity waves.\\
19 & & Denner (2016) \citep{denner2016frequency} & IVP & 2D, Viscous & Small amplitude pure capillary waves on a free-surface\\
20 & & Vartdal \& Osnes (2019) \citep{vartdal2019linear}  & IVP & 2D, Viscous & Small-amplitude waves in multiple superposed viscous fluids\\
\hline
\end{tabular}
\label{tab:lit_survey}
\end{sidewaystable} 
\newpage
\normalsize
	 For a stratified fluid, numerous studies have been concerned with the initial value problem of the generation of internal waves from a point or a line source; stationary, moving, or oscillating source, dipoles; moving or oscillating cylinders and spheres; and topography and tidal flows. These are as summarised in \citet{sutherland2010internal}. For an inviscid, uniformly stratified fluid assumed to be unbounded along all directions (i.e. without a free surface), the initial value problem for a wave number $\mathbf{k}$ was solved by \citet{lighthill1996internal} using the stationary-phase method with arbitrary initial perturbations and a solenoidal distribution of initial velocities. For three-dimensional internal gravity waves, he showed that the amplitude decreases as $t^{-3/2}$ at large times. \citet{wu1967two} studied the generation of internal and surface waves by an impulsive point source (starting at $t = 0$). Using Fourier (in space) and Laplace (in time) transforms, they found steady-state surface wave solutions for large times and at large distances and the attenuation rate of internal gravity waves. \citet{voisin1991internal} studied the internal waves generated by monochromatic and impulsively started point source using a Green's function formulation. Using asymptotic theory, he studied the behavior of Green's function for the case of a non-Boussinesq point source at small and large times. Later, the same technique was applied to the case of a pulsating spherical source and to moving point sources \cite{voisin1994internal}. \citet{simakov1993initial} subsequently extended the same for various boundary conditions. The Cauchy-Poisson initial value problem in an inviscid, axisymmetric, stratified fluid with a free surface has been studied by \citet{debnath1989cauchy}. For initial free-surface perturbations, they found only the contribution of surface and internal waves at long times and large distances.  Similarly, \citet{rollins1992cauchy} found the contribution only from the surface and internal-inertial waves with the inclusion of a rotating background. Later, a similar study with a Cartesian 2D domain has been reported by \citet{ghosh2000waves} where the author used the stationary phase method to derive an equation for the behavior of the interface at large times. Although the study considers an infinite depth domain, the implementation of the no-penetration boundary condition ($w = 0$) at $z \to -\infty$ neglects the presence of a continuous spectrum. 
  
     For an unbounded, viscous, stratified liquid, a similarity solution was provided by \citet{thomas1972similarity} for oscillating localized disturbances using transformation of coordinates. These results have been validated for large distances with particle displacements from experimental measurements. When the fluid is bounded between two flat plates \cite{gurski2004slow} showed, analytically, the existence of a countable infinite set of non-oscillatory normal modes in addition to the internal wave modes. Table \ref{tab:lit_survey} summarises representative literature focused on analytical solutions to IVPs (and also normal mode analysis) for stratified and unstratified cases in the linear approximation.
 
    Our aim in the present study is a analytical and simulational investigation of the effect of vortical and free-surface perturbations on the evolution of surface and internal waves on a stratified pool of liquid, taking into account gravity as well as surface tension. By considering effects of stratification, our results extend the classical results due to \cite{lamb1932} and \cite{yih1960gravity,prosperetti1976viscous,prosperetti1982small} to vortical perturbations on a viscous, stratified liquid pool. We allow deformation of the free surface and assume that it is is not covered by a monolayer \cite{lucassen1970properties}. Consequently it is devoid of elastic or viscous properties, eliminating the possibility of ``longitudinal'' or ``transverse'' capillary waves \cite{lucassen1968longitudinal,felderhof2006dynamics}. The capillary-gravity waves which however do appear in our analysis correpond to the ones appearing at the free surface of a pure viscous liquid, possessing constant surface tension. Such waves are of mixed character and cannot be classified as being purely transverse or longitudinal. The theoretical results presented here signficantly extend the earlier results by \citet{cortelezzi1981small} as well as \citet{prosperetti1982small}, who presented the linear theory for the effect of vorticity initial conditions on a unstratified, viscous liquid layer with a free surface. The inclusion of viscosity, especially in the unstratified case, is important as this allows the temporal evolution of vorticity eigenmodes, within the linear approximation. Our analytical results may also be viewed as being complementary to those in \citet{rednikov2000rayleigh} who studied (via normal mode as well as and a weakly, nonlinear analysis), internal and dilatational surface waves due to instabilities arising from a top heated air layer, overlying a liquid layer of finite depth. Incidentally, these authors \citep{rednikov2000rayleigh} did not allow for deformation of the free surface (page $59$, section $2$ in \cite{rednikov2000rayleigh}); thus the capillary waves that they obtain are purely dilatational in contrast to the ones that we report here. The study is organised as follows: in the next section, we present analytical solutions to the IVP for a range of initial conditions on a stratified and unstratified liquid pool. The predictions are rationalised by comparing against the temporal spectrum in all cases. These predictions are then tested against numerical simulations of the Navier-Stokes equations with density stratification, conducted using the open source code Basilisk \cite{popinet2014basilisk}. We discuss these comparisons and conclude with a summary of salient observations and future scope. 

\section{Perturbations on a stratified  liquid layer: initial value problem}
\subsection{Governing equations: Boussinesq approximation}
\begin{figure}
    \centering
    \includegraphics[scale=0.3]{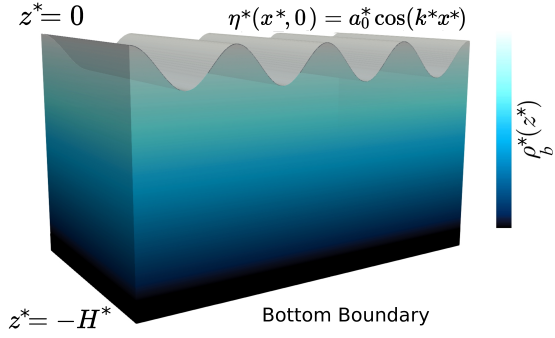}
    \caption{A schematic of the geometry. In the base state, there is quiescent fluid with a flat free-surface. The base state density profile $\rho_b^{*}(z^{*})$ varies linearly with depth.}
    \label{fig0}
\end{figure}
As depicted in fig.\ref{fig0}, consider a pool of depth $H^*$ comprising of quiescent, density-stratified fluid (uniform stratification) bounded at the top with a free surface and at the bottom by a wall. Assuming density to be a (linear) function of temperature $T$ (and independent of other thermodynamic variables like salinity and pressure), we obtain the Boussinesq set of equations \citep{vallis2017atmospheric,foster1970drag} governing variations in velocity, pressure and density inside the pool viz. 
\begin{subequations}\label{eqn1}
        \begin{align}
	&\bm{\nabla^*}\cdot\mathbf{\tilde{u}^*} = 0,\quad\quad\frac{D\mathbf{\tilde{u}^*}}{Dt^*}=  -\frac{1}{\rho_{\text{ref}}}\bm{\nabla^*}\tilde{p}^* + \frac{\tilde{\rho}^*}{\rho_{\text{ref}}}\mathbf{g} + \frac{\mu }{\rho_{\text{ref}}}\nabla^{*2} \mathbf{\tilde{u}^*} \tag{\theequation a,b}\\	
        & \frac{D\tilde{\rho}^*}{Dt^*} = \kappa \nabla^{*2} \tilde{\rho}^* \tag{\theequation c}
    \end{align}
\end{subequations}
where $\mathbf{\tilde{u}^*}  = (\tilde{u}^*,\tilde{w}^*)$, $\mathbf{g} = (0,-g)$, $\rho_{\text{ref}}$ is a reference density and $\kappa$ is the thermal/mass diffusivity. The base state whose stability is being investigated comprises of a quiescent viscous fluid with a density profile $\rho^*_b(z)$ and uniform viscosity $\mu$, with pressure field satisfying the hydrostatic balance $\dfrac{dp_b^*}{dz^*} = - \rho_b^*(z^*)g$. Note that all base state quantities are indicated with the subscript `$b$' while dimensional quantities have an asterisk. Equations \ref{eqn1}a,b and \ref{eqn1}c are supplemented with boundary conditions at the free surface ($z^* = \eta^*$) and the wall ($z^*=-H^*$). These are
\begin{subequations}
\begin{align}
	& \frac{D\eta^*}{Dt^*} = \left(\mathbf{\tilde{u}^*}\cdot\mathbf{e}_z\right)_{z^* = \eta^*} \label{4}\\
	& \left(\mathbf{t}\cdot\tilde{\underline{\underline{\sigma}}}^*\cdot\mathbf{n}\right)_{z^* = \eta^*} = 0  \label{5}\\
	& \left(\mathbf{n}\cdot\tilde{\underline{\underline{\sigma}}}^*\cdot\mathbf{n}\right)_{z^* = \eta^*} = -\tau\left(\bm{\nabla^*}\cdot\mathbf{n}\right)_{z^*=\eta^*} \label{6}\\
	& \tilde{\rho}^*(x^*,z^*=\eta^*,t^*) = \rho_b^*(\eta^*) \label{7}\\
	 & \mathbf{\tilde{u}^*}(x^*,z^*=-H^*,t^*) = 0 \label{8}\\
	 & \tilde{\rho}^*(x^*,z^*=-H^*,t^*) = \rho_b^*(-H^*) \label{9}
\end{align}
\end{subequations}
\noindent where $\mathbf{t}$ and $\mathbf{n}$ are the unit tangent and normal to the perturbed interface, $\tilde{\underline{\underline{\sigma}}}^{*}$ is the stress tensor and $\mathbf{e}_z$ is a unit normal along the z-direction.  Eqns. \ref{4}-\ref{8} are the kinematic boundary condition, the zero shear stress condition, the jump in normal stresses due to surface tension $\tau$ at the free surface, constant density condition at the interface and the zero-velocity condition at the bottom wall respectively. Decomposing all dependent variables as a sum of base and perturbation variables viz.
\begin{eqnarray}
	\mathbf{\tilde{u}^*} = \mathbf{0} + \mathbf{u}^*, \; \tilde{p}^*(\mathbf{x}^*,t^*) = p_b^*(z^*) + p^*(\mathbf{x}^*,t^*), \; \tilde{\rho}^*(\mathbf{x}^*,t^*) = \rho_b^*(z^*) + \rho^*(\mathbf{x}^*,t^*).\nonumber \label{9}
\end{eqnarray}
Substituting in eqns. \ref{eqn1}a,b,c and (\ref{4}-\ref{8}), linearising in the perturbed variables (disturbances are small compared to the base state) we obtain in component form 
\begin{subequations}
\begin{align}
        &   \frac{\partial u^*}{\partial x^*} + \frac{\partial w^*}{\partial z^*} = 0 \label{10} \\
	& \frac{\partial u^*}{\partial t^*}  = -\frac{1}{\rho_{\text{ref}}}\frac{\partial p^*}{\partial x^*} + \frac{\mu}{\rho_{\text{ref}}}\nabla^{*2}u^*  \label{11} \\
	&  \frac{\partial w^*}{\partial t^*} = -\frac{1}{\rho_{\text{ref}}}\frac{\partial p^*}{\partial z^*} - \frac{\rho^*}{\rho_{\text{ref}}}g + \frac{\mu}{\rho_{\text{ref}}}\nabla^{*2}w^* \label{12} \\
	&     \frac{\partial \rho^*}{\partial t^*} + \left(\frac{d\rho_b^*}{dz^*}\right)w^* = \kappa\nabla^{*2}\rho^* \label{13}
\end{align}
\end{subequations}
with the corresponding linearised boundary conditions obtained from using Taylor series expansion about $z^*=0$,
\begin{subequations}
\begin{align}
	&\frac{\partial\eta^*}{\partial t^*} = w^*(x^*,0,t^*) \label{14}\\
	& \frac{\partial u^*}{\partial z^*}(x^*,0,t^*) + \frac{\partial w^*}{\partial x^*}(x^*,0,t^*) = 0 \label{15}\\
	& \rho_b^*(0)g\eta^* - p^*(x^*,0,t^*) + 2\mu\frac{\partial w^*}{\partial z^*}(x^*,0,t^*) - \tau\left(\frac{\partial^2\eta^*}{(\partial x^*)^2}\right) = 0 \label{16}\\
	& \rho^*(x^*,0,t^*) = 0 \label{17}\\
	& u^*(x^*,-H^*,t^*)  = w^*(x^*,-H^*,t^*) = \rho^*(x^*,-H^*,t^*)=0 \label{18} 
\end{align}
\end{subequations}
where we have used the Newtonian constitutive relation $\tilde{\underline{\underline{\bm{\sigma}}}}^{*} = -\tilde{p}^{*}\;\underline{\underline{\mathbf{I}}} + \mu\left[\bm{\nabla}\tilde{\mathbf{u}}^{*} + \left(\bm{\nabla}\tilde{\mathbf{u}}^{*}\right)^{\intercal}\right]$. Note that we have used $p_b^*(0)=0$ in \ref{16}.

\subsection{Non-dimensional equations}
We now non-dimensionalise eqns. (\ref{10}-\ref{13}) and boundary conditions (\ref{14}-\ref{18}) using the length, velocity, time, density and pressure scales $L$, $U$, $L/U$, $\rho_{ref}$ and $\rho_{ref}U^2$ respectively to obtain (unstarred variables are non-dimensional)
\begin{subequations}
\begin{align}
 & \frac{\partial u}{\partial x} + \frac{\partial w}{\partial z} = 0 \label{19}\\
&\frac{\partial u}{\partial t} = -\frac{\partial p}{\partial x} + \frac{1}{Re}\left(\frac{\partial^2}{\partial x^2} + \frac{\partial^2}{\partial z^2}\right)u \label{20}\\
& \frac{\partial w}{\partial t} = -\frac{\partial p}{\partial z} + \frac{1}{Re}\left(\frac{\partial^2}{\partial x^2} + \frac{\partial^2}{\partial z^2}\right)w - \left(\frac{gL}{U^2}\right)\rho \label{21}\\
& \frac{\partial \rho}{\partial t} + \left(\frac{d\rho_b}{dz}\right)w = \frac{1}{Re Pr}\left(\frac{\partial^2}{\partial x^2} + \frac{\partial^2}{\partial z^2}\right)\rho \label{22}
\end{align}
\end{subequations}
and the boundary conditions
\begin{subequations}
 \begin{align}
&\frac{\partial\eta}{\partial t} = w(x,0,t) \label{23}\\
& \frac{\partial u}{\partial z}(x,0,t)+\frac{\partial w}{\partial x}(x,0,t) = 0 \label{24}\\
& \rho_b(0)\eta - \left(\frac{U^2}{gL}\right)p(x,0,t) + \frac{2}{Re}\left(\frac{U^2}{gL}\right)\frac{\partial w}{\partial z}(x,0,t) - \frac{1}{Bo}\left(\frac{\partial^2\eta}{\partial x^2}\right) = 0 \label{25}\\
& \rho(x,0,t) = 0 \label{26}\\
& u(x,-H,t)  = w(x,-H,t) = \rho(x,-H,t) =  0 \label{27}
\end{align}
\end{subequations}
The non-dimensional numbers are the Reynolds number $Re \equiv \dfrac{\rho_{\text{ref}}L U}{\mu}$, the Bond number $Bo \equiv \dfrac{\rho_{\text{ref}} g L^2}{\tau}$, the Richardson number $Ri \equiv \dfrac{-\frac{g}{\rho_{\text{ref}}}\left(\frac{d\rho_b^*}{dz^*}\right)L^2}{U^2}$, the Prandtl number $Pr \equiv \dfrac{\mu}{\rho_{\text{ref}}\;\kappa}$ and the ratio $H \equiv H^{*}/L$. We choose the length and the velocity scale $L$ and $U$ to be $1/k$ and $\sqrt{g/k}$ respectively viz. the length-scale and phase speed of a surface gravity wave. Therefore, the ratio $\frac{gL}{U^2}$ (can be defined as inverse Froude number squared) which appears in equation \eqref{25} turns out to be unity. Also, the Richardson number $Ri$ here becomes $N^2/\omega^2$, i.e. buoyancy frequency non-dimensionalized with deep-water surface wave frequency.

All equations and boundary conditions are Fourier and Laplace transformed in $x$ and $t$ respectively. As $k$ is used non-dimensionalising lengths, the Fourier-Laplace transformed variables (indicated with a overbar) are thus independent of $k$. Eliminating pressure  $\bar{p}(z,s)$ through cross differentiation and defining the perturbation stream-function $\bar{\psi}(z,s)$, we obtain after some algebra the following equation.
\begin{eqnarray}
&\Big[(D^2 - 1)-sRePr\Big]\Big[(D^2-1) - sRe\Big](D^2-1)\bar{\psi}(z,s) - Ri\;Re^2Pr\bar{\psi}(z,s)\nonumber \\ &= \textbf{i}Re^2Pr\hat{\rho}(z,t=0)+Re\Big[(D^2-1)-sRePr\Big]\hat{\omega}(z,t=0)\label{29}
\end{eqnarray}
where $D \equiv \partial/\partial z$, is the differentiation operator and where $\mathbf{i} \equiv \sqrt{-1}$. Eqn. \ref{29} is the initial-boundary-value problem that we need to solve in order to determine the evolution of the perturbation streamfunction $\bar{\psi}(s,z)$. This is supplemented by boundary conditions which in the Fourier-Laplace domain are:
\begin{subequations}
\begin{align}
&(D^2+1)\bar{\psi}(s,z=0) = 0\label{34a}\\
&s\Big((D^2 - 3) - s Re\Big)D\bar{\psi}(s,z=0) - Re(1+Bo^{-1})\bar{\psi}(s,z=0)= Re(1+Bo^{-1})\textbf{i}\hat{\eta}(t=0) - s Re D\hat{\psi}(t=0,z=0)\label{35a}\\
& \Big[s-\frac{1}{Re}\big(D^2 - 1\big)\Big]\big(D^2-1\big)\bar{\psi}(s,z=0) = -\hat{\omega}(z=0,t=0)\label{36a} \\
&\bar{\psi}(s,-H) = 0,\hspace{0.3cm}D\psi(s,-H) = 0, \hspace{0.3cm}  \Big[s-\frac{1}{Re}\big(D^2 - 1\big)\Big]\big(D^2 - 1 \big)\bar{\psi}(s,-H) = -\hat{\omega}(z = -H,t=0)\label{38-0}\tag{8d,e,f}
\end{align}
\end{subequations}
We remind the reader that equation \ref{34a} represents the zero shear stress condition at the linearised free-surface $z=0$. Equation \ref{35a} is the normal stress boundary condition at $z=0$, also taking into account the kinematic boundary condition. Eqns. \ref{36a} and 8f enforce the vanishing of density perturbations at the free surface and the bottom-wall (when depth is finite) respectively. Eqns. 8d and 8e are the no-penetration and no-slip conditions at the wall. The stratified and the unstratified ($Ri=0,Pr \rightarrow \infty$ and $\hat{\rho}=0$) limits differ in the order of the equation governing $\bar{\psi}(s,z)$. In the former case, eqn. \ref{29} is sixth order (see eqn. $104$ in chapter II in \cite{chandrasekharhydrodynamic} for a similar sixth order equation) while for the latter, it is a fourth order equation.  Consequently, the boundary conditions for solving eqn. \ref{29} are chosen accordingly e.g. for the infinite-depth, stratified layer, we choose to impose conditions \ref{34a}, \ref{35a} and \ref{36a} at $z = 0$ alongwith $\bar{\psi} \rightarrow \text{finite}$ as $z\rightarrow-\infty$. On the other hand, for the unstratified, finite depth layer, we implement boundary conditions \ref{34a} and \ref{35a} at $z=0$ and 8d and 8e at $z=-H$ and so on.\\

In solving equation \ref{29} for the stratified case, the homogenous part to the solution maybe chosen to be of the form the form $\sim C_i e^{\lambda_i z}$ and using the method of variation of parameters, we can write the general solution as
\begin{align}
\bar{\psi}(s,z) = \sum_{i=1}^6\exp{(\lambda_iz)}\Bigg[C_i(s) + \int^z \frac{Adj(\mathscr{W})_{i,6}(s,z')\mathscr{R}(s,z')}{|\mathscr{W}(s,z')|}dz'\Bigg]\label{30}
\end{align}
Here, $\mathscr{R}(s,z)$ is the RHS of equation \ref{29}, $|\mathscr{W}(s,z)|$ is the sixth-order Wronskian (supplementary material) constructed from the homogeneous part to the solution while Adj$(\mathscr{W})$ represents the adjugate of the Wronskian matrix, the subscripts indicating the elements of this matrix. It is shown in the supplementary material that expression \ref{30} maybe rewritten as (\citet{riley2006mathematical}, page 510):
\begin{equation}
\bar{\psi}(s,z) = \sum_{i=1}^6\exp{(\lambda_iz)}\Bigg[C_i(s) + \mathscr{E}_{i}(s)I_i(s,z)\Bigg],\; \text{with}\;\mathscr{E}_i(s) \equiv \frac{1}{\displaystyle \prod_{j=1,j\neq i}^{6} (\lambda_i - \lambda_j)},\;I_i(s,z) \equiv \int \exp(-\lambda_i z)\mathscr{R}(s,z)dz.\label{31}
\end{equation}
In the following sub-section(s), we determine expressions for $C_i(s)$ subject to various initial conditions for specific limits like a stratified or an unstratified layer, of finite or infinite depth. These are then employed to obtain explicit expression(s) for the perturbation $\hat{\psi}(z,t)$ post Laplace inversion.
\subsection{Linearised analytical predictions}\label{sec:C}
First consider the case when the liquid layer is stratified and of finite depth. By substituting equation \ref{31} in the boundary conditions given by equations \ref{34a}-\ref{38-0}, we can obtain a linear set of equations for $C_i(s)$ ($i=1,2\ldots 6$). These maybe expressed symbolically as,
\begin{equation}
    \sum_{i=1}^{6}\mathscr{D}_{ji}(s)C_i(s) = \mathscr{R}_j(s), \quad j=1,2\ldots 6\label{33}
\end{equation}
where, $\mathscr{D}_{ji}(s)$ is the coefficient matrix (i.e. the matrix which yields the dispersion relation if one were to do normal mode analysis). $C_i(s)$ are the coefficients in equation \ref{31} and $\mathscr{R}_j(s)$ is a column matrix of remaining terms in the boundary conditions \ref{34a}-\ref{38-0}, taken to the right hand side. Eqns. \ref{33} may be solved for the $C_i(s)$ to obtain,
\begin{equation}
    C_i(s) = \frac{1}{|\mathscr{D}(s)|}\sum_{j=1}^{6}Adj(\mathscr{D}(s))_{ij}\mathscr{R}_j(s). \label{34}
\end{equation}
Substituting expression \ref{34} into \ref{31} and transferring from the Laplace to time domain, we obtain formally
\begin{align}
\hat{\psi}(t,z) = \dfrac{1}{2\pi\mathbf{i}}\int_{\gamma-\mathbf{i}\infty}^{\gamma+\mathbf{i}\infty}\exp{(st)}ds\sum_{i=1}^6\exp{(\lambda_i(s)z)}\Bigg[\sum_{j=1}^6\frac{1}{|\mathscr{D}(s)|}Adj(\mathscr{D}(s))_{ij}\mathscr{R}_j(s) + \mathscr{E}_{i}(s)I_i(s,z)\Bigg].\label{35}
\end{align}
Here onwards, we obtain explicit expressions for the perturbation stream-function $\hat{\psi}(z,t)$ using eqn. \ref{35} by choosing explicit initial conditions and using Cauchy residue theorem to evaluate the corresponding Bromwich contour integral. One can similarly obtain corresponding expressions for the free-surface perturbation $\hat{\eta}(t)$ and the density perturbation $\hat{\rho}(z,t)$. In order to  see that eqn. \ref{35} depends explicitly on initial conditions, note that it has $I_i(s,z)$ on the right hand side which in turn depends on $\mathscr{R}(s,z)$ see expression \ref{31}, the latter explicitly depending on the initial density and vorticity perturbation viz. $\hat{\rho}(z,t=0)$ and $\hat{\omega}(z,t=0)$ respectively. In addition, $\mathscr{R}_j(s)$ in eqn. \ref{35} also contain initial conditions. In the following sub-section, we choose these initial conditions corresponding to vorticity and free-surface perturbations for which \ref{35} yields exact solutions. The next two sub-sections presents these analytical solutions in the context of a stratified layer and an unstratified layer respectively while comparing these to the discrete temporal spectrum of the problem.
\subsubsection{\textbf{Stratified layer ( $\text{Ri}\neq0$) of finite depth}: \textbf{vorticity and free-surface perturbations}}
\begin{figure}[h]
    \centering
    \subfloat[Viscous spectrum]{\includegraphics[scale=0.42]{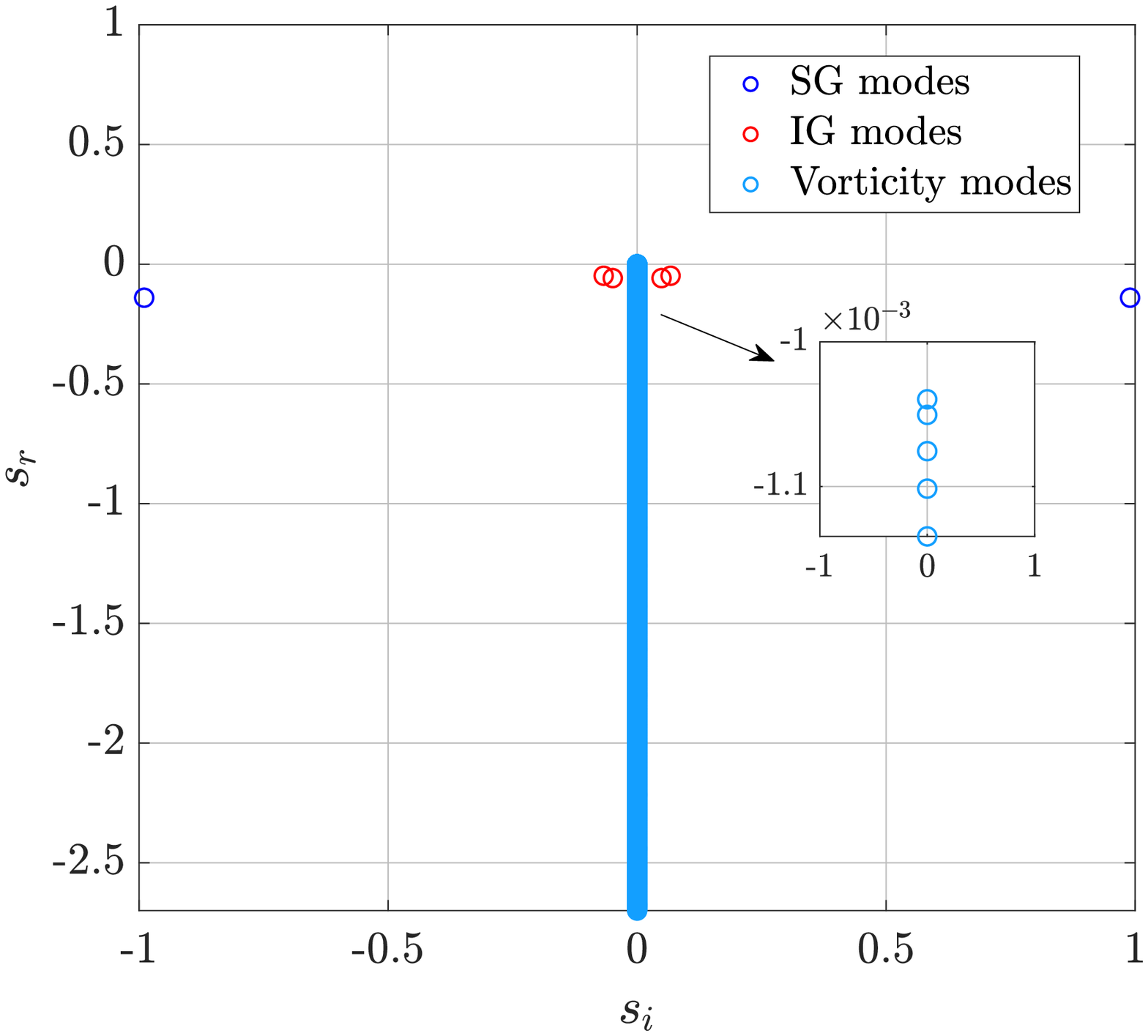}}
    \subfloat[Inviscid spectrum]{\includegraphics[scale=0.42]{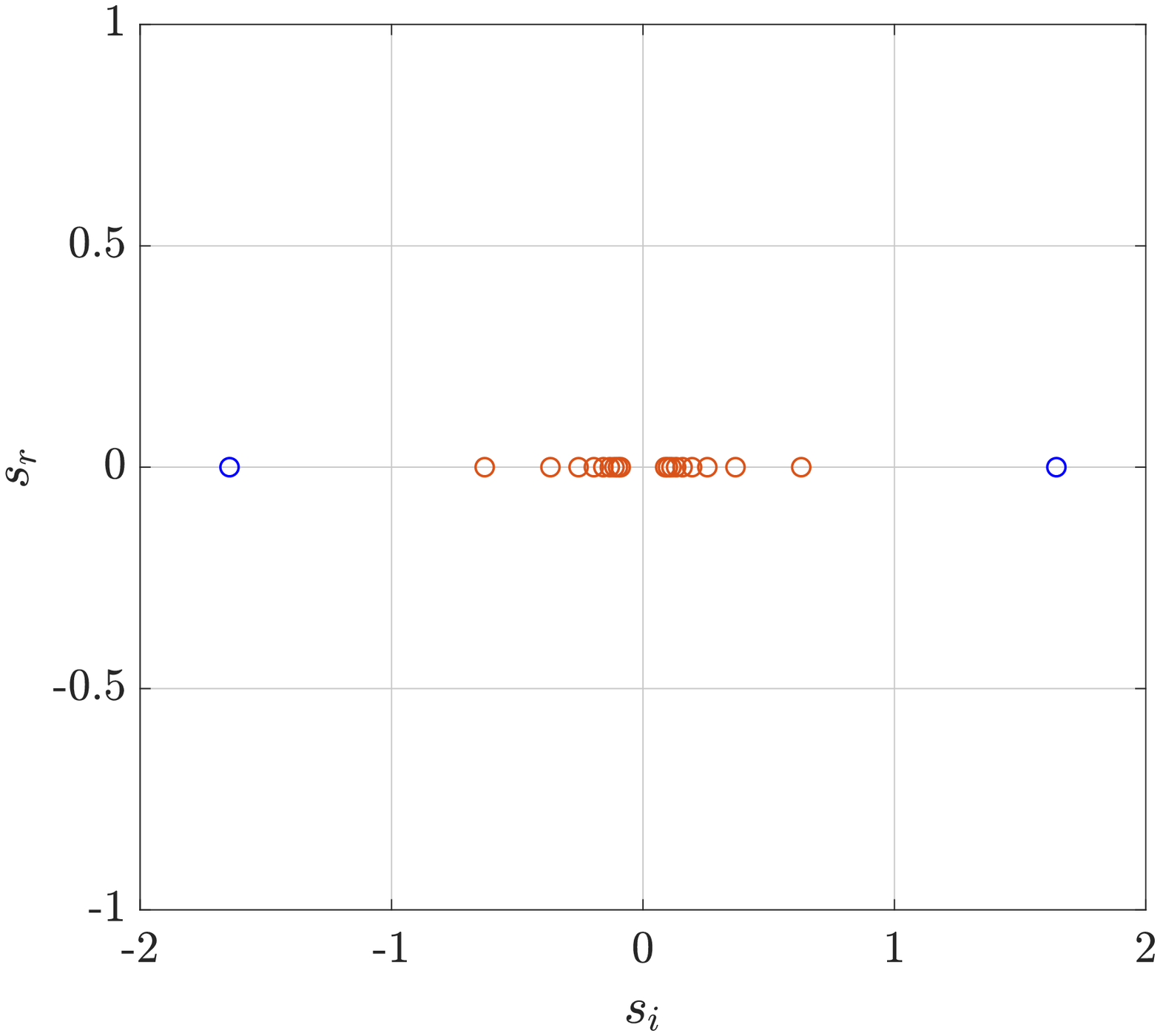}}
    \caption{Temporal eigenvalue spectrum for a stratified, pool of finite depth: Panel (a) The viscous discrete spectrum modes obtained by numerically solving the determinant eqn. $\mathcal{D}(s) = 0$ (see eqn. $33$). Note that the imaginary axis $s_i$ has been plotted horizontally instead of vertically. Deep blue circles correspond to under-damped surface gravity (SG) modes. Circles in red represent damped internal gravity (IG) modes. Circles in sky-blue (also in the inset) correspond to damped vorticity modes. We plot the corresponding inviscid spectrum in panel (b). In the inviscid limit, all the vorticity modes of panel (a) (sky blue circles) collapse to the origin in panel (b). In this limit, the IG modes are countably infinite \citep{yih1960gravity}, only some of which are depicted in panel (b). In the viscous case in panel (a) we could numerically find only four IG modes. The parameters for the plots correspond to case $c$ in table \ref{tab_dns}.}
    \label{fig2}
\end{figure}

The algebra for specific initial conditions is lengthy but straightforward and the reader is referred to the accompanying supplementary material where exact expressions are derived. We focus on the specific case of vortical initial perturbation on a stratified pool of finite depth here.  Assuming an initial localised, vorticity perturbation of the form $\hat{\omega}(z,t=0) = \Omega  \exp\left(-\dfrac{(z-z_d)^2}{d^2}\right)$, where $z_d \in ( -H, 0)$ and no initial density ($\hat{\rho}(z,t=0)=0$) or free-surface perturbations ($\hat{\eta}(t=0)=0$), we wish to obtain the expression of $\hat{\psi}(t,z)$ for this initial condition from eqn. \ref{35}. Equation \ref{35} may be converted into a contour integral on the complex $s$ plane and the integral evaluated using Cauchy residue theorem. The final expression is then written as the summation of the residues at all the zeros of the dispersion matrix $(|\mathscr{D}(s)|)$ as,
\begin{equation}
\hat{\psi}(z,t) = \sum_{n = 1}^\infty\frac{1}{|\mathscr{D}(s_n)|'}\exp{(s_nt)} \sum_{i=1}^6\exp{(\lambda_i(s_n)z)}\Bigg[\sum_{j=1}^6Adj(\mathscr{D}(s_n))_{ij}\mathscr{R}_j(s_n)\Bigg].\label{36}
\end{equation}
It should be noted that no poles are found for the second term on the right hand side of expression \ref{35} and thus this term does not appear in eqn. \ref{36}.  Similarly the expression for the perturbed free-surface is
\begin{align}
\frac{\hat{\eta}(t)}{\hat{\eta}(0)} = 1&-\frac{\textbf{i}}{\hat{\eta}(0)}\sum_{n = 1}^\infty \exp{(s_nt)}\sum_{i=1}^6\Bigg[\frac{1}{s_n|\mathscr{D}(s_n)|'}\sum_{j=1}^6Adj(\mathscr{D}(s_n))_{ij}\mathscr{R}_j(s_n)\Bigg]- \frac{\textbf{i}}{\hat{\eta}(0)}\sum_{i=1}^6\Bigg[\frac{1}{|\mathscr{D}(0)|}\sum_{j=1}^6Adj(\mathscr{D}(0))_{ij}\mathscr{R}_j(0)\Bigg].\nonumber\\\label{37}
\end{align}
A similar equation for the density evolution can also be obtained, although this is not provided here (see supplementary material). For understanding the temporal evolutions predicted in eqns. \ref{36} and \ref{37} qualitatively, it is useful to plot the temporal eigenvalue spectrum for this case. Plotted in fig. \ref{fig2}, panel (a) are the roots of $\mathscr{D}(s)$ (eqn. \ref{34}) calculated numerically in Mathematica 
\citep{Mathematica} on the complex $s$ plane (note that the vertical axis is real while the horizontal is imaginary). The physical parameters for this plot are chosen corresponding to case $(c)$ in table \ref{tab_dns}. As seen, there are two underdamped surface gravity (SG) modes (deep-blue circles) arising from viscous corrections to the inviscid, irrotational capillary-gravity waves with dispersion relation $\omega^2 = \left(gk + \frac{\tau k^3}{\rho}\right)\tanh(kH)$. Note that despite the inclusion of surface tension, we refer to the surface modes as SG (surface-gravity) modes rather than capillary-gravity modes. In addition, in fig. \ref{fig2}a there is a countably infinite set of purely damped modes (sky blue circles, also see inset) which we term vorticity modes as they do not owe their origin to the presence of the free surface. Also shown in this figure are the internal-gravity modes (red circles). The infinite summations in eqns. \ref{36} and \ref{37} reflect the contributions from these discrete modes shown in fig. \ref{fig2}, panel a. For reference, panel (b) in the same figure provides the spectrum for the inviscid case. Here we observe the absence of the vorticity modes, while the IG modes become countably infinite in agreement with the inviscid predictions of \cite{yih1960gravity}.
\subsubsection*{\textbf{Stratified, deep water limit}: $\bm{H\rightarrow\infty}$}
Here we examine the deep water limit of a stratified, pool of viscous liquid. In this limit the governing equation \ref{29} remains unchanged while the free surface boundary conditions in eqns. \ref{34a}, \ref{35a} and \ref{36a} also remain valid in this limit. However boundary conditions \ref{38-0} need to be replaced by suitable finiteness conditions as $z\rightarrow{-\infty}$. It is shown in the supplementary material in this limit $H\rightarrow\infty$, the perturbation stream-function may be written in the same form as eqn. \ref{36}, although now the $C_i(s)$ have different expressions compared to the finite depth case in eqn. \ref{34}. It is of interest to ask, what are the features of  the infinite-depth counterpart of the spectrum in fig. \ref{fig4b}.  We note that this was studied earlier by Yih \cite{yih1960gravity}, demonstrating that in the inviscid, stratified, infinite depth limit, if there is no density discontinuity in the domain, there is no discrete spectrum and only a continuous one (see discussion in \cite{yih1960gravity}, page $496$ second paragraph.)  
\subsubsection{\textbf{Unstratified layer ( $\textbf{Ri}= 0$) \textbf{of finite depth}: \textbf{vorticity} \& free-surface perturbations}}\label{sec:E}
\begin{figure}[h]
    \centering
    \subfloat[Viscous spectrum]{\includegraphics[scale=0.42]{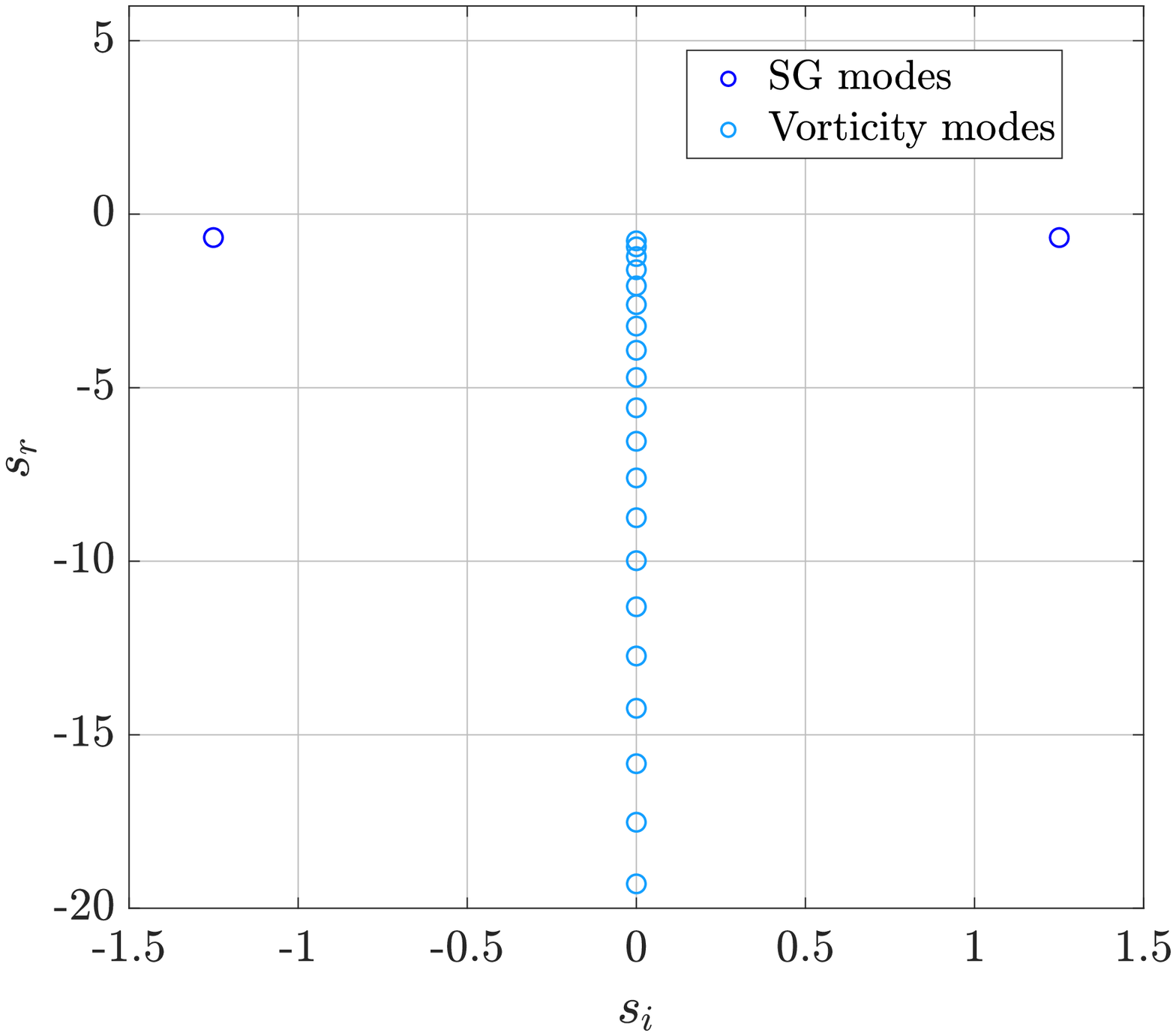}\label{fig4a}}
    \subfloat[Inviscid spectrum]{\includegraphics[scale=0.42]{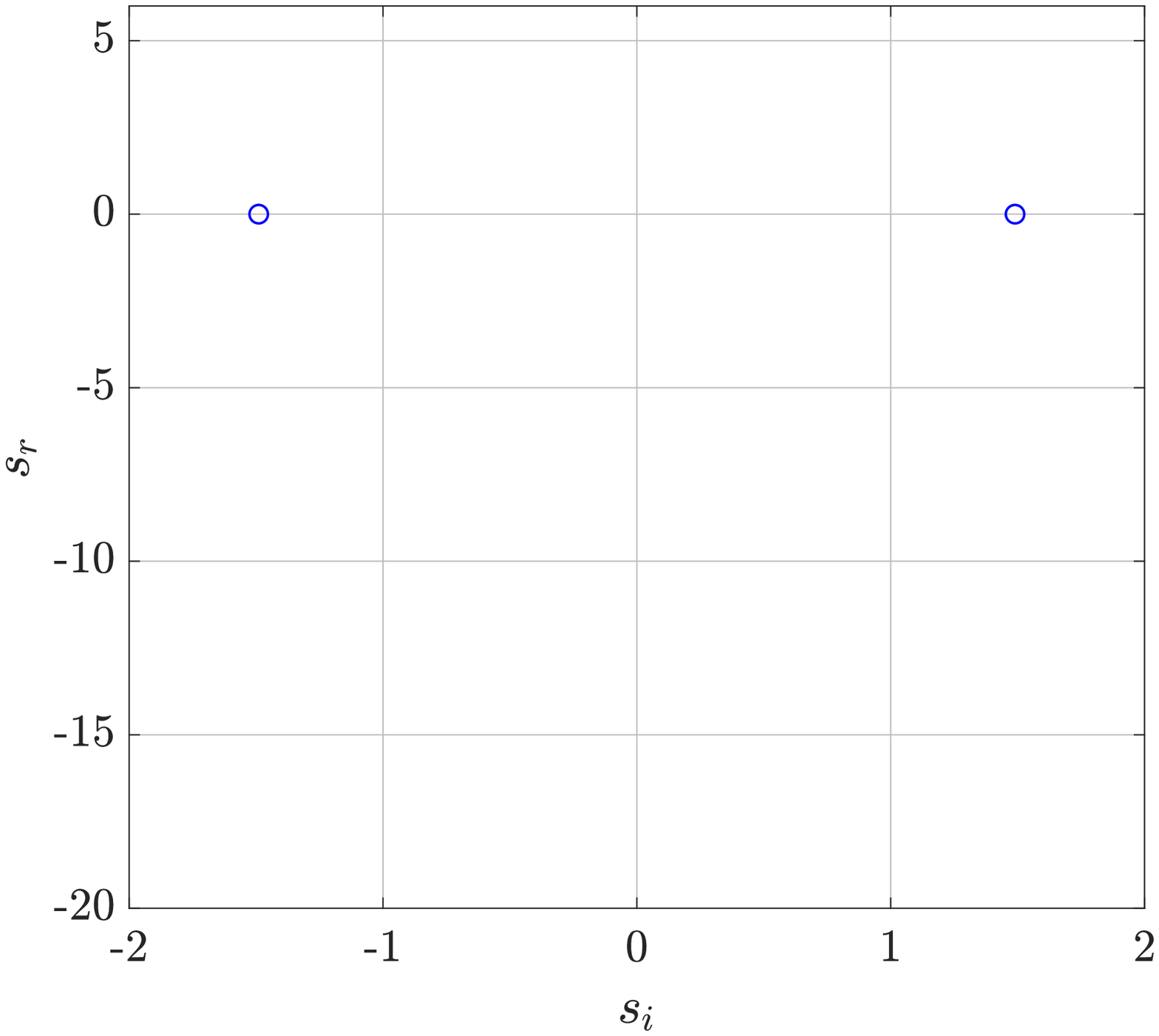}\label{fig4b}}
    \caption{The discrete spectrum modes obtained by numerically solving $\mathcal{D}(s) = 0$ for an unstratified layer corresponding to Case (a) in table \ref{tab_dns}. The symbols indicate roots of $\mathscr{D}(s)$ in the complex Laplace domain $s$. (a) The two modes indicated as deep blue circles corresponded to under-damped capillary-gravity modes while those in sky-blue correspond to purely damped, vorticity modes. (b) In the invscid limit, the deep-blue circles correspond to two roots from the capillary-gravity dispersion relation $\omega^2 = \left(gk + \frac{\tau k^3}{\rho}\right)\tanh(kH)$}
    \label{fig4}
\end{figure}

Having discussed the spectrum for the stratified case, we now turn to the unstratified limit where the Richardson number is $\text{Ri}= 0$. We recall that for the unstratified case, eqn. \ref{29} is a fourth order equation. Analogous to the earlier section, the expression for the perturbation stream-function after Laplace inversion viz. $\hat{\psi}(z,t)$ is obtained as (note the sum over four roots rather than six in the unstratified case here)  
\begin{align}
\hat{\psi}(z,t) = \dfrac{1}{2\pi\mathbf{i}}\int_{\gamma-\textbf{i}\infty}^{\gamma+\textbf{i}\infty}\exp{(st)}ds\sum_{i=1}^4\exp{(\lambda_i(s)z)}\Bigg[\sum_{j=1}^4\frac{1}{|\mathscr{D}(s)|}Adj(\mathscr{D}(s))_{ij}\mathscr{R}_j(s) + \mathscr{F}_{i}(s)I_i(s,z)\Bigg].\label{38}
\end{align}
Like earlier, for a localised vortical initial condition (with no free-surface perturbation) $\hat{\omega}(z,t=0) = \Omega  \exp(-(z-z_d)^2/d^2)$, where $z_d \in ( -H, 0)$, we use Cauchy residue theorem to evaluate the Bromwich contour integral for eqn. \ref{38} to obtain the perturbation stream-function. This is
\begin{equation}
\hat{\psi}(z,t) = \sum_{n = 1}^\infty \sum_{i=1}^4\exp{(\lambda_i(s_n)z)}\Bigg[\frac{1}{|\mathscr{D}(s_n)|'}\sum_{j=1}^4Adj(\mathscr{D}(s_n))_{ij}\mathscr{R}_j(s_n)\Bigg]\exp{(s_nt)}.\label{39}
\end{equation}
On the other hand, for a free-surface initial perturbation $\hat{\eta}(0)$ and zero vorticity initial condition, the free-surface at a later time $t$ is predicted to be
\begin{align}
\frac{\hat{\eta}(t)}{\hat{\eta}(0)} = 1-&\frac{\textbf{i}}{\hat{\eta}(0)}\sum_{n = 1}^\infty \sum_{i=1}^4\Bigg[\frac{1}{s|\mathscr{D}(s_n)|'}Adj(\mathscr{D}(s_n))_{ij}\mathscr{R}_j(s_n)\Bigg]\exp{(s_nt)} - \frac{\textbf{i}}{\hat{\eta}(0)}\sum_{i=1}^4\Bigg[\frac{1}{|\mathscr{D}(0)|}Adj(\mathscr{D}(0))_{ij}\mathscr{R}_j(0)\Bigg]\label{40}
\end{align}
Here $n$ runs over all the (simple) roots of $\mathscr{D}(s)$ and a prime indicates differentiation. In order the understand these analytical predictions, it is once again useful to visualise the temporal spectrum from the dispersion relation $\mathscr{D}(s)=0$ on the complex $s$ plane. This is shown for the present case (unstratified, pool of finite depth) in fig. \ref{fig4a}. The two deep blue circles represent the surface-gravity modes which like earlier, are shifted slightly off the imaginary axis due to to viscous damping. The light-blue circles once again represent the purely damped, vorticity modes. The inviscid limit of this spectrum is depicted in fig. \ref{fig4b} where there are only two deep blue circles while the the vorticity modes collapse to the origin as discussed below.

In order to understand the vorticity modes slightly better, we note that in the absence of background flow and in the unstratified, linearised limit, the transport of perturbation vorticity occurs purely diffusively. This is seen from the diffusion equation governing the equation of perturbation vorticity: $\frac{\partial }{\partial t}\left(\bm{\nabla}\times \bm{\hat{u}}\right) = -\nu\bm{\nabla}^2\left(\bm{\nabla}\times \bm{\hat{u}}\right)$. As indicated by the viscous spectrum in fig. \ref{fig4a}, there are a countably infinite set of purely damped, vorticity modes which play an important role in this diffusive process. When we however take the inviscid ($\nu\rightarrow 0$) limit, the perturbation vorticity equation becomes trivial predicting $\frac{\partial }{\partial t}\left(\bm{\nabla}\times \bm{\hat{u}}\right)=0$. This implies that these vorticity modes (sky blue circles) in figure \ref{fig4a}, all have zero frequency in the inviscid limit explaining their disappearance in fig. \ref{fig4b}. The physical implication is that any initial distribution of perturbation vorticity stays frozen in time in the inviscid limit and can evolve only when viscosity is included in the model. It is known from prior studies \cite{farsoiya2017axisymmetric}, that for zero vorticity initial condition accompanied by  a free-surface deformation i.e. $\hat{\eta}(t=0)$ and for small value of liquid viscosity, the contribution from these vorticity modes is numerically quite small. In contrast, our choice of a localised vorticity initial perturbation here viz. $\hat{\omega}(z,t=0) = \Omega  \exp\left(-\dfrac{(z-z_d)^2}{d^2}\right)$ will be made in such a way that this perturbation will have a relatively large projection on the vorticity modes of fig. \ref{fig4a}. This prediction will be numerically tested in the next section where in addition, the cases of free-surface initial perturbations will also be considered.

 \subsubsection*{\textbf{Unstratified, deep water limit}: $\bm{H\rightarrow\infty}$}
Analogous to the earlier section, the unstratified, viscous, deep water limit is also interesting. Here prior studies have identified the presence of a viscous continuous spectrum \cite{lamb1932, prosperetti1976viscous} in addition to the discrete one. Notably, this continuous spectrum comprises the vorticity modes which were also present in the finite depth case, but constituted a discrete spectrum there. For the case of free-surface perturbations only (zero initial vorticity perturbation), it was shown by \cite{prosperetti1976viscous} (see eqn. $33$ in \cite{prosperetti1976viscous} with $u_0=0$) that the presence of the continuous spectrum may be understood through the multi-valuedness of the dispersion relation. Here in dimensional (Laplace) variables, the point $s=-\nu k^2$ turns out to be branch-point of the viscous, infinite depth dispersion relation necessitating a branch-cut in the Laplace domain, necessary for contour integration. It was already remarked upon by Lamb \cite{lamb1932} (see the discussion around eqn. $26$ in \cite{lamb1932}, section $349$) that in the infinite depth limit, the continuous spectrum (CS) modes do not decay with depth but instead posses a sinusoidal, oscillatory character persisting upto $z\rightarrow -\infty$. A general vortical initial condition may have large projection on these continuous spectrum modes depending on the depth where the initial vorticity distribution is concentrated. We will return to this point when we compare our simulations with analytical predictions in the next section.
\section{Numerical Simulations}
In this section, we briefly describe the numerical simulations of a density-stratified, viscous liquid layer with a free-surface. As discussed earlier, we impose small amplitude, initial vortical and free surface perturbations, tracking their spatio-temporal evolution. The simulations are carried out using the open-source code Basilisk \citep{popinet2014basilisk}. This code numerically solves the Navier-Stokes equations and has been benchmarked quite extensively against a wide range of interfacial flow problems e.g. \cite{farsoiya2023direct, sanjay2022taylor,kayal2022dimples,singh2019test,basak2021jetting}. Basilisk is used to solve the following equations
\begin{eqnarray}
&& \bm{\nabla}\cdot\mathbf{u}=0,\quad\frac{\partial\mathbf{u}}{\partial t}+\bm{\nabla}\cdot(\mathbf{u}\otimes\mathbf{u}) = 
-\frac{1}{\rho_{\mathrm{ref}}}\bm{\nabla} p + \frac{\mu}{\rho_{\mathrm{ref}}}\nabla^2 \mathbf{u} + \mathbf{F}_b  \label{eq41}
 + \frac{T}{\rho_{\mathrm{ref}}} \kappa \delta_s \mathbf{n}\\
&&\frac{\partial f}{\partial t} + \bm{\nabla}\cdot(f \mathbf{u}) = 0 \label{eq42}
\end{eqnarray}
In equations \ref{eq41} and \ref{eq42}, $\mathbf{u}$, $p$ and $f$ are the velocity, pressure and volume fraction fields respectively and $\mathbf{F}_b$ represents a body force related to buoyancy (see below). $T$ is the coefficient of surface tension, $\kappa$ is the curvature associated with a particular point at the interface with $\mathbf{n}$ as the unit normal vector at that point, while $\rho_{\mathrm{ref}}$ and $\mu$ denote the reference density and dynamic viscosity of the fluid in the computational domain given by the expressions
\begin{eqnarray}
&& \rho_{\mathrm{ref}} = f\rho_{\mathrm{ref}}^{(\mathcal{L})} + (1 - f)\rho^{(\mathcal{G})}\\
&& \mu = f\mu^{(\mathcal{L})} + (1-f)\mu^{(\mathcal{G})}
\end{eqnarray}
where the superscripts $\mathcal{L}$ and $\mathcal{G}$ refer to values in the liquid and gas phases respectively. $f$ is unity in the liquid and zero in the gas phase and cells with $0<f<1$ contain the interface (approximated as a free-surface in the theory). In order to incorporate density stratification in the liquid, a body force term $\mathbf{F}_b$ is included in the momentum equation \ref{eq41} incorporating the Boussinesq approximation viz.
\begin{eqnarray}
&& \mathbf{F}_b = b\mathbf{\hat e}_z = \left[f\left(-\frac{\rho}{\rho_{\mathrm{ref}}}\right)g + (1 - f)(-g)\right]\mathbf{\hat e}_z.
\end{eqnarray}
The density field $\rho$ is updated at each time step by numerically solving a tracer diffusion equation following \cite{BasiliskRepoAntoon},
\begin{eqnarray}
&& \frac{\partial \rho}{\partial t} + \bm{\nabla}\cdot(\rho \mathbf{u}) = \nu_{T} \nabla^2 \rho
\end{eqnarray}
where $\nu_{T}$ is the thermal diffusivity of the fluid.  As the code utilizes the one fluid formulation of the VoF algorithm, it solves for both the gas and liquid phases necessarily. We however ignore the gas phase in our analytical calculations. In order to facilitate comparison with theory, we maintain high density ratio value between the liquid phase and the gas phase (i.e. $\rho_{\mathrm{ref}}^{(\mathcal{L})}/\rho^{(\mathcal{G})} = 1000$) in the numerical simulations. The boundary conditions imposed on the field variables at the four walls of the domain box are elucidated in the Table II.

\section{Results and comparison with theory}
In this section, we compare the analytical results from section 2 with the results from direct numerical simulations (DNS). In the first subsection, we present the results of an unstratified liquid layer perturbed at the interface, without and with initial vorticity (cases a and b in Table I respectively). In the next subsection, we discuss simulations of a stratified liquid layer (without interface perturbation) with initial perturbation vorticity at two different values of the Brunt-Vaisala frequency (refer cases c and d in Table I respectively). Table 2 describes the boundary conditions used for different parameters in DNS calculations. The thermodynamic and simulation parameters in table \ref{tab_dns} have been chosen to correspond to high density ratio like that of air-water. This is important for the dynamic effect of air to be negligible in the simulations, allowing us to compare with analytical predictions for a stress-free interface (free-surface). However, the dynamic viscosity of the liquid has been deliberately chosen to be signficantly higher than water (i.e. we choose $\mu^{\mathcal{L}}=2\;\text{dyn-s/cm}^{2}$ compared to $10^{-3}$ in same units for water). This is partly motivated by our observations in \cite{farsoiya2017axisymmetric} that increasing the viscosity of the liquid, increases the numerical contribution from the vorticity modes. The wavelength $\lambda = \frac{2\pi}{k}$ has been chosen small, $1$ cm and $5$ cms respectively for cases (a), (b) and (c), (d) respectively while the surface tension is chosen to be smaller than that for air-water. Gravity is set to $g=981\;\text{cm/s}^2$ for all simulations. For our choice of parameters, the capillary length scale is $\approx 1.7$ mm. 
\begin{table}
\footnotesize
	\centering
	\caption{Simulation parameters (in C.G.S. units).}
	\label{tab_dns}
	\begin{tabular}{cc|ccccccccc|cccc|cccc|c}
	    \hline
		Case & Grid & $H^{*}$ & $k$  & $N$ & $\tau$ & $\mu^{(\mathcal{L})}$ & $\mu^{(\mathcal{G})}$ & $\rho_{ref}^{(\mathcal{L})}$ & $\rho^{(\mathcal{G})}$ & $\nu_{T}$ & $Ri$ & $Re$ & $Pr$ & $Bo$ & $a_0$ & $\Omega$ & $z_d$ & $d$ & Initial perturbation(s)\\
    	\hline
		(a) & 1024 $\times$ 1024 & 2 & $2\pi$ & -  & 30 & 2 & 0.000185 & 1 & 0.001 & - & - & 0.9943 & - & 0.8283 & 0.0079 & 0 & - & - & Free suface only\\
		(b) & 1024 $\times$ 1024 & 2 & $2\pi$ & - & 30 & 2 & 0.000185 & 1 & 0.001 & - & - & 0.9943 & - & 0.8283 & 0.0079 & 0.05 & -1.3 & 0.1 & Free-surface and Vorticity\\
		(c) & 1024 $\times$ 1024 & 10 & $\frac{2\pi}{5}$ & 3 & 30 & 2 & 0.000185 & 1 & 0.001 & 0.001 & 0.0073 & 11.1171 & 2000 & 20.7075  & 0 & 0.0079 & -5 & 1 & Vorticity only\\
		(d) & 1024 $\times$ 1024 & 10 & $\frac{2\pi}{5}$ & 10 & 30 & 2 & 0.000185 & 1 & 0.001 & 0.001 & 0.0811 & 11.1171 & 2000 & 20.7075  & 0 & 0.0079 & -5 & 1 & Vorticity only\\
		\hline
	\end{tabular}
 \\
	\caption{Boundary conditions in DNS.}
	\begin{tabular}{cccccc}
        \hline
        Wall & \multicolumn{2}{c}{$\textbf{u}$} & $p$ & $f$ & $b$ \\ 
        \hline 
        \multicolumn{1}{c}{} & ${u}_x$ & ${u}_z$ & \multicolumn{1}{c}{} & \multicolumn{1}{c}{} \\ \hline
        Top  & N & D  & N &  N  & $-g$\\ 
        Bottom & D & D  & N &  N  & $-g -N^2 H$\\ 
        Left  & D & N  & N &  N  &  N  \\ 
        Right  & D & N  & N &  N  & N  \\ 
        \hline
        \multicolumn{3}{c}{D: Dirichlet} & \multicolumn{2}{c}{N: Neumann}\\ 
        \hline
    \end{tabular}
\normalsize
\end{table}
\subsection*{Discussion}
\begin{figure}
    \centering
    \subfloat[Case (a)]{\includegraphics[scale=0.35]{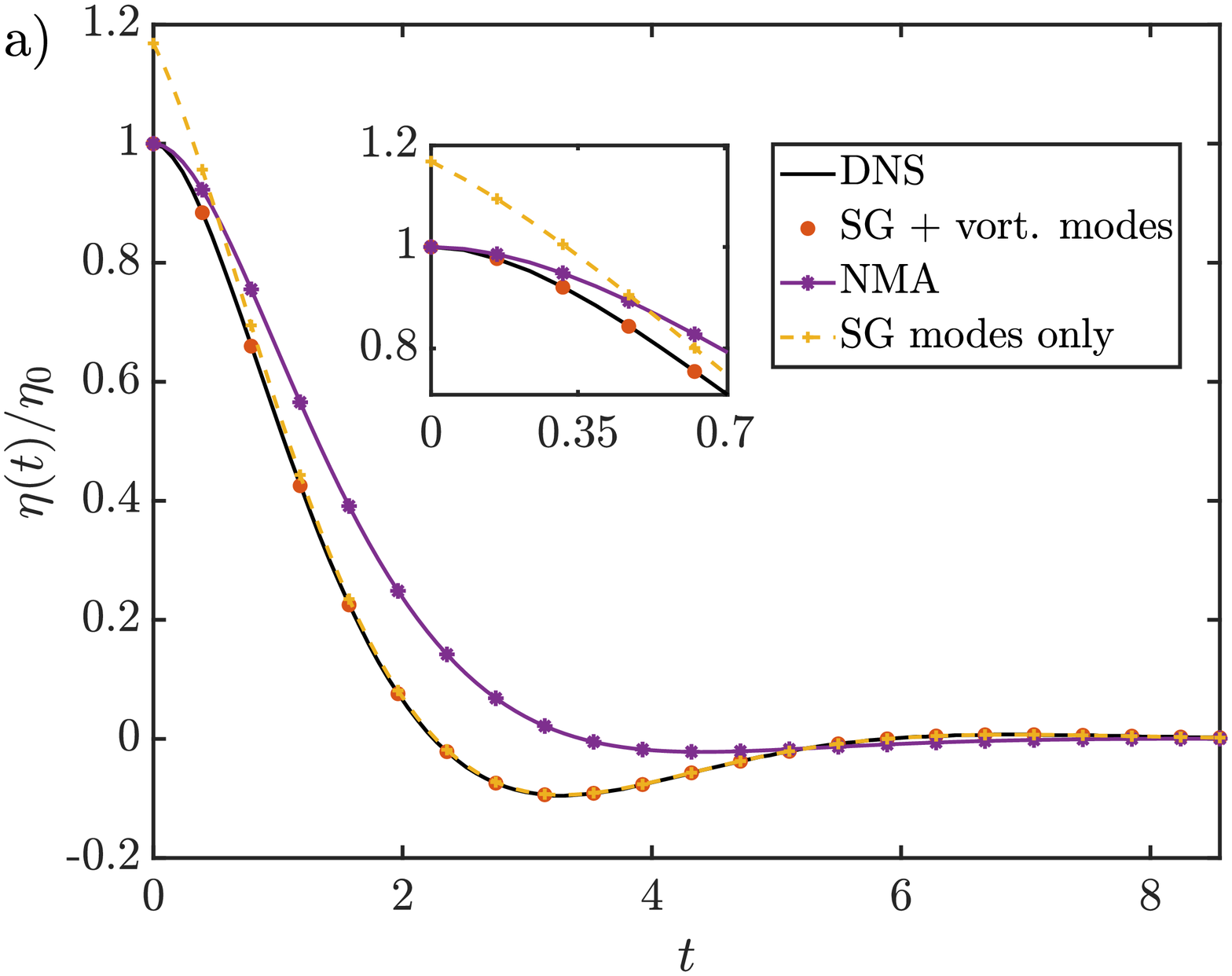}\label{fig:figure2a}}
    \subfloat[Case (b)]{\includegraphics[scale=0.35]{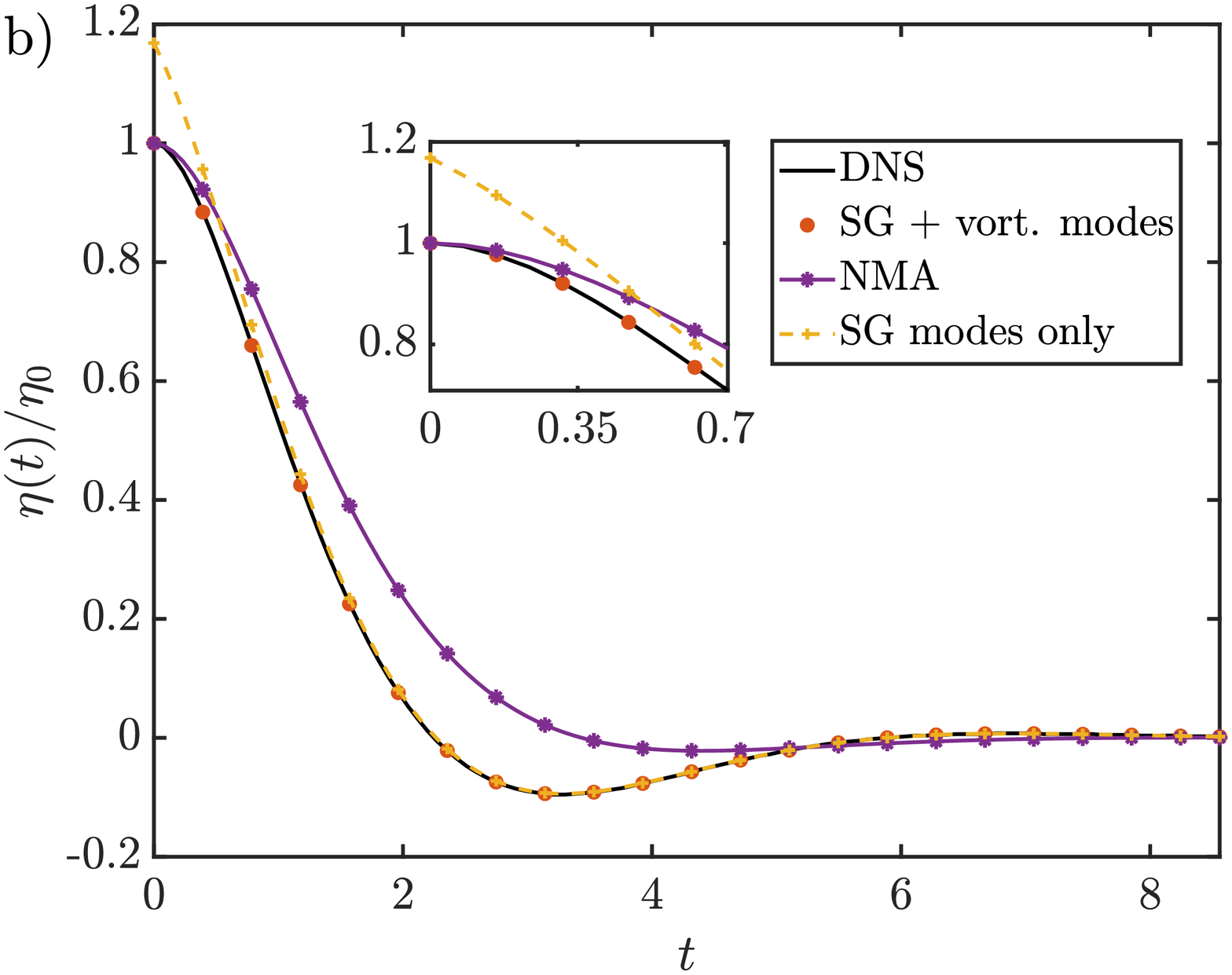}\label{fig:figure2b}}
    \caption{Free-surface amplitude variation as a function of time for the cases (a) and (b), respectively. The inset is a zoomed in version of the plot at the initial time. The horizontal axis is the non-dimensional time ($t^*\sqrt{g^*k^*}$) and the vertical axis is the non-dimensional amplitude.}
    \label{fig:figure2}
\end{figure}

Figures \ref{fig:figure2a} and \ref{fig:figure2b} represent the time evolution of the free-surface perturbation (normalised by its peak initial value) for case (a) and (b) in table \ref{tab_dns}. Both simulations represent an unstratified pool with the free surface deformed as a Fourier mode of wavenumber $k=2\pi$ (i.e. wavelength $1$ cm). The two cases however differ in the absence (case a) or presence (case b) of initial perturbation vorticity in the bulk of the pool. The most notable aspect of both figures is that they look indistinguishable despite having different initial conditions in terms of vorticity. We will discuss this important point later on in this section.

The curves in yellow (`SG modes only') in both figures \ref{fig:figure2a} and \ref{fig:figure2b} represent the contribution to the free-surface evolution from the two SG modes only (i.e. the deep blue circles depicted in fig. \ref{fig4a}). A common observation in both figures is that this approximation (`SG modes only') is unable to satisfy the initial conditions at the free-surface viz. $a(0)=a_0$ and $\dot{a}(0)=0$, see the mismatch between these and the DNS signal at $t=0$. This is understood as follows: the contribution to the free-surface displacement comes not only from the two SG modes which, by definition, are `surface modes' but also from the vorticity modes. Independent of whether we impose zero vorticity initial condition in case (a) or an initial, localised vortical perturbation in case (b), these vorticity modes are excited in both cases. Consequently, the contribution of these vorticity modes are necessary (in addition to the contribution from SG modes) for satisfying the conditions $a(0)=a_0$ and $\dot{a}(0)=0$. This is seen clearly in figs. \ref{fig:figure2a} and \ref{fig:figure2b}, where the curve labelled as `SG + vort. modes' (i.e. considering both contributions) satisfies both initial conditions and displays excellent agreement with the DNS signal. At slightly large time, however, the contribution from the vorticity modes becomes insignificant and the surface response from SG modes only, becomes identical to the DNS response. 

\begin{figure}
    \centering
    \subfloat[Case (a)]{\includegraphics[scale=0.40]{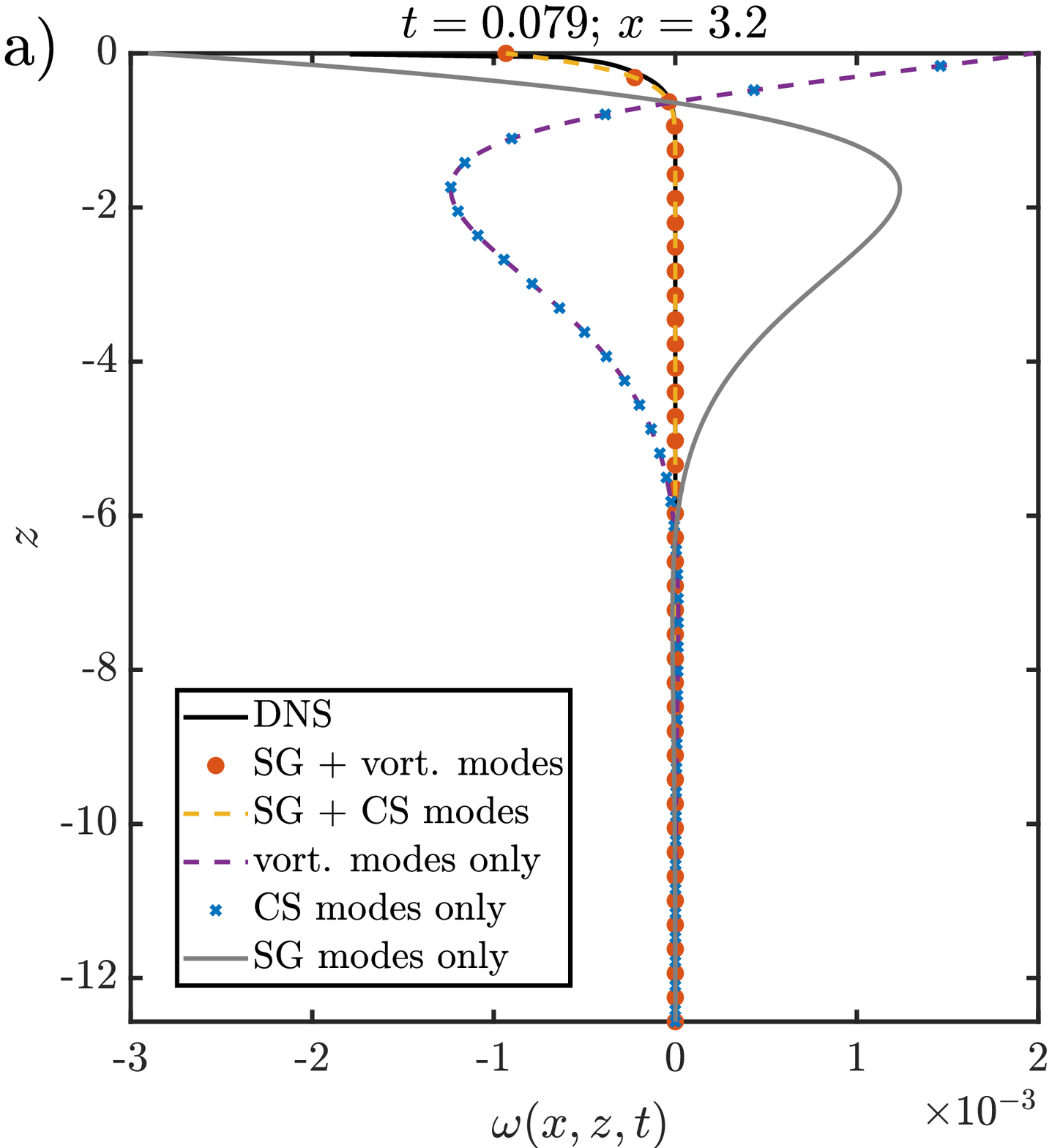}\label{fig:figure3a}}
    \subfloat[Case (b)]{\includegraphics[scale=0.40]{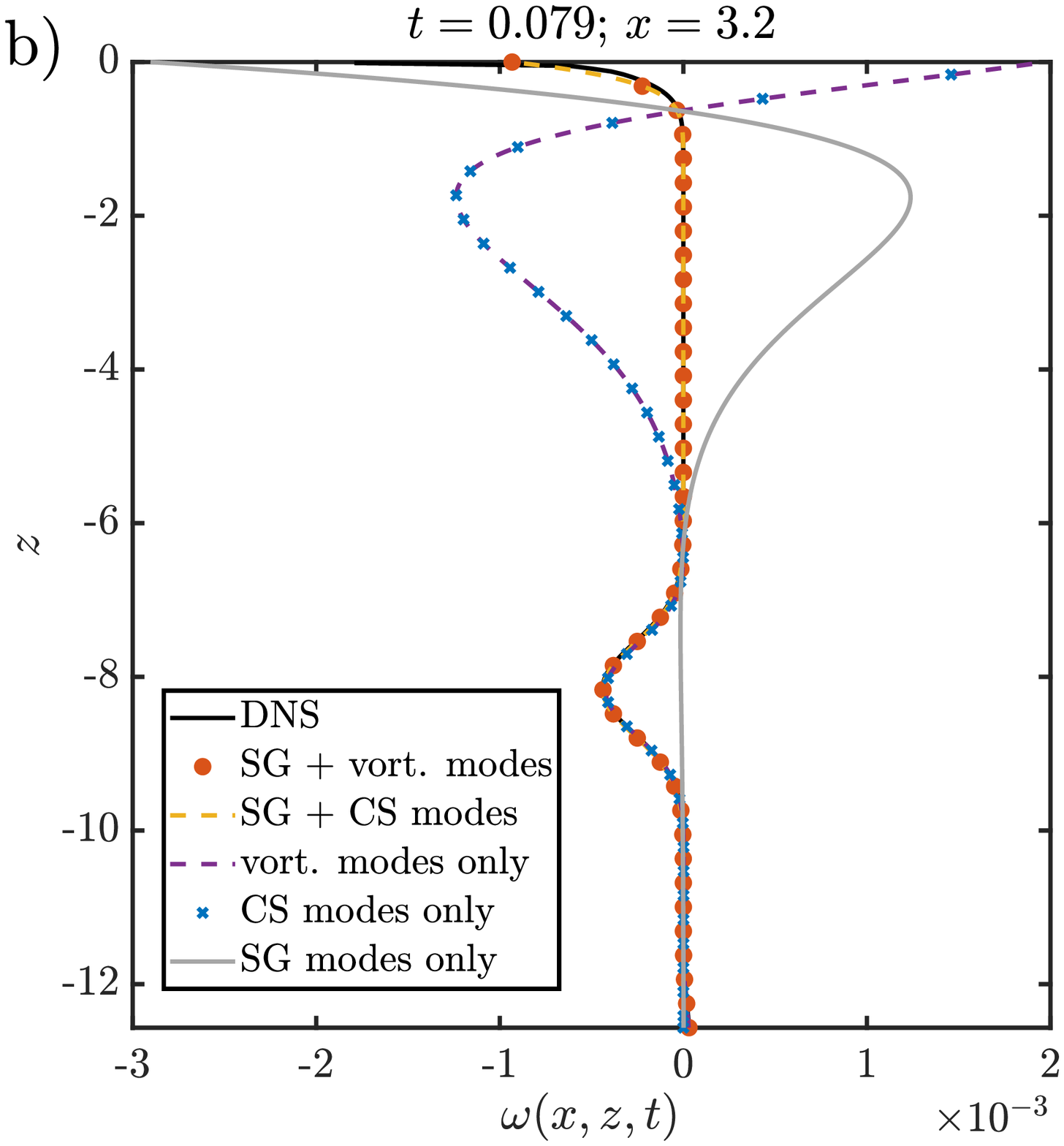}\label{fig:figure3b}}
    \caption{Vorticity plotted as a function of the vertical coordinate at a particular time $t(=0.079)$ and horizontal coordinate $x(=3.2)$ for the cases (a) and (b), respectively. The horizontal axis is the non-dimensional vorticity ($\omega^*/\sqrt{g^*k^*}$) and the vertical axis is the non-dimensional vertical coordinate ($z^*k^*$) }
    \label{fig:figure3}
\end{figure}

Just as the vorticity modes can contribute to the free-surface displacement, the two SG modes in fig. \ref{fig4a} have vorticity associated with them. This vorticity field decays exponentially along the depth of the pool (negative z-direction). An estimate of the depth of penetration of this vorticity field is given by the real part of $\left(k^2 + \tilde{s}/\nu^{\mathcal{(L)}}\right)^{-1/2}$ (considering dimensional variables, $s=\frac{\tilde{s}}{\sqrt{gk}}$), which for the parameters corresponding to case (a) and (b) of table \ref{tab_dns} is approximately $0.10\;\text{cm}$. Note that in evaluating this, we have chosen the value of $s$ to be one of the two complex conjugate pairs (SG modes) in fig. \ref{fig4a}. It is apparent that this vorticity penetration depth is substantially less than the wavelength of the Fourier mode at the free-surface, which is $1$ cm for this case. It is thus intuitively clear that a linear superposition of the two SG modes only, can neither produce a zero initial vorticity field (case (a)) nor satisfy the free-surface initial conditions as already discussed. This is verified in fig. \ref{fig:figure3a} where the vorticity profile at $x=3.2$ is plotted at $t=0.079$ as a function of the depth coordinate $z$. One observes that the curve labelled as `vort. modes only' in this figure is a poor descriptor of the DNS vorticity profile (black curve). Incidentally, the inability of the vorticity modes only (or alternatively the SG modes only) to satisfy the initial conditions is consistent with our earlier results in \cite{farsoiya2020azimuthal} where azimuthal perturbations on a hollow cylindrical filament were studied and similar conclusions, for zero vorticity initial conditions, were reported. 

For case (b) in table \ref{tab_dns} the chosen Gaussian shaped, initial vorticity perturbation is centred at $z_d=1.3$ cm and has a thickness of $0.1$ cm. At this depth ($z_d=1.3$ cm), the vorticity of the SG modes is practically zero (see grey curve in fig. \ref{fig:figure3b}) and the contribution from the countably infinite vorticity modes of fig. \ref{fig4a}, becomes crucial in order to express this initial vortical patch. This is verified in fig. \ref{fig:figure3b} where we note the small `hump' in vorticity in the DNS signal at $z\approx-8$ arising from the initial perturbation. It is seen that the vorticity from the SG modes is zero at this depth, whereas the contribution from the vorticity modes are not. Interestingly, although the curve labelled as `vorticity modes only' in this figure provides an excellent approximation to the perturbation 
 vorticity field at large depth seen in DNS, the approximation becomes quite inaccurate close to the free-surface $z=0$. Here at early time $t > 0$, the contribution from both SG and vorticity modes nearly (but not entirely) cancel each other, producing a (boundary) vorticity layer adjacent to the free surface (see DNS signal in black). The overall conclusion is that in order to obtain a good match with the DNS vorticity field close to the free surface, we require contributions from both sets of modes. These conclusions remain unaltered when we refer back to case (a) in table \ref{tab_dns}, which corresponds to zero vorticity initial condition. Observing the DNS signal in fig. \ref{fig:figure3a}, we note that the perturbation vorticity field in this case is only adjacent the free surface (and zero in the bulk liquid). Expressing this vorticity field near the free-surface accurately requires contributions from vorticity and SG modes consistent with observations by \citep{prosperetti1976viscous}.  

We had mentioned earlier \citep{lamb1932,prosperetti1976viscous} that in the unstratified, infinite depth limit the spectrum has an additional continuous part and that these continuous spectrum (CS) modes have an oscillatory variation in the z direction. Such behaviour may also been seen among a subset of the discrete spectrum modes in the finite depth case as we argue below. In dimensional variables, the diffusion equation governing the perturbation vorticity field admits solutions of the form $\exp\left[z\left(k^2 + \tilde{s}/\nu^{\mathcal{(L)}}\right)^{1/2}\right]$  with $\mathcal{\text{Re}}\left(k^2 + \tilde{s}/\nu^{\mathcal{(L)}}\right)^{1/2} > 0,\; z < 0$. For the countably infinite set of vorticity modes in fig. \ref{fig4a} for which $\tilde{s}$ is real, negative and which additionally satisfy $|\tilde{s}| >> \nu^{\mathcal{(L)}} k^2$, such a solution to the diffusion equation becomes an oscillatory functions of $z$ behaving  as $\sim\exp\left(z\sqrt{\tilde{s}/\nu^{\mathcal{(L)}}}\right),\;\tilde{s}$ being real and negative. For comparison, the contribution from the CS modes in the infinite depth case (from the branch cut) are also plotted in figs. \ref{fig:figure3a} and \ref{fig:figure3b}. We observe that these are indistinguishable from the contribution due to the vorticity modes in the finite depth case. Note that the non-dimensional depth $H$ for case (a) and (b) in table \ref{tab_dns} is finite but large.

The acronym NMA in figures \ref{fig:figure2a} and \ref{fig:figure2b} represents what we term as the normal mode approximation (eqn. $3.3$ also used earlier in \citep{farsoiya2017axisymmetric}).  This is an approximation based on the observation that there are only two SG modes in the spectrum. This may be used to satisfy the free surface initial conditions $a(0)=a_0$ and $\dot{a}(0)=0$, leading to a predictive expression for the free surface evolution viz. $\frac{a(t)}{a(0)} = \frac{1}{s_1 - s_2}\left(s_1\exp{s_2t}-s_2\exp{s_1t}\right)$, where $s_1,s_2$ are the two complex conjugate pair of SG modes in fig. \ref{fig4a}. We note that by construction the NMA satisfies the initial conditions at the free surface; although it shows a rather large mismatch with the DNS signal in figures. \ref{fig:figure2a} and \ref{fig:figure2b}, at intermediate time. The curves labelled as `SG + vort. modes' in these figs. represent the solution to the complete IVP taking into account contributions from both sets of modes. Expectedly, we note the excellent agreement of this with DNS. Returning to our observation of figures \ref{fig:figure2a} and \ref{fig:figure2b} being identical, we note that this is an outcome of the vorticity perturbation being placed sufficiently deep. As a consequence, this perturbation vorticity at large depth (which is well described by a linear superposition of the vorticity modes only), evolves purely diffusively nearly independent of the evolution of the transient vorticity layer generated at the free-surface. This behaviour is to be contrasted with the observations of \citep{prosperetti1982small} where surface waves produced due to a submerged vortex filament (extending along the $y$ direction) and an initial flat free-surface was studied analytically. In contrast to our present study where a single Fourier mode of wave number $k$ is excited along the x-direction (both in case (a) and case (b) in table \ref{tab_dns}), the initial condition of \cite{prosperetti1982small} excites a continuous spectrum of wavenumbers $k$ at $t=0$. Waves at the free surface are observed subsequently due to the presence of the vortex filament in the pool below (see figs. $1$ and $2$ in \cite{prosperetti1982small}). The submergence depth of the vortex filament in their case was $0.1$ and $0.01$ cms respectively, both of which are significantly smaller than the wavelength of the slowest capillary-gravity surface wave excited in their system. In contrast, our submergence depth of the Gaussian patch in case (b) ($1.3$ cm) exceeds the wavelength of the surface wave excited at $t=0$.

\begin{figure}
    \centering
    \subfloat[Case (a)]{\includegraphics[scale=0.35]{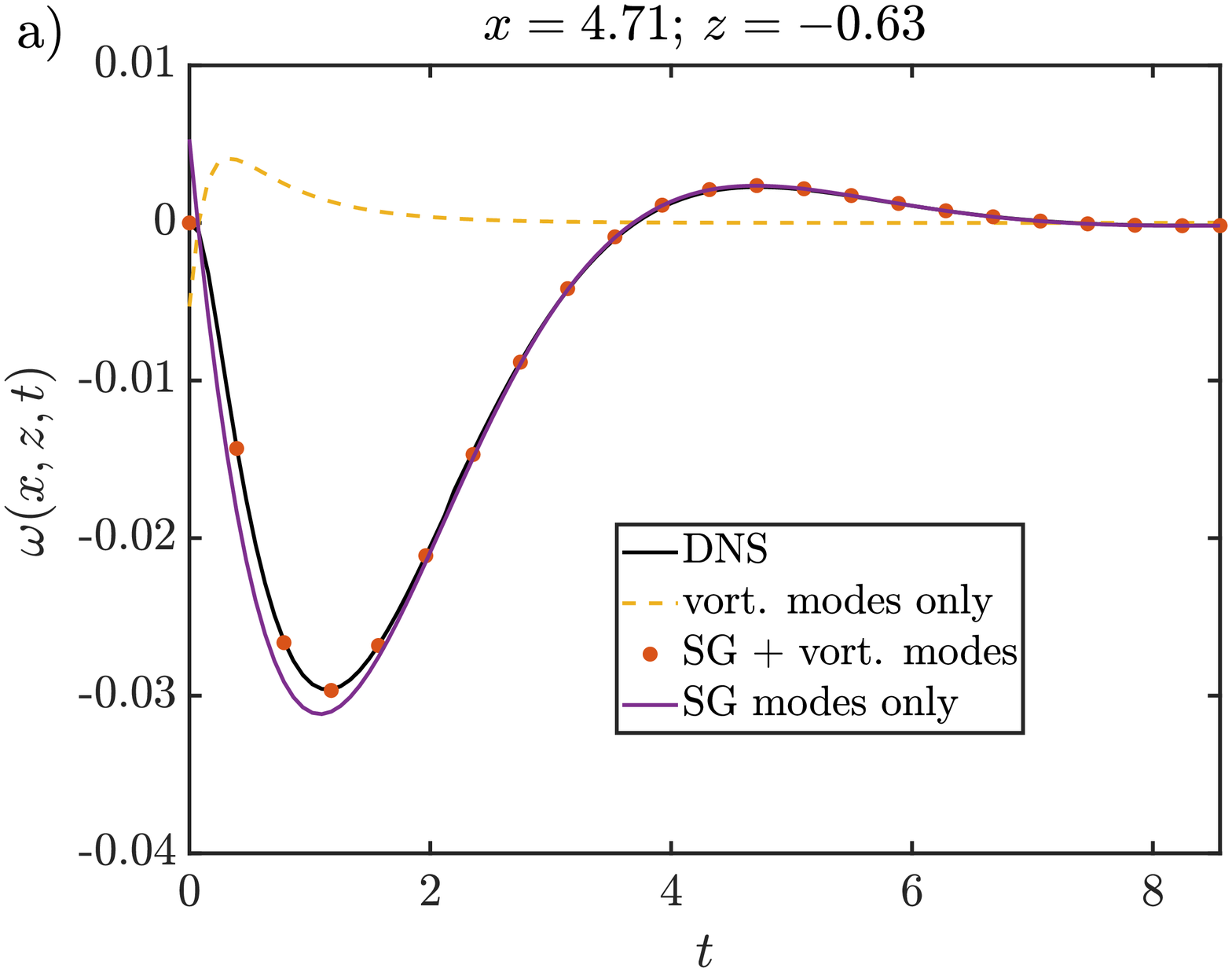}\label{fig:figure7a}}
    \subfloat[Case (b)]{\includegraphics[scale=0.35]{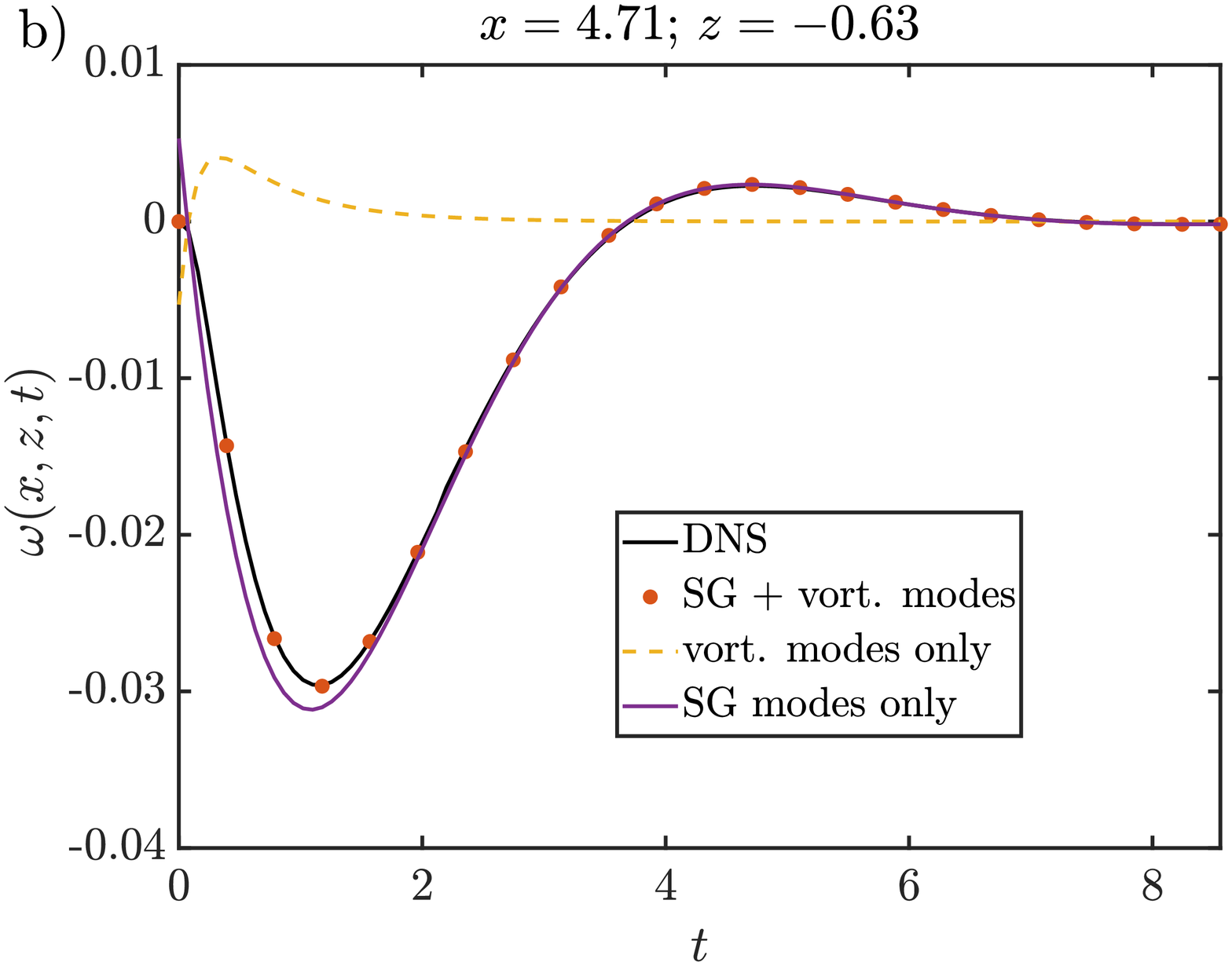}\label{fig:figure7b}}
    \caption{Vorticity, at a location in space ($x = 4.71, z = -0.63$) close to the interface, plotted as a function of time for the cases (a) and (b), respectively. Here, the vertical axis is the non-dimensional vorticity ($\omega^*/\sqrt{g^*k^*}$) and the horizontal axis the non-dimensional time ($t^*\sqrt{g^*k^*}$).}
    \label{fig:figure7}
\end{figure}

\begin{figure}
    \centering
    \subfloat[Case (a)]{\includegraphics[scale=0.35]{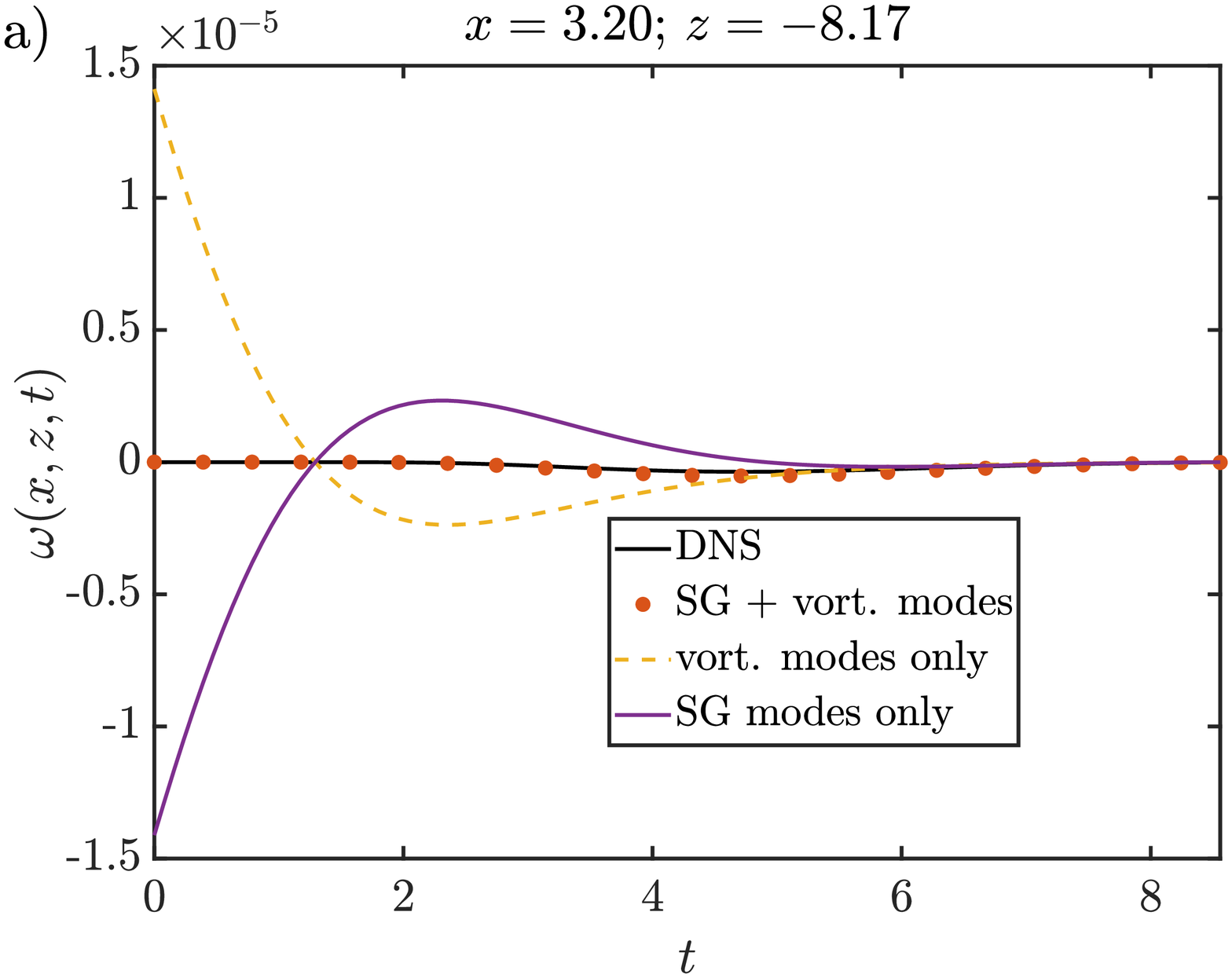}\label{fig:figure8a}}
    \subfloat[Case (a)]{\includegraphics[scale=0.35]{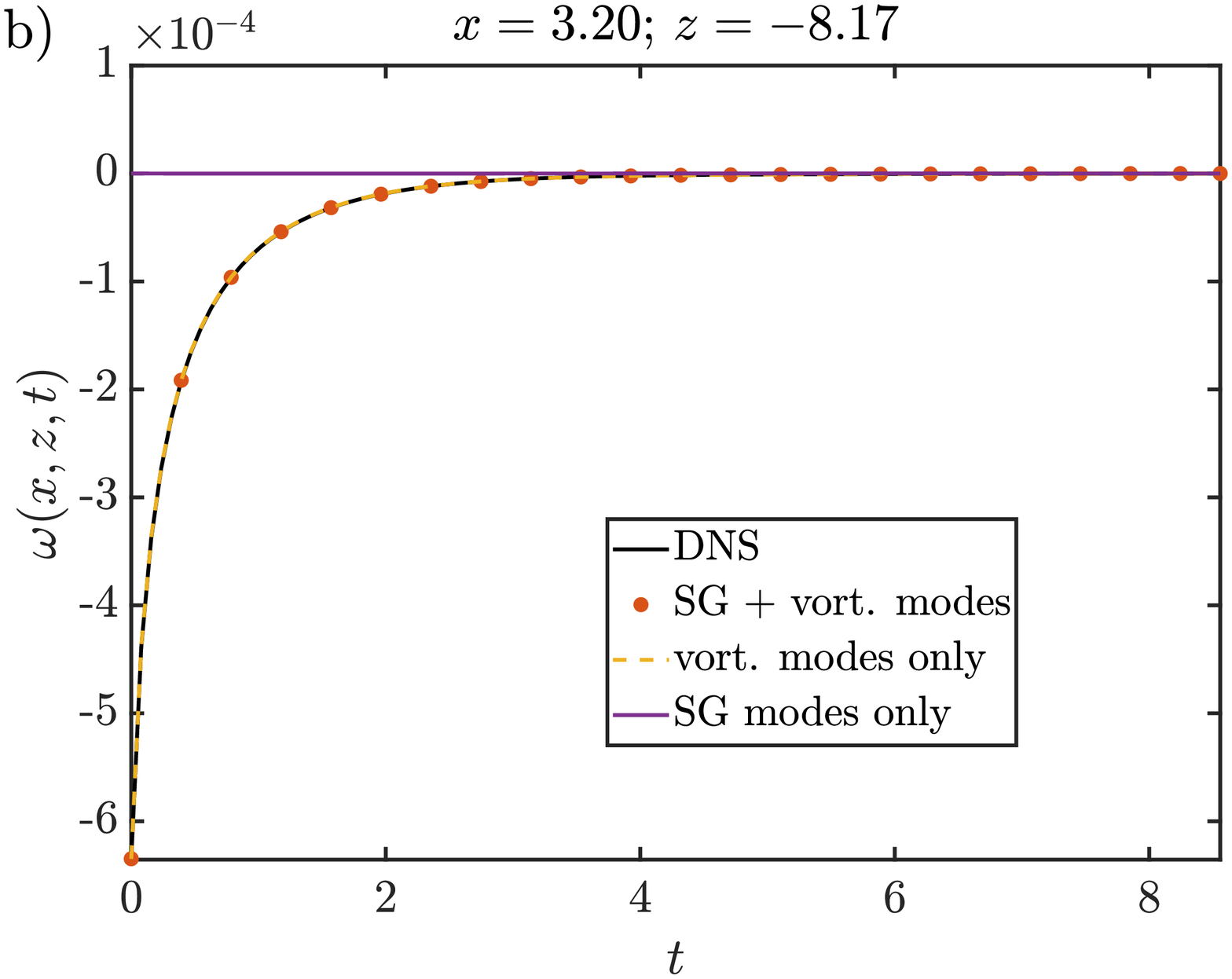}\label{fig:figure8b}}
    \caption{Vorticity, at a location in space ($x = 4.71, z = -8.17$) deep inside the pool, plotted as a function of time for the cases (a) and (b), respectively. Here, the vertical axis is the non-dimensional vorticity ($\omega^*/\sqrt{g^*k^*}$) and the horizontal axis the non-dimensional time ($t^*\sqrt{g^*k^*}$).}
    \label{fig:figure8}
\end{figure}

Figures \ref{fig:figure7a} and \ref{fig:figure7b} show the vorticity at a particular spatial location $x (=4.71)$ and $z(=-0.63)$, plotted as a function of time for cases (a) and (b) respectively of table \ref{tab_dns}. Note that the vertical coordinate of this location is quite close to the free-surface and thus we expect the vorticity field in case (a) and case (b) to look similar as is easily verified from the two figures. Considering the contribution of SG modes once again leads to slightly erroneous results at initial times. When the depth is choosen to be farther from the free-surface, as in figs. \ref{fig:figure8a} and \ref{fig:figure8b}, a clear difference is seen in the temporal evolution of the perturbation vorticity. For case (a) described in fig. \ref{fig:figure8a}, at $z=-8.17$ the perturbation vorticity field is effectively zero at all time and this is acheived analytically only by taking the mutually cancelling contributions from the SG modes (yellow) and the vorticity modes (purple). In contrast, for case (b) in figure \ref{fig:figure8b} the perturbation vorticity field is markedly different from zero at this depth (due to the initial Gaussian vorticity perturbation). As expected, the burden of describing the time variation of this vorticity is taken up solely by the vorticity modes without any contribution from the SG modes.

\begin{figure}
    \centering
    \includegraphics[scale=0.35]{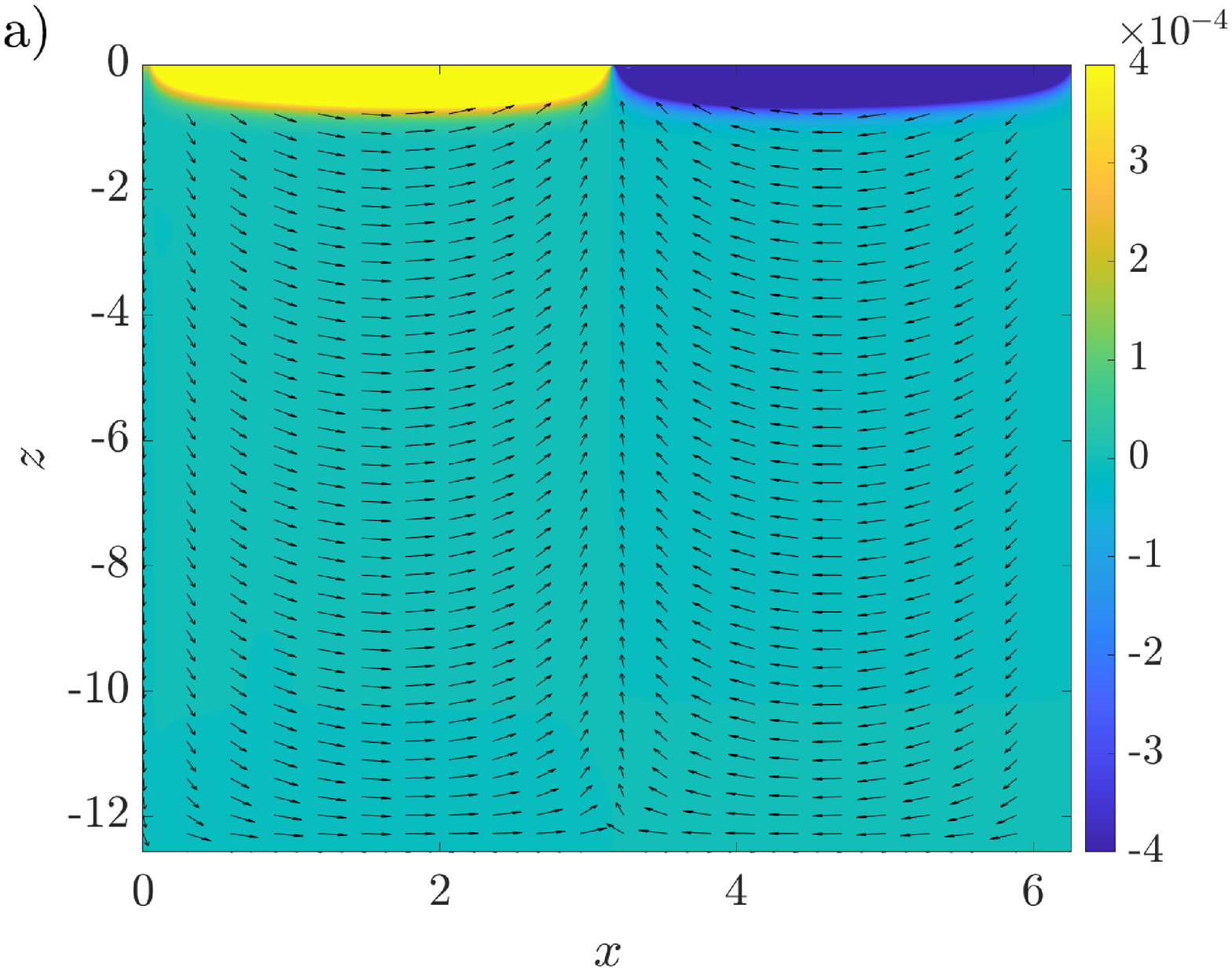}\label{fig:figure9a}
    \includegraphics[scale=0.35]{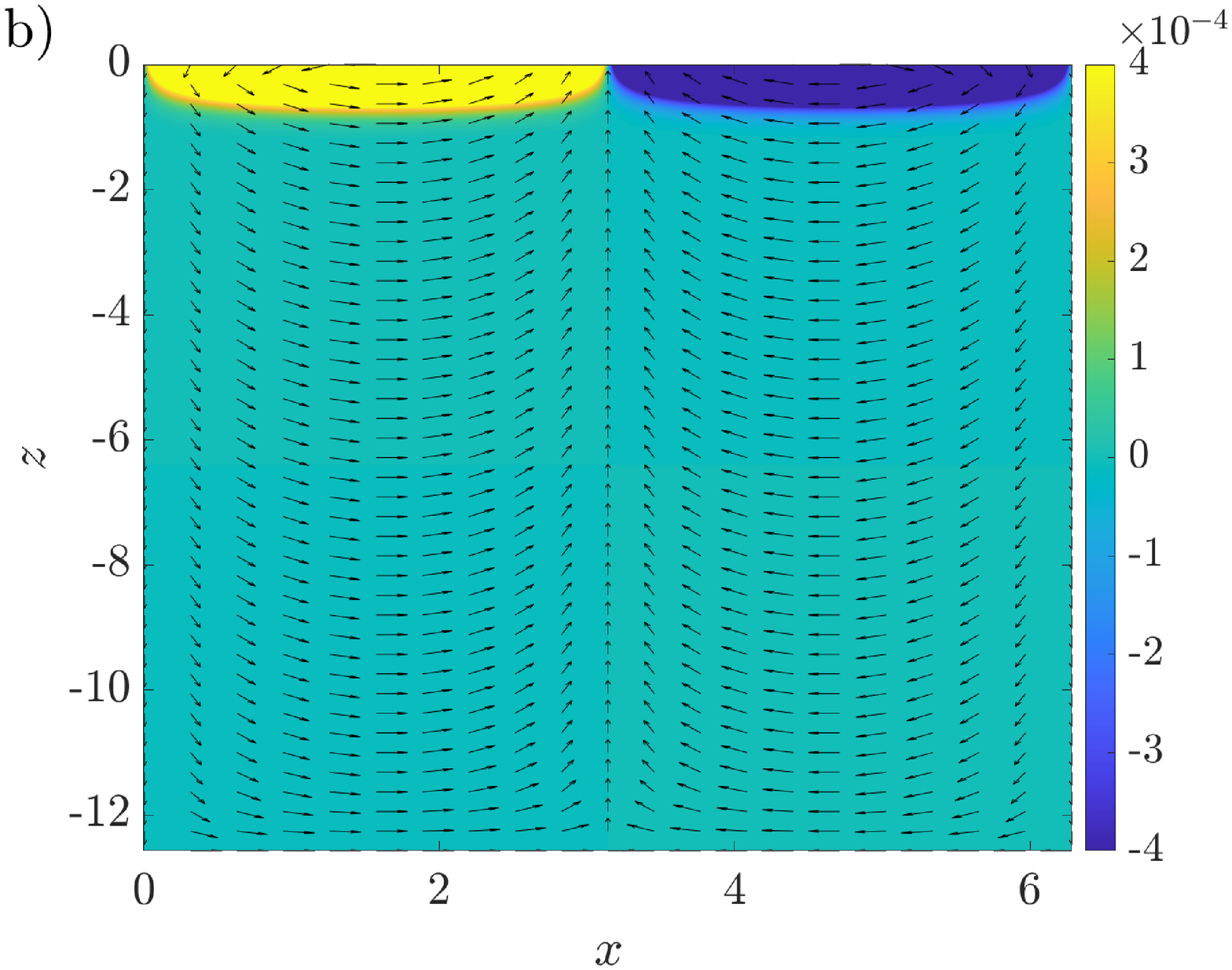}\label{fig:figure9b}
    \includegraphics[scale=0.35]{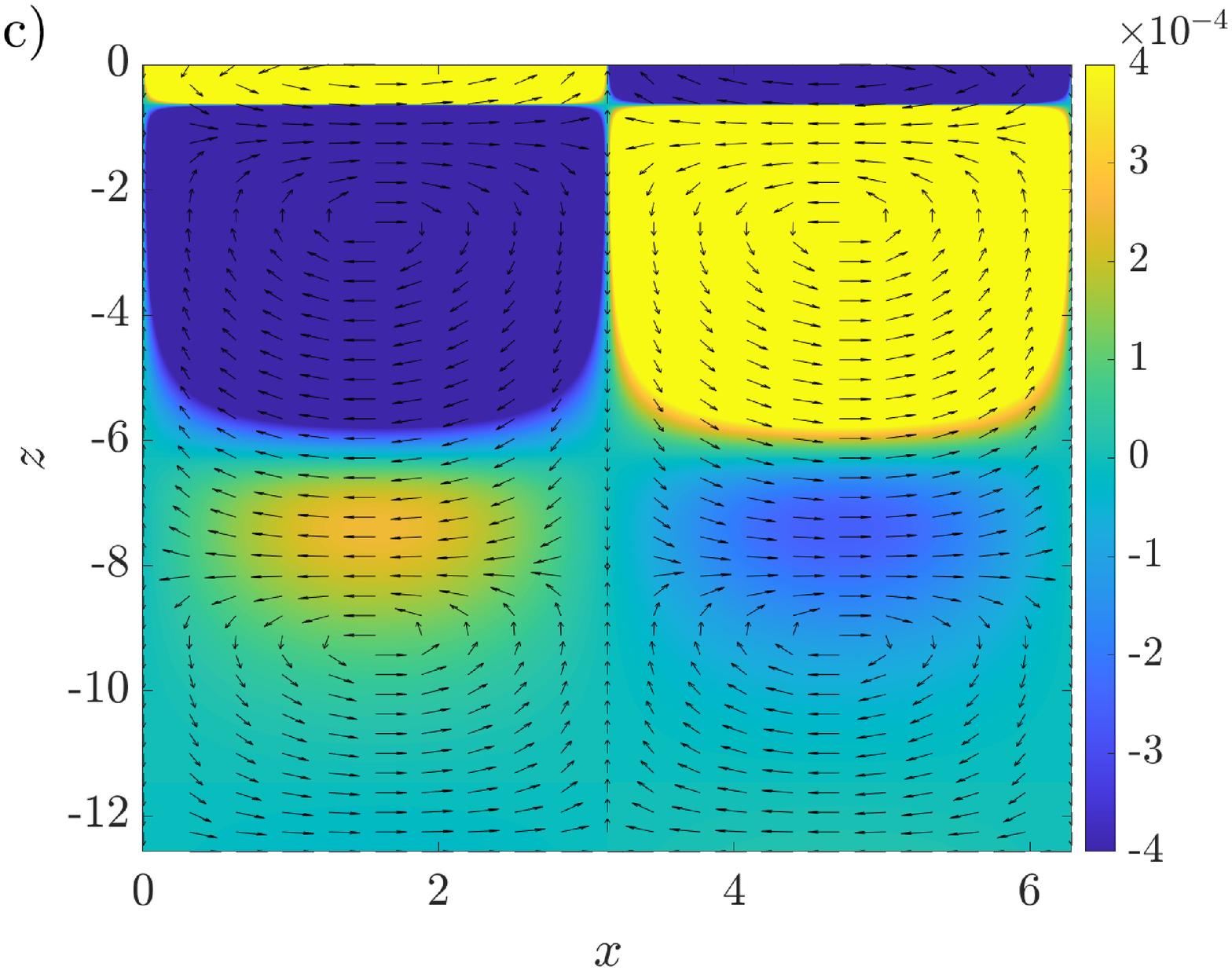}\label{fig:figure9c}
    \includegraphics[scale=0.35]{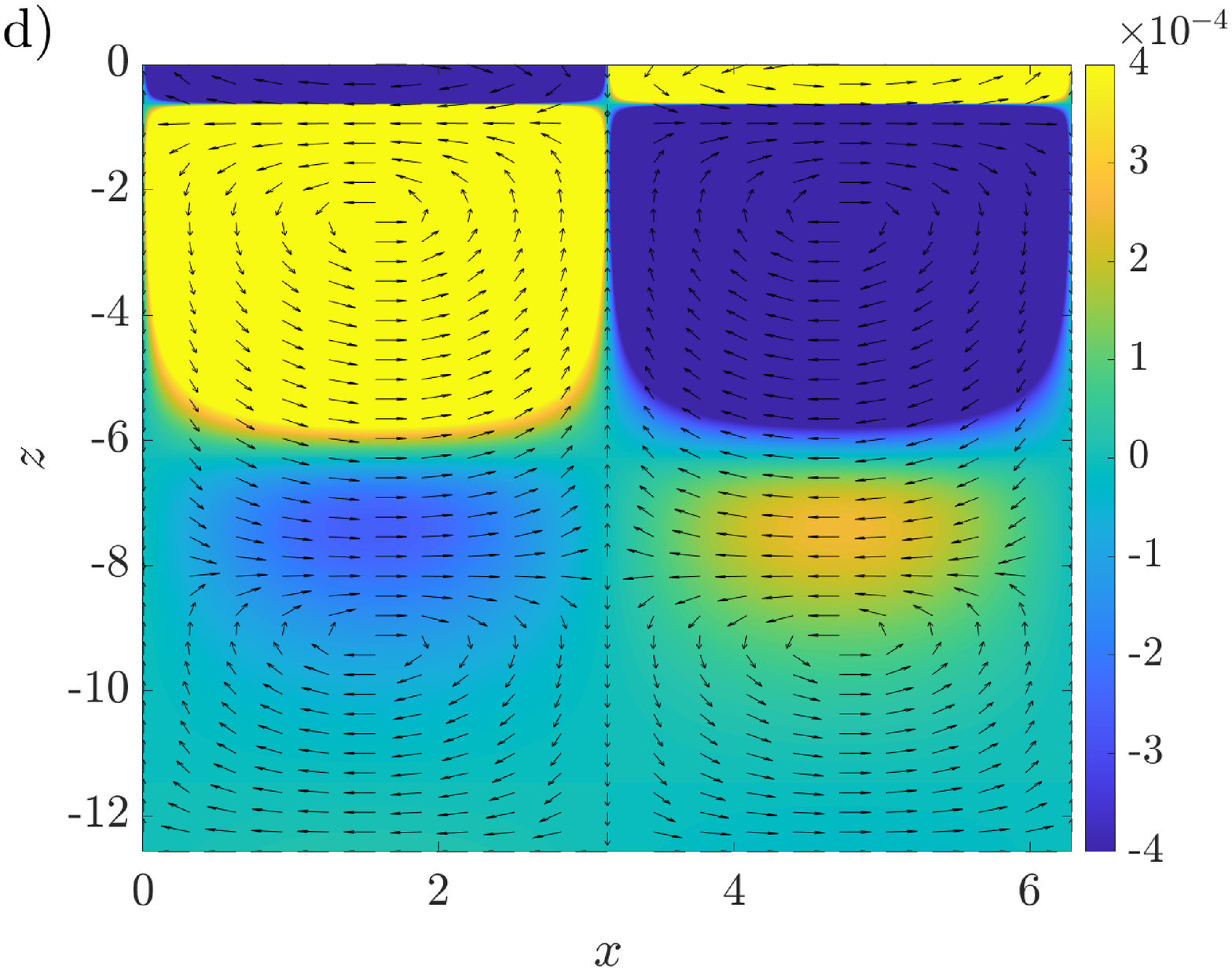}\label{fig:figure9d}
    \caption{Vorticity contours corresponding to case (a) at time $t = 0.079$ from: a) DNS b) Total IVP (SG + visc. modes) c) IVP (SG modes only) d) IVP (vort. modes only). The arrows at different locations indicate the instantaneous velocity at these locations. The horizontal and vertical axes are non-dimensional $x$ ($=x^*k^*$) and $z$ ($=z^*k^*$) coordinates, respectively}
    \label{fig:figure9}
\end{figure}

\begin{figure}[h]
    \centering
    \includegraphics[scale=0.35]{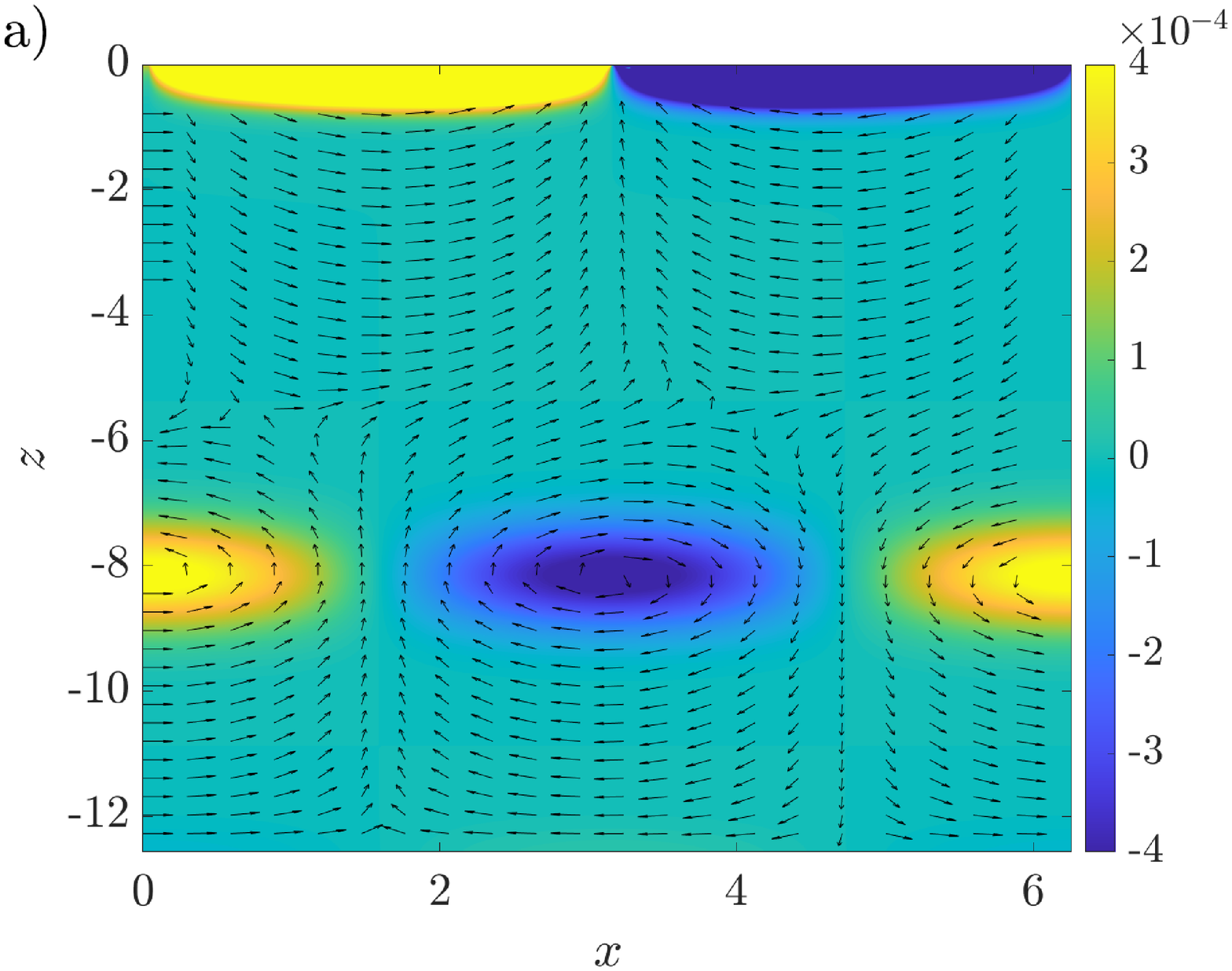}\label{fig:figure10a}
    \includegraphics[scale=0.35]{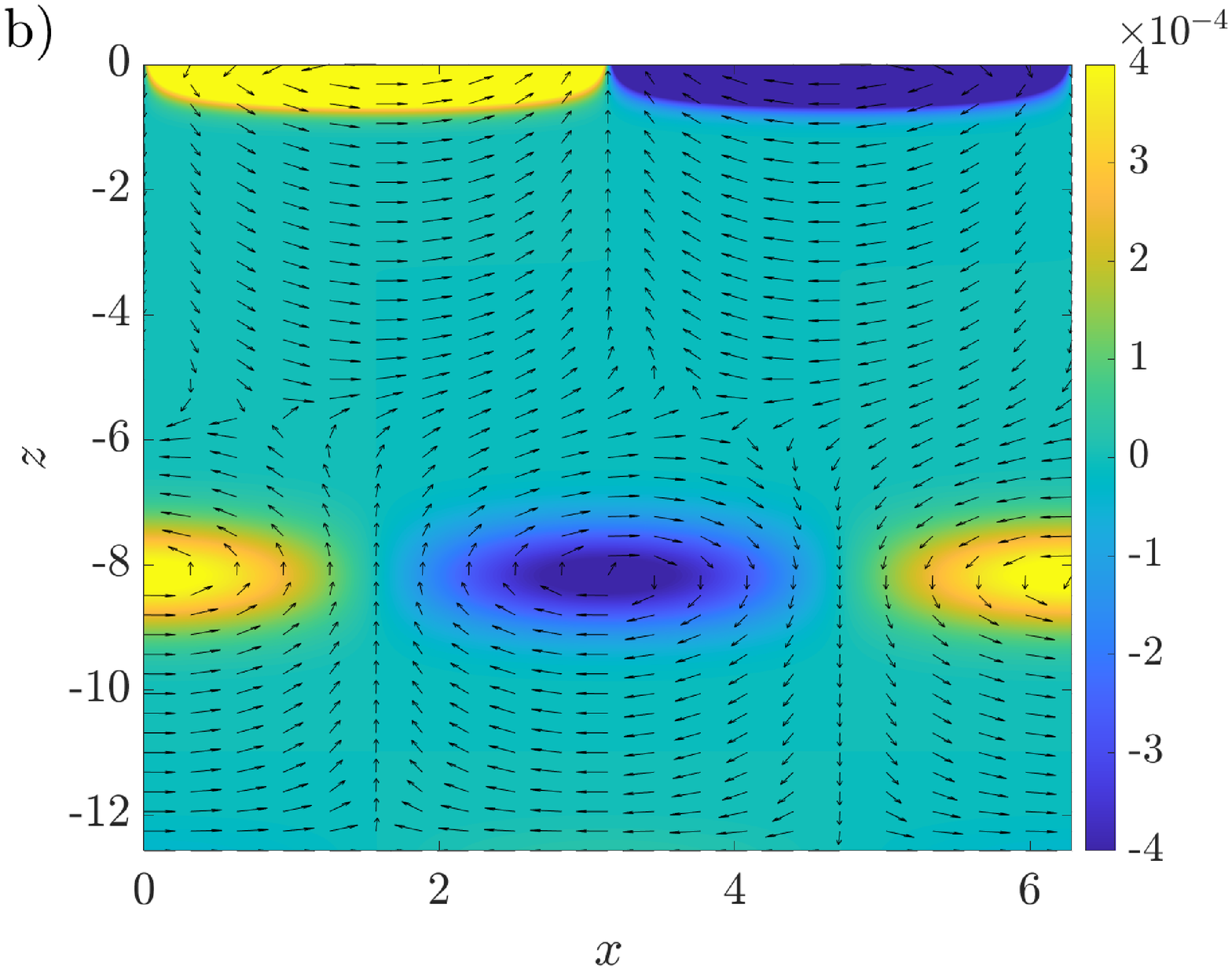}\label{fig:figure10b}
    \includegraphics[scale=0.35]{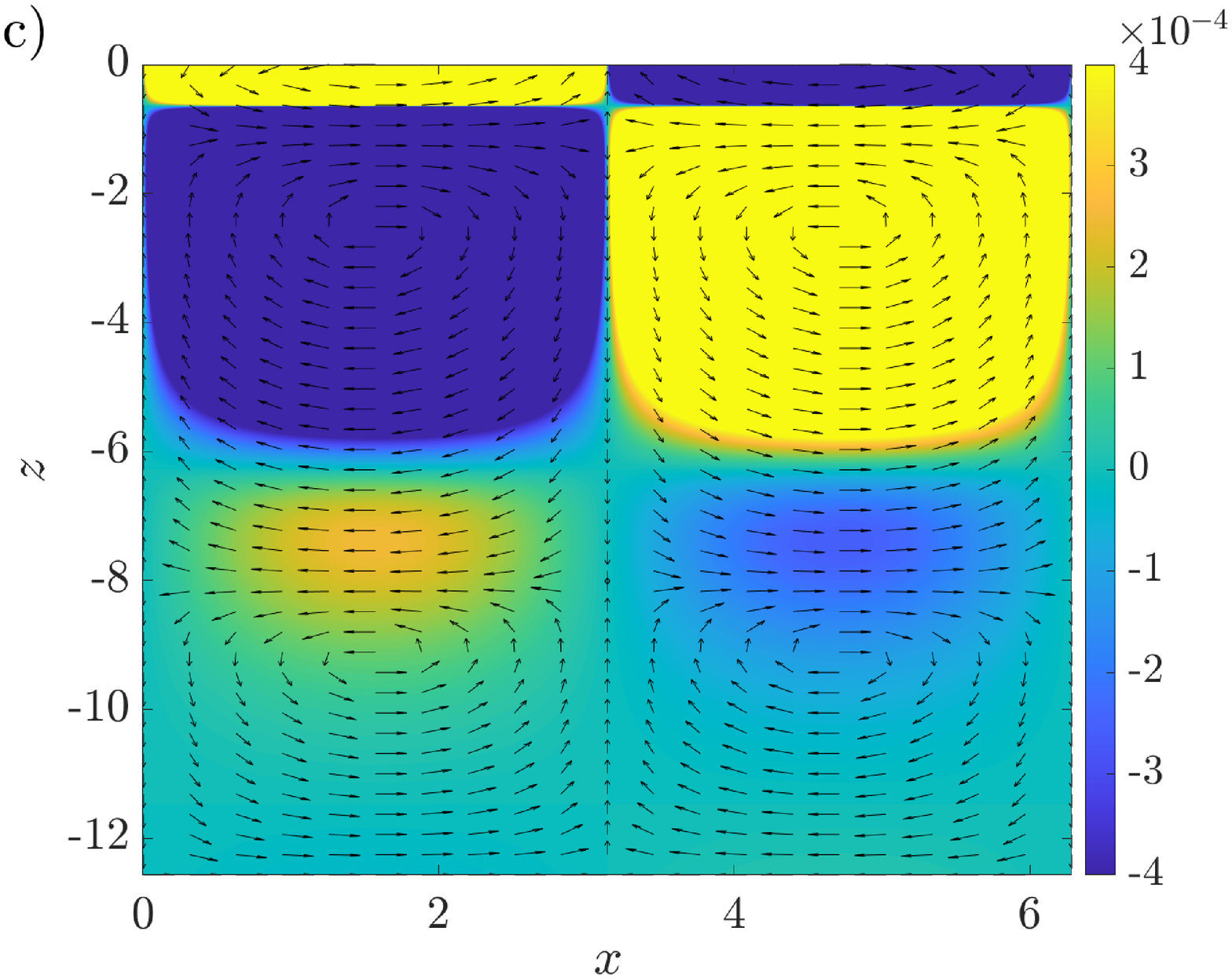}\label{fig:figure10c}
    \includegraphics[scale=0.35]{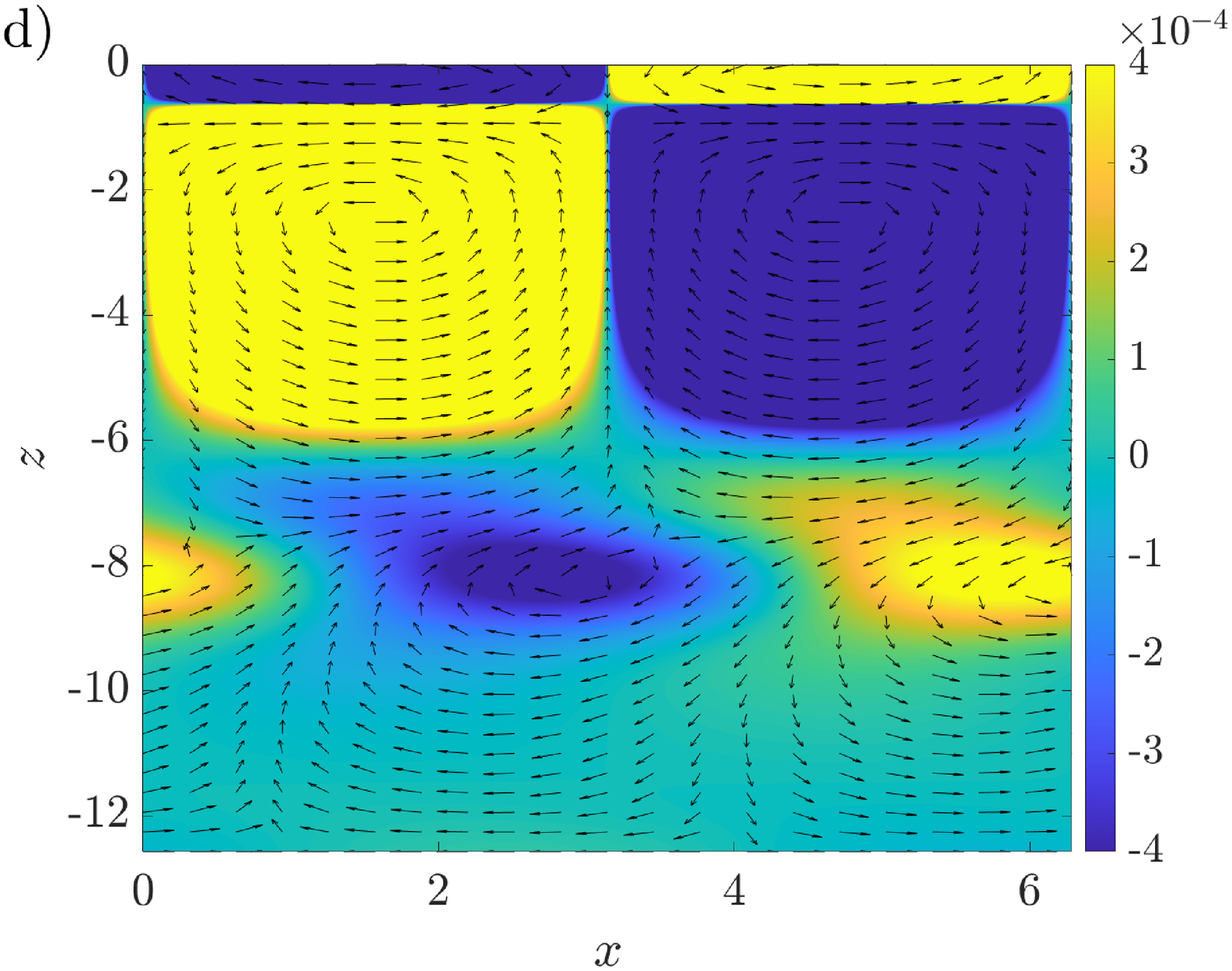}\label{fig:figure10d}
    \caption{Vorticity contours corresponding to case (b) at time $t = 0.079$ from: a) DNS b) Total IVP (SG + visc. modes) c) IVP (SG modes only) d) IVP (vort. modes only). The horizontal and vertical axes are non-dimensional in the same way as fig. \ref{fig:figure9}.}
    \label{fig:figure10}
\end{figure}

A global picture of the perturbation vorticity as well as flow field, at  early time $t (=0.079)$ once again for cases (a) and (b) in table \ref{tab_dns} are provided in figure \ref{fig:figure9}, panels (a) - (d) and figure \ref{fig:figure10}, panels (a) - (d) respectively. In these figures, the horizontal axis spans one wavelength and the vertical axis spans two wavelengths (in non-dimensional sense). Panels (a) and (b) in fig. \ref{fig:figure9} indicate a good agreement between DNS results and IVP predictions. Panels (c) and (d) in the same figure show flow structures due to contributions from  SG modes and vorticity modes, respectively, nearly cancel each other leading to what is observed in the upper panels. 
A similar behaviour is observed in figures \ref{fig:figure10}, panels (a) - (d), representing case (b) in table \ref{tab_dns}. Note that in contrast to fig. \ref{fig:figure9}, panels (c) and (d) in this figure are visibly as-symmetrical and the contribution from vorticity modes resolves almost entirely, the initial perturbation vorticity.
\begin{figure}[h]
    \centering
    \subfloat[Vorticity $\omega(x,z,t)$]{\includegraphics[scale=0.5]{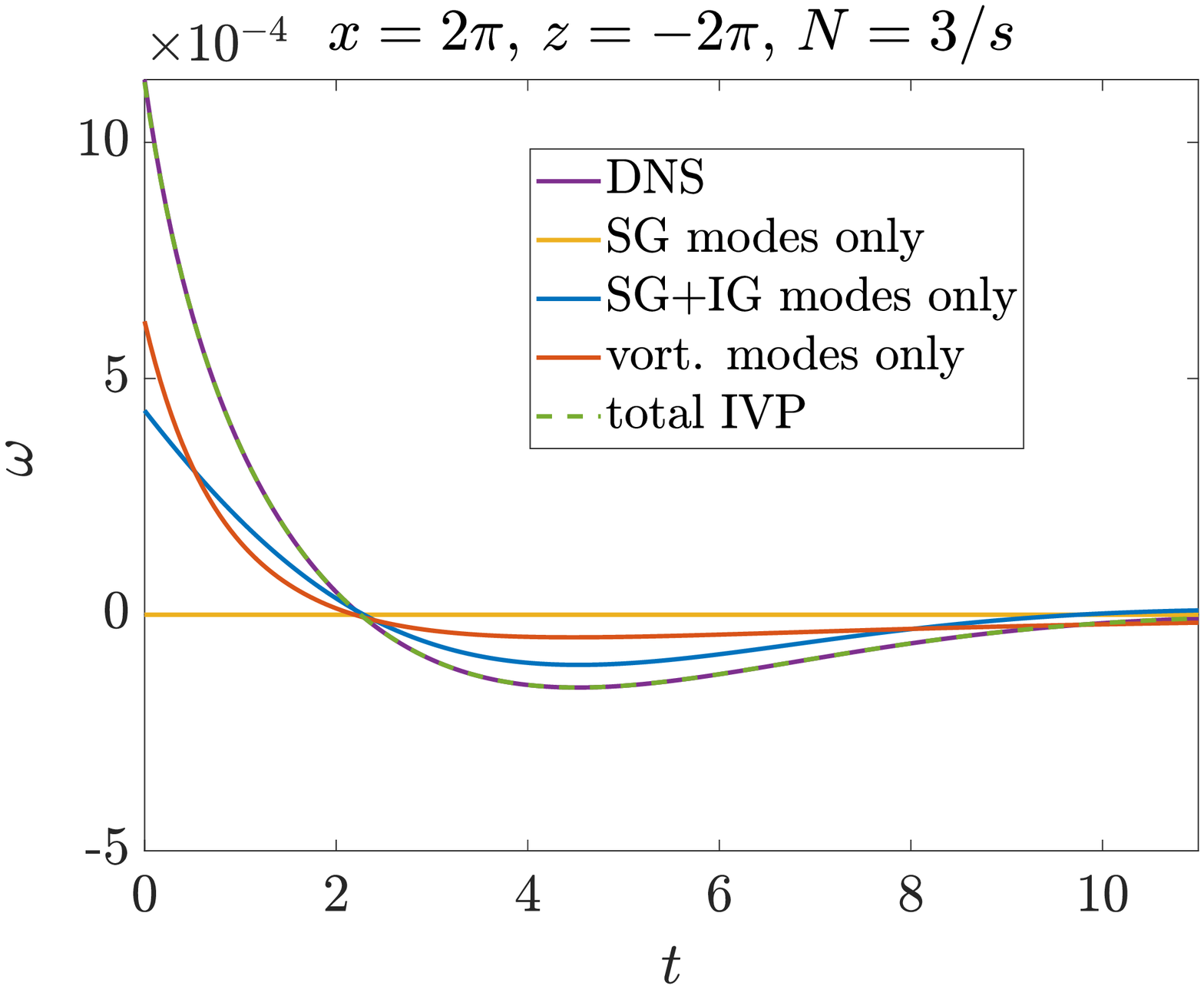}\label{fig:figure11a}}\\
    \subfloat[$t=0$]{\includegraphics[scale=0.4]{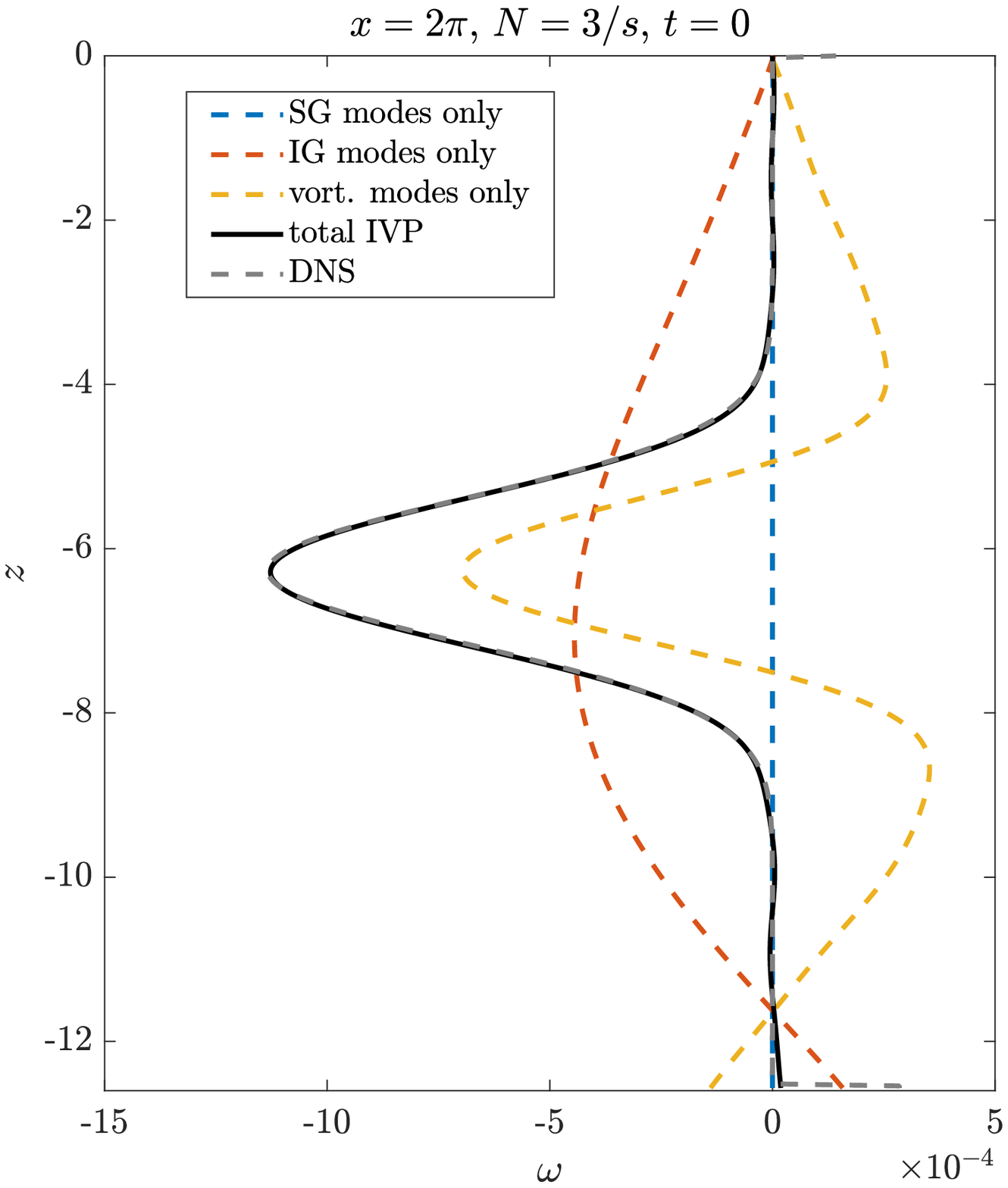}\label{fig:figure11b}}
    \subfloat[$t=3.91$]{\includegraphics[scale=0.4]{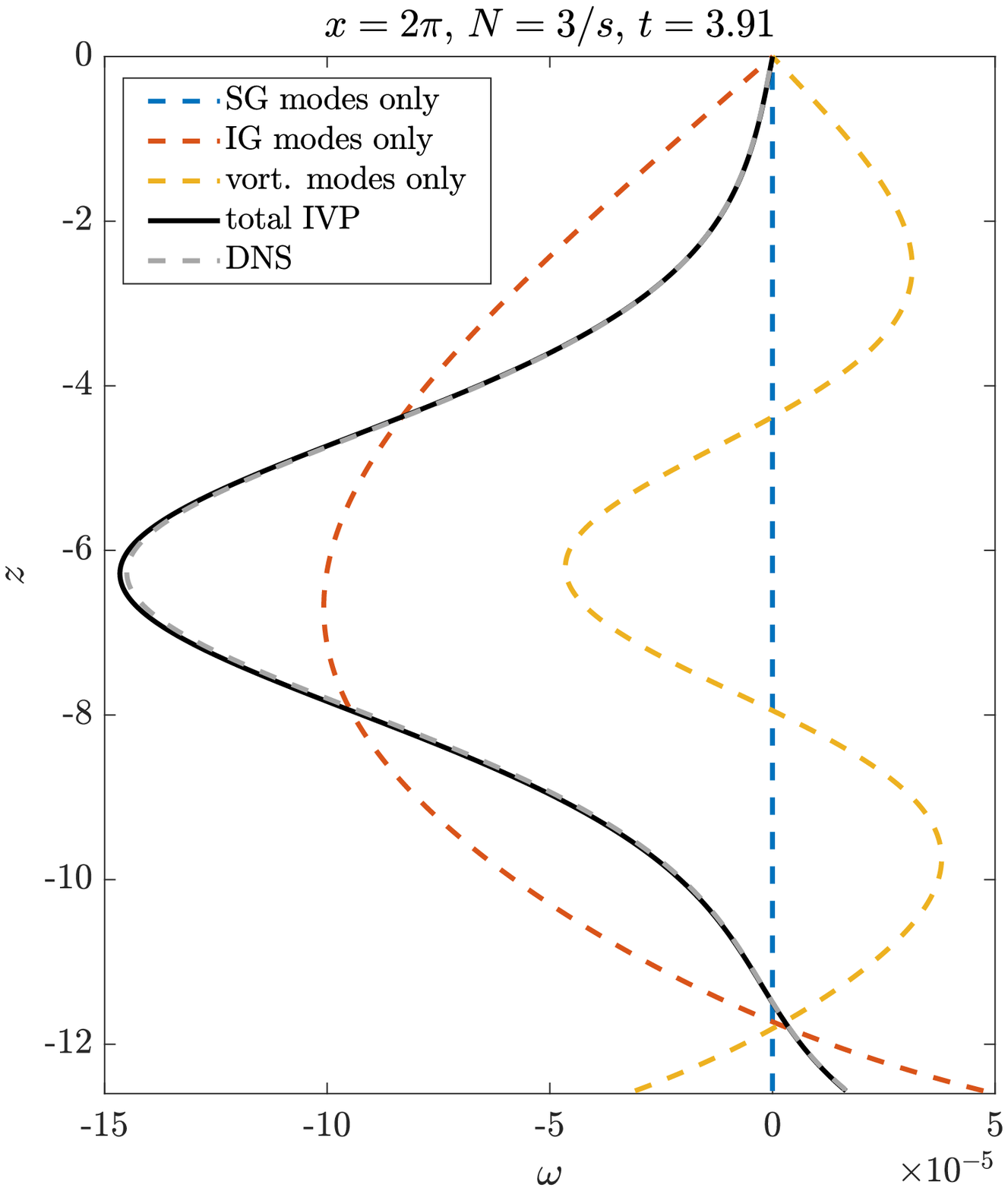}\label{fig:figure11c}}
    \caption{Vorticity plots corresponding to case (c) in table \ref{tab_dns} at $t = 0$ from DNS versus analytical predictions: Panel (a) vorticity vs time, panels (b) and (c) vorticity vs z at $t=0$ and $t=3.91$.}
    \label{fig:figure11}
\end{figure}
\begin{figure}[htbp]
    \centering
    \includegraphics[scale=0.4]{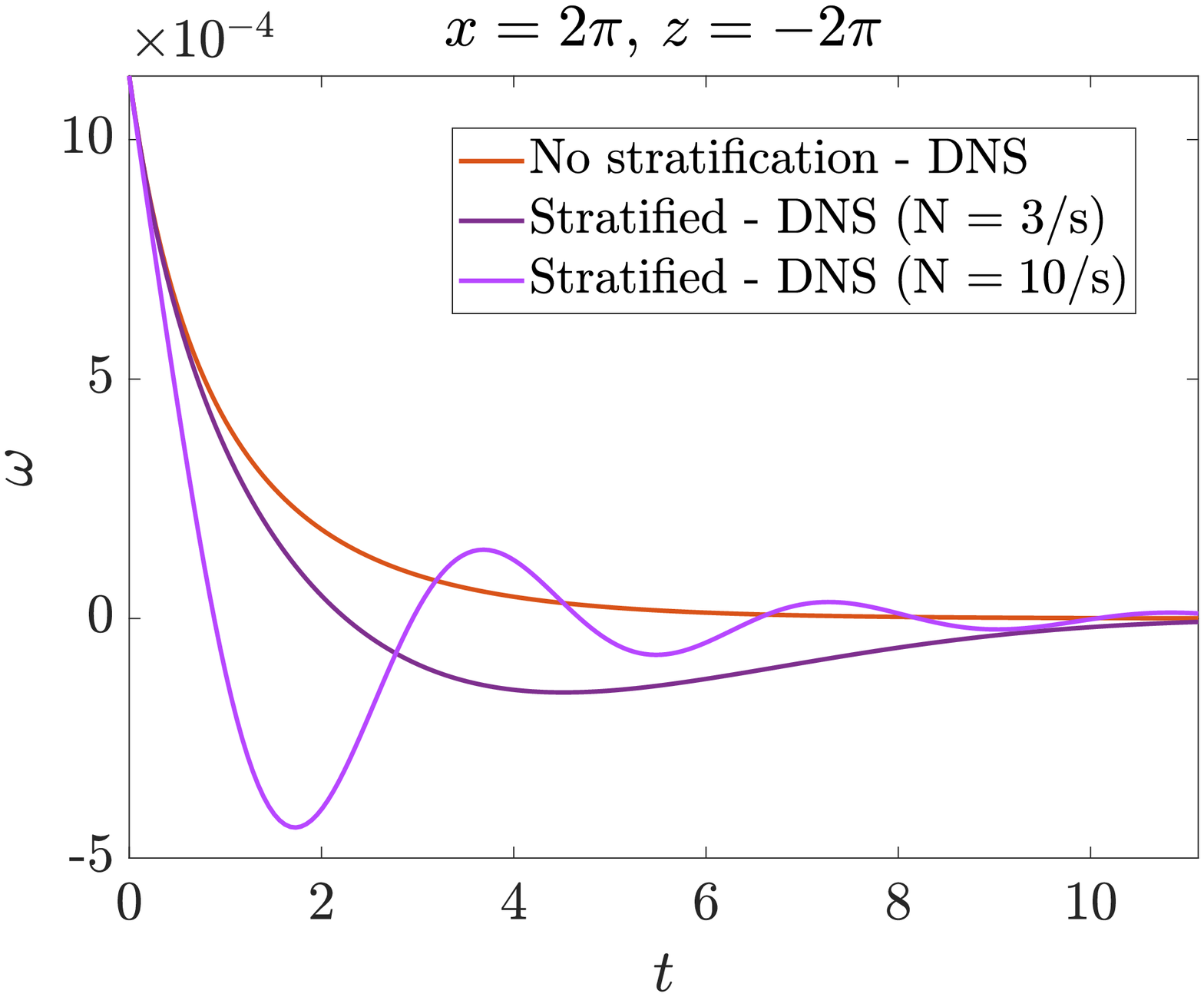}
    \caption{Vorticity vs time corresponding to cases (c) and (d) in table \ref{tab_dns} from DNS.}
    \label{fig:figure12}
\end{figure}
Turning now to case (c) in table \ref{tab_dns}, which represents a stratified case (Brunt-Vaisala frequency $N=3\text{s}^{-1}$) with only vorticity initial perturbation. Figs. \ref{fig:figure11}depict the vorticity at a particular location $x=2\pi, z=-2\pi$. It is seen from this figure that the initial vorticity perturbation has projections on both vorticity as well as internal gravity modes (IG). Note that in this case, we do not impose any free-surface disturbance and thus the SG modes are never excited. It is seen that the vorticity at this location changes sign going to zero at large time. In order to obtain the vorticity signal seen in DNS, contributions from both SG as well as IG modes are found to be equally important. Figures \ref{fig:figure11b} and \ref{fig:figure11c} show the perturbation vorticity field at $t=0$ and $t=3.91$ respectively. Once again, we see the same qualitative features as \ref{fig:figure11a} viz. that in order to match the DNS vorticity profile accurately, contributions from both IG and vorticiy modes are to be taken into account. Finally, figure \ref{fig:figure12} shows the effect of increasing stratification (case (d) compared to case (c)) on the evolution of the vorticity perturbation as a function of time. Due to increased stratification in case (d), a significant oscillatory response in the local vorticity value is seen for case (d).

\section{Conclusions} In this study, we have solved the linearised initial-value problem (IVP) for a stratified pool of viscous liquid, corresponding to free-surface, vorticity and density initial conditions. The temporal spectrum has been studied carefully in various limits (viscous versus inviscid, stratified versus unstratified and finite versus infinite depth) relating this to the response predicted from our analytical solution to the IVP. We have compared some of our analytical predictions against Direct Numerical Simulations using the open source code Basilisk \citep{popinet2014basilisk}, obtaining excellent agreement in all cases. The analytical solutions permits individual examination of various parts of the spectrum  and their respective contributions. In the unstratified, finite depth limit, we find that the countably infinite set of vorticity modes are crucial for satisfying initial conditions as well as for describing any perturbation vorticity imposed at depths, where the contribution from the surface-gravity (SG) modes is effectively zero. However, when this perturbation vorticity field (Gaussian) is chosen to be at sufficient depth, this field evolves nearly independently of the transient vorticity field produced at the free-surface. Consequently, we find that when the perturbation vorticity field is introduced at large depth, it produces a negligible surface signature, but nevertheless requires the vorticity modes to describe its temporal evolution. In contrast, the vorticity at the free surface has near equal contributions from both SG as well as vorticity modes and neglecting any of these leads to a large error. In the infinite depth limit, the contribution from the continuous spectrum modes to the time evolution of the vorticity field is found to be numerically indistinguishable from that of the countably infinite, vorticity modes in the finite, but large depth case. Considering a stratified viscous pool of liquid without any initial surface perturbation and an initial Gaussian shaped localised perturbation vorticity patch, we find that the time evolution of the patch has nearly equal contributions from the vorticity and the internal gravity modes. Our study with a careful comparison of the temporal spectrum in each case to the IVP response, extends previous results for an unstratified pool due to \cite{prosperetti1976viscous, prosperetti1982small} and to the viscous case for a stratified pool \citep{yih1960gravity}. Several analytical predictions are tested against fully nonlinear, numerical simulations obtaining excellent agreement in all cases. This study will have several applications in studies of thermo-capillary instabilities on a liquid layer.
\newpage
\section*{Acknowledgements:} We thank Dr. Palas Kumar Farsoiya for discussions during the early stages of this study. Financial support from DST-SERB (Govt. of India) grants \#MTR/2019/001240, \#MTR/2021/000706 and \#SPR/2021/000536 are gratefully acknowledged. The tenure of SB at IIT Bombay, was supported by the Prime Minister's Research Fellowship and is thankfully acknowledged. We acknowledge the support by IIT Madras of the "Geophysical Flows Lab" research initiative under the Institute of Eminence framework.
\bibliography{apssamp}
\end{document}


\maketitle
\newcommand{\A}{\mathcal{A}}
\newcommand{\B}{\mathcal{B}}
\newcommand{\C}{\mathcal{C}}
\newcommand{\D}{\mathcal{D}}
\newcommand{\F}{\mathcal{F}}
\newcommand{\G}{\mathcal{G}}
\newcommand{\mk}{{\mathup{K}}}
\newcommand{\mi}{{\mathup{I}}}
\newcommand{\mj}{{\mathup{J}}}
\newcommand{\my}{{\mathup{Y}}}
	
\newcommand{\M}{\mathcal{M}}
\newcommand{\I}{^{\mathcal{I}}}
\newcommand{\Id}{_{\mathcal{I}}}
\newcommand{\Ou}{^{\mathcal{O}}}
\newcommand{\Oud}{_{\mathcal{O}}}

\newcommand{\OI}{{\Omega}\I}
\newcommand{\SI}{{\Psi}\I}
\newcommand{\OO}{{\Omega}\Ou}
\newcommand{\SO}{{\Psi}\Ou}
\newcommand{\hs}{\hat{s}}
\newcommand{\hr}{\hat{r}}
\newcommand{\CO}{\tilde{\Omega}}
\newcommand{\CP}{\tilde{\Psi}}
\newcommand{\CPh}{\tilde{\phi}}	
\newcommand{\tP}{\tilde{\mathcal{P}}}
\newcommand{\ak}{\tilde{a}_k}
\newcommand*{\dt}[1]{%
\accentset{\mbox{\large\bfseries .}}{#1}}
\newcommand{\erf}{\mathrm{Erf}\,}
\newcommand{\erfc}{\mathrm{Erfc}\,}

\newtcbox{\mymath}[1][]{%
	nobeforeafter, math upper, tcbox raise base,
	enhanced, colframe=blue!30!black,
	colback=blue!30, boxrule=1pt,
	#1}

\newcommand\hcancel[2][black]{\setbox0=\hbox{$#2$}%
	\rlap{\raisebox{.45\ht0}{\textcolor{#1}{\rule{\wd0}{1pt}}}}#2}

\section{Perturbations on a stratified  water layer}
\subsection{Governing equations: Boussinesq approximation}
Consider a pool of depth $H$ comprising of quiescent, density stratified fluid (uniform stratification) bounded at the top with a free surface and at the bottom by a wall. Assuming density to be a (linear) function of temperature $T$ (and independent of other thermodynamic variables like salinity and pressure), we obtain the Boussinesq set of equations governing variations in velocity, pressure and density inside the pool viz. \citep{foster1970drag}
\begin{eqnarray}
	&&\bm{\nabla}\cdot\mathbf{\tilde{u}^*} = 0 \label{1}\\
	&&\frac{D\mathbf{\tilde{u}^*}}{Dt^*} = -\frac{1}{\rho_{\text{ref}}}\bm{\nabla}\tilde{p}^* + \frac{\tilde{\rho}^*}{\rho_{\text{ref}}}\mathbf{g} + \frac{\mu }{\rho_{\text{ref}}}\nabla^2 \mathbf{\tilde{u}^*} \label{2}\\
    && \frac{D\tilde{\rho}^*}{Dt^*} = \kappa \nabla^2 \tilde{\rho}^* \label{3}
\end{eqnarray}
where $\mathbf{\tilde{u}^*}  = (\tilde{u}^*,\tilde{w}^*)$, $\mathbf{g} = (0,-g)$, $\rho_{\text{ref}}$ is a reference density at a reference temperature $T_{\text{ref}}$ and $\kappa$ is the thermal diffusivity. The base state whose stability will be studied comprises of quiescent viscous fluid with a density profile $\rho^*_b(z)$ and uniform viscosity $\mu$, with pressure field satisfying the hydrostatic balance $\dfrac{dp_b^*}{dz^*} = - \rho_b^*(z^*)g$. Note that all base state quantities are indicated with the subscript `$b$' while dimensional quantities have an asterisk. Equations \ref{1}-\ref{3} are supplemented with boundary conditions at the free surface ($z^* = \eta^*$) and the wall ($z^*=-H^*$). These are
\begin{eqnarray}
	&& \frac{D\eta^*}{Dt^*} = \left(\mathbf{\tilde{u}^*}\cdot\mathbf{e}_z\right)_{z^* = \eta^*} \label{4}\\
	&& \left(\mathbf{t}\cdot\tilde{\underline{\underline{\sigma}}}^*\cdot\mathbf{n}\right)_{z^* = \eta^*} = 0  \label{5}\\
	&& \left(\mathbf{n}\cdot\tilde{\underline{\underline{\sigma}}}^*\cdot\mathbf{n}\right)_{z^* = \eta^*} = -\tau\left(\bm{\nabla}\cdot\mathbf{n}\right)_{z^*=\eta^*} \label{6}\\
	&& \tilde{\rho}^*(x^*,z^*=\eta^*,t^*) = \rho_b^*(\eta^*) \label{7}\\
	 && \mathbf{\tilde{u}^*}(x^*,z^*=-H^*,t^*) = 0 \label{8}
\end{eqnarray}
where $\mathbf{t}$ and $\mathbf{n}$ are the unit tangent and normal to the perturbed interface.  Eqns. (\ref{4}-\ref{8}) are the kinematic boundary condition, the zero shear stress condition, the jump in normal stresses due to surface tension $\tau$ at the free surface, constant density condition at the interface and the zero-velocity condition at the bottom wall respectively. Decomposing all dependent variables as a sum of base and perturbation variables viz.
\begin{eqnarray}
	\mathbf{\tilde{u}^*} = \mathbf{0} + \mathbf{u}^*, \; \tilde{p}^*(\mathbf{x}^*,t^*) = p_b^*(z^*) + p^*(\mathbf{x}^*,t^*), \; \tilde{\rho}^*(\mathbf{x}^*,t^*) = \rho_b^*(z^*) + \rho^*(\mathbf{x}^*,t^*),\; z^* = 0 + \eta^*(x^*,t^*).\nonumber \\\label{9}
\end{eqnarray}
Substituting in eqns. (\ref{1}-\ref{3}) and (\ref{4}-\ref{8}), linearising in the perturbed variables using Taylor series expansion about $z^*=0$, we obtain in component form
\begin{eqnarray}
    &&     \frac{\partial u^*}{\partial x^*} + \frac{\partial w^*}{\partial z^*} = 0 \label{10} \\
	&& \frac{\partial u^*}{\partial t^*}  = -\frac{1}{\rho_{\text{ref}}}\frac{\partial p^*}{\partial x^*} + \frac{\mu}{\rho_{\text{ref}}}\nabla^2u^*  \label{11} \\
	&&  \frac{\partial w^*}{\partial t^*} = -\frac{1}{\rho_{\text{ref}}}\frac{\partial p^*}{\partial z^*} - \frac{\rho^*}{\rho_{\text{ref}}}g + \frac{\mu}{\rho_{\text{ref}}}\nabla^2w^* \label{12} \\
	&&     \frac{\partial \rho^*}{\partial t^*} + \left(\frac{d\rho_b^*}{dz^*}\right)w^* = \kappa\nabla^2\rho^* \label{13}
\end{eqnarray}
with the corresponding linearised boundary conditions
\begin{eqnarray}
	&&\frac{\partial\eta^*}{\partial t^*} = w^*(x^*,0,t^*) \label{14}\\
	&& \frac{\partial u^*}{\partial z^*}(x^*,0,t^*) + \frac{\partial w^*}{\partial x^*}(x^*,0,t^*) = 0 \label{15}\\
	&& \rho_b^*(0)g\eta^* - p^*(x^*,0,t^*) + 2\mu\frac{\partial w^*}{\partial z^*}(x^*,0,t^*) - \tau\left(\frac{\partial^2\eta^*}{\partial (x^*)^2}\right) = 0 \label{16}\\
	&& \rho^*(x^*,0,t^*) = 0 \label{17}\\
	&& u^*(x^*,-H^*,t^*)  = w^*(x^*,-H^*,t^*) = \rho^*(x^*,-H,t^*)=0 \label{18}
\end{eqnarray}
where we have used the Newtonian constitutive relation $\tilde{\underline{\underline{\bm{\sigma}}}} = -\tilde{p}\;\underline{\underline{\mathbf{I}}} + \mu\left(\bm{\nabla}\tilde{\mathbf{u}} + \bm{\nabla}\tilde{\mathbf{u}}^{\intercal}\right)$. Note that we have used $p_b^*(0)=0$ in \ref{16}. 

We now non-dimensionalise eqns. (\ref{10}-\ref{13}) and boundary conditions (\ref{14}-\ref{18}) using the length, velocity and time scales $L$, $U$, $L/U$, $\rho_{ref}U^2$ and $\rho_{ref}$ respectively to obtain
\begin{eqnarray}
 && \frac{\partial u}{\partial x} + \frac{\partial w}{\partial z} = 0 \label{19}\\
&&\frac{\partial u}{\partial t} = -\frac{\partial p}{\partial x} + \frac{1}{\textrm{Re}}\left(\frac{\partial^2}{\partial x^2} + \frac{\partial^2}{\partial z^2}\right)u \label{20}\\
&& \frac{\partial w}{\partial t} = -\frac{\partial p}{\partial z} + \frac{1}{\textrm{Re}}\left(\frac{\partial^2}{\partial x^2} + \frac{\partial^2}{\partial z^2}\right)w - \dfrac{gL}{U^2}\rho \label{21}\\
&& \frac{\partial\rho}{\partial t} - \textrm{Ri}w = \frac{1}{\textrm{Re} \textrm{Pr}}\left(\frac{\partial^2}{\partial x^2} + \frac{\partial^2}{\partial z^2}\right)\rho \label{22}
\end{eqnarray}
and the boundary conditions
 \begin{eqnarray}
&&\frac{\partial\eta}{\partial t} = w(x,0,t) \label{23}\\
&& \frac{\partial u}{\partial z}(x,0,t)+\frac{\partial w}{\partial x}(x,0,t) = 0 \label{24}\\
&& \rho_b(0g\eta - \left(\dfrac{U^2}{gL}\right)p(x,0,t) + \left(\dfrac{U^2}{gL}\right)\frac{2}{Re}\frac{\partial w}{\partial z}(x,0,t) - \frac{1}{Bo}\left(\frac{\partial^2\eta}{\partial x^2}\right) = 0 \label{25}\\
&& \rho(x,0,t) = 0 \label{26}\\
&& u(x,-H,t)  = w(x,-H,t) = \rho(x,-H,t) =  0 \label{27}
\end{eqnarray}
The non-dimensional numbers are defined by $\textrm{Re} \equiv \frac{\rho_{ref}LU}{\mu}$, $\textrm{Bo} \equiv \dfrac{\rho_{ref}gL^2}{\tau}$, $\textrm{Ri}\equiv\dfrac{\frac{-g}{\rho_{ref}}\frac{d\rho_b^*}{dz^*}L^2}{U^2}$ and $\textrm{Pr} \equiv \dfrac{\mu}{\rho_{ref}\kappa}$. We chose $L$ and $U$ to be $1/k$ and $\sqrt{g/k}$ respectively viz. the length-scale and phase speed of a surface gravity wave. Note that $gL/U^2$ is the inverse Froude number squared which is unity here due to the choise of scales. Fourier and Laplace transforming along $x$ and $t$ respectively
\begin{eqnarray}
	&&\bigg[\bar{u}(k,z,s),\bar{w}(k,z,s),\bar{p}(k,z,s), \bar{\eta}(k,s),\bar{\rho}(k,z,s)\bigg]^{\intercal} \nonumber \\
	 &&\equiv \frac{1}{\sqrt{2\pi}}\int_{0}^{\infty}dt \exp[-st]\int_{-\infty}^{\infty}dk\exp[-ikx]\bigg[u(x,z,t),w(x,z,t),p(x,z,t), \eta(x,t),\rho(x,z,t)\bigg]^{\intercal} \nonumber \\
	 &&\label{28}
\end{eqnarray}
Eliminating $\bar{p}(k,z,s)$ by cross differentiation and using the continuity eqn.  we obtain an equation for $\bar{w}(k,z,s)$ viz.
\begin{align}
\nonumber
\Big[s-&\frac{1}{\textrm{RePr}}(D^2 - k^2)\Big]\Big[s-\frac{1}{\textrm{Re}}(D^2-k^2)\Big](D^2-k^2)\bar{w}(k,z,s) - k^2\textrm{Ri}\bar{w}(k,z,s)\\ \nonumber &= k^2\left(\frac{gL}{U^2}\right)\hat{\rho}(k,z,t=0)+\Big[s-\frac{1}{\textrm{RePr}}(D^2-k^2)\Big](D^2-k^2)\hat{w}(k,z,t=0)
\end{align}
Here $D = d/dz$, $\textrm{Ri} = N^2L^2/U^2$ is the Richardson number. Re-writing the above equation using the perturbation stream-function $\bar{\psi}$ and the out-of-plane (z) component of vorticity $\hat{\omega}$, 
\begin{align}
\nonumber
\Big[(D^2 - 1)-s\textrm{RePr}\Big]&\Big[(D^2-1) - s\textrm{Re}\Big](D^2-1)\bar{\psi}(s,z) - \textrm{Ri Re}^2\textrm{Pr}\bar{\psi}(s,z) \\ &= \textbf{i}\textrm{Re}^2\textrm{Pr}\hat{\rho}(z,t=0)+\textrm{Re}\Big[(D^2-1)-s\textrm{RePr} \Big]\hat{\omega}(z,t=0)\label{29}
\end{align}
Note that as we are non-dimensionalising our equations by some reference wavenumber $k$, all $k$ are set to unity in the following equations. The expressions for $\bar{u}$, $\bar{w}$ and $\bar{p}$ can be found from the stream-function and vorticity formulation and equation \ref{20} as
\begin{equation}
\nonumber
\bar{u} = D\bar{\psi}, \hspace{1cm} \bar{w} = -\textbf{i}\bar{\psi},
\end{equation}
\begin{equation}
\nonumber
 \bar{p} =  -\textbf{i}\Big[Re^{-1}(D^2 - 1)-s\Big]D\bar{\psi} - \textbf{i}D\hat{\psi}(z,t=0)
\end{equation}
\subsection{Solution of sixth order equation}
The homogeneous solutions of equation \ref{29} are of the form $\sim C_i e^{\lambda_i z}$. Using variation of parameters, we may write the general solution to equation \ref{29} as \citep{riley1999mathematical}
\begin{align}
\bar{\psi}(s,z) = \sum_{i=1}^6\exp{(\lambda_iz)}\Bigg[C_i(s) + \int^z \frac{Adj(\mathscr{W})_{i,6}(s,z')\mathscr{R}(s,z')}{|\mathscr{W}(s,z')|}dz'\Bigg]\label{30}
\end{align}
Here, $\mathscr{R}(s,z)$ is the RHS of equation \ref{29} and $|\mathscr{W}(s,z)|$ is the sixth-order Wronskian constructed from the homogeneous solutions of equation \ref{29}. The Wronskian matrix is
\begin{equation*}
\mathscr{W}_{j,i} = \begin{bmatrix}
    e^{\lambda_1 z} & e^{\lambda_2 z} & e^{\lambda_3 z} & e^{\lambda_4 z} & e^{\lambda_5 z} & e^{\lambda_6 z}\\
   \lambda_1 e^{\lambda_1 z} & \lambda_2 e^{\lambda_2 z} & \lambda_3 e^{\lambda_3 z} & \lambda_4 e^{\lambda_4 z} & \lambda_5 e^{\lambda_5 z} & \lambda_6 e^{\lambda_6 z}\\
     \lambda_1^2 e^{\lambda_1 z} & \lambda_2^2 e^{\lambda_2 z} & \lambda_3^2 e^{\lambda_3 z} & \lambda_4^2 e^{\lambda_4 z} & \lambda_5^2 e^{\lambda_5 z} & \lambda_6^2 e^{\lambda_6 z}\\
      \lambda_1^3 e^{\lambda_1 z} & \lambda_2^3 e^{\lambda_2 z} & \lambda_3^3 e^{\lambda_3 z} & \lambda_4^3 e^{\lambda_4 z} & \lambda_5^3 e^{\lambda_5 z} & \lambda_6^3 e^{\lambda_6 z}\\
      \lambda_1^4 e^{\lambda_1 z} & \lambda_2^4 e^{\lambda_2 z} & \lambda_3^4 e^{\lambda_3 z} & \lambda_4^4 e^{\lambda_4 z} & \lambda_5^4 e^{\lambda_5 z} & \lambda_6^4 e^{\lambda_6 z}\\
     \lambda_1^5 e^{\lambda_1 z} & \lambda_2^5 e^{\lambda_2 z} & \lambda_3^5 e^{\lambda_3 z} & \lambda_4^5 e^{\lambda_4 z} & \lambda_5^5 e^{\lambda_5 z} & \lambda_6^5 e^{\lambda_6 z}\\
\end{bmatrix}
\end{equation*}
and its determinant is $|\mathscr{W}(s,z)|$. The Wronskian determinant may be written as:
\begin{equation*}
|\mathscr{W}(s,z)| = \exp{\left(\sum_{i=1}^6\lambda_i z\right)} \begin{vmatrix}
    1 & 1 & 1 & 1 & 1 & 1\\
   \lambda_1  & \lambda_2  & \lambda_3  & \lambda_4  & \lambda_5  & \lambda_6 \\
     \lambda_1^2  & \lambda_2^2  & \lambda_3^2  & \lambda_4^2  & \lambda_5^2  & \lambda_6^2 \\
      \lambda_1^3  & \lambda_2^3  & \lambda_3^3  & \lambda_4^3  & \lambda_5^3  & \lambda_6^3 \\
      \lambda_1^4  & \lambda_2^4  & \lambda_3^4  & \lambda_4^4  & \lambda_5^4  & \lambda_6^4 \\
     \lambda_1^5  & \lambda_2^5  & \lambda_3^5  & \lambda_4^5  & \lambda_5^5  & \lambda_6^5 \\
\end{vmatrix}
= \exp{\left(\sum_{i=1}^6\lambda_i z\right)} \mathscr{E}_{\lambda}(s)
\end{equation*}

Similarly each element in the last column of the adjugate matrix of $\mathscr{W}_{i,j}$, is  given by $Adj(\mathscr{W})_{i,6}(s,z)$. They are:
\begin{equation*}
Adj(\mathscr{W})_{1,6} = - \exp{\left(\sum_{i=1}^6\lambda_i - \lambda_1\right)z} \begin{vmatrix}
    1 & 1 & 1 & 1 & 1\\
   \lambda_2  & \lambda_3  & \lambda_4  & \lambda_5  & \lambda_6 \\
    \lambda_2^2  & \lambda_3^2  & \lambda_4^2  & \lambda_5^2  & \lambda_6^2 \\
     \lambda_2^3  & \lambda_3^3  & \lambda_4^3  & \lambda_5^3  & \lambda_6^3 \\
     \lambda_2^4  & \lambda_3^4  & \lambda_4^4  & \lambda_5^4  & \lambda_6^4 \\
\end{vmatrix}
= \exp{\left(\sum_{i=1}^6\lambda_i - \lambda_1 z\right)} \mathscr{E}_{\lambda,1}(s)
\end{equation*}
\begin{equation*}
Adj(\mathscr{W})_{2,6} =  e^{\big(\sum_{i=1}^6\lambda_i - \lambda_2\big)z} \begin{vmatrix}
    1 & 1 & 1 & 1 & 1\\
   \lambda_1  & \lambda_3  & \lambda_4  & \lambda_5  & \lambda_6 \\
     \lambda_1^2  & \lambda_3^2  & \lambda_4^2  & \lambda_5^2  & \lambda_6^2 \\
      \lambda_1^3  & \lambda_3^3  & \lambda_4^3  & \lambda_5^3  & \lambda_6^3 \\
      \lambda_1^4  & \lambda_3^4  & \lambda_4^4  & \lambda_5^4  & \lambda_6^4 \\
\end{vmatrix}
= e^{\big(\sum_{i=1}^6\lambda_i - \lambda_2\big)z} \mathscr{E}_{\lambda,2}(s)
\end{equation*}
\begin{equation*}
Adj(\mathscr{W})_{3,6} = - e^{\big(\sum_{i=1}^6\lambda_i - \lambda_3\big)z} \begin{vmatrix}
    1 & 1 & 1 & 1 & 1\\
   \lambda_1  & \lambda_2  & \lambda_4  & \lambda_5  & \lambda_6 \\
     \lambda_1^2  & \lambda_2^2   & \lambda_4^2  & \lambda_5^2  & \lambda_6^2 \\
      \lambda_1^3  & \lambda_2^3   & \lambda_4^3  & \lambda_5^3  & \lambda_6^3 \\
      \lambda_1^4  & \lambda_2^4   & \lambda_4^4  & \lambda_5^4  & \lambda_6^4 \\
     
\end{vmatrix}
=e^{\big(\sum_{i=1}^6\lambda_i - \lambda_3\big)z} \mathscr{E}_{\lambda,3}(s)
\end{equation*}
\begin{equation*}
Adj(\mathscr{W})_{4,6} =  e^{\big(\sum_{i=1}^6\lambda_i - \lambda_4\big)z} \begin{vmatrix}
    1 & 1 & 1 & 1 & 1\\
   \lambda_1  & \lambda_2  & \lambda_3  & \lambda_5  & \lambda_6 \\
     \lambda_1^2  & \lambda_2^2  & \lambda_3^2  & \lambda_5^2  & \lambda_6^2 \\
      \lambda_1^3  & \lambda_2^3  & \lambda_3^3   & \lambda_5^3  & \lambda_6^3 \\
      \lambda_1^4  & \lambda_2^4  & \lambda_3^4  & \lambda_5^4  & \lambda_6^4 \\
    
\end{vmatrix}
= e^{\big(\sum_{i=1}^6\lambda_i - \lambda_4\big)z} \mathscr{E}_{\lambda,4}(s)
\end{equation*}
\begin{equation*}
Adj(\mathscr{W})_{5,6} = - e^{\big(\sum_{i=1}^6\lambda_i - \lambda_5\big)z} \begin{vmatrix}
    1 & 1 & 1 & 1 & 1\\
   \lambda_1  & \lambda_2  & \lambda_3  & \lambda_4  & \lambda_6 \\
     \lambda_1^2  & \lambda_2^2  & \lambda_3^2  & \lambda_4^2  & \lambda_6^2 \\
      \lambda_1^3  & \lambda_2^3  & \lambda_3^3  & \lambda_4^3  & \lambda_6^3 \\
      \lambda_1^4  & \lambda_2^4  & \lambda_3^4  & \lambda_4^4  & \lambda_6^4 \\
\end{vmatrix}
=e^{\big(\sum_{i=1}^6\lambda_i - \lambda_5\big)z} \mathscr{E}_{\lambda,5}(s)
\end{equation*}
\begin{equation*}
Adj(\mathscr{W})_{6,6} =e^{\big(\sum_{i=1}^6\lambda_i - \lambda_6\big)z} \begin{vmatrix}
    1 & 1 & 1 & 1 & 1 \\
   \lambda_1  & \lambda_2  & \lambda_3  & \lambda_4  & \lambda_5 \\
     \lambda_1^2  & \lambda_2^2  & \lambda_3^2  & \lambda_4^2  & \lambda_5^2   \\
      \lambda_1^3  & \lambda_2^3  & \lambda_3^3  & \lambda_4^3  & \lambda_5^3 \\
      \lambda_1^4  & \lambda_2^4  & \lambda_3^4  & \lambda_4^4  & \lambda_5^4  \\
\end{vmatrix}
= e^{\big(\sum_{i=1}^6\lambda_i - \lambda_6\big)z} \mathscr{E}_{\lambda,6}(s)
\end{equation*}
Here, $\mathscr{E}_\lambda$ and $\mathscr{E}_{\lambda,i}$ are the determinants in the above terms. Therefore, equation \ref{30} can be rewritten as:
\begin{align}
\nonumber
\bar{\psi}(s,z) &= \sum_{i=1}^6\exp{(\lambda_iz)}\Bigg[C_i(s) + \frac{\mathscr{E}_{\lambda,i}(s)}{\mathscr{E}_{\lambda}(s)}\int \exp(-\lambda_i z)\mathscr{R}(s,z)dz\Bigg] \\ &= \sum_{i=1}^6\exp{(\lambda_iz)}\Bigg[C_i(s) + \mathscr{E}_{i}(s)I_i(s,z)\Bigg]\label{31}
\end{align}
Here, $I_i(s,z)$ is the integral and $\mathscr{E}_i$ is the ratio of determinants. It can be shown by factorization that:
\begin{equation*}
\mathscr{E}_i(s) = \frac{1}{\displaystyle \prod_{j=1,j\neq i}^{6} (\lambda_i - \lambda_j)},\hspace{1cm}I_i(s,z) = \int \exp(-\lambda_i z)\mathscr{R}(s,z)dz
\end{equation*}
If the stratification is absent, $Ri = 0$, and we take density perturbations also to be absent. Therefore, the governing equation \ref{29} modifies to:
\begin{align}
\Big[(D^2-1)-&sRe\Big](D^2-1)\bar{\psi}(z) = Re\hat{\omega}(z,t=0)\label{32}
\end{align}
Therefore, assuming solutions of the form $\sim \exp(\lambda_i z)$, we can modify equation \ref{31} as:
\begin{align}
\bar{\psi}(s,z) = \sum_{i=1}^4\exp{(\lambda_iz)}\Bigg[C_i(s) + \mathscr{F}_{i}(s)I_i(s,z)\Bigg]\label{33}
\end{align}
where, 
\begin{equation*}
\mathscr{F}_i(s) = \frac{1}{\displaystyle \prod_{j=1,j\neq i}^{4} (\lambda_i - \lambda_j)},\hspace{1cm} I_i(s,z) = Re\int_0^z \exp(-\lambda_i z)\hat{\omega}(z,t=0)dz
\end{equation*}
\subsection{Boundary Conditions}
After applying Fourier and Laplace transforms and using  $\sqrt{g/k}$ as the velocity scale $U$ and the inverse of wave-number $1/k$ as length scale $L$
\begin{equation}
\nonumber
s\bar{\eta}(s) = \bar{w}(s,z=0) + \hat{\eta}(t=0)
\end{equation}
\begin{equation}
\nonumber
(D^2+1)\bar{w}(s,z=0) = 0
\end{equation}
\begin{equation}
\nonumber
\frac{2}{Re}D\bar{w}(s,z=0)+ (1+Bo^{-1})\bar{\eta}(s)= \frac{1}{Re}\Big(D^2 - 1 - s Re\Big)D\bar{w}(s,z=0) + D\hat{w}(z=0,t=0)
\end{equation}
\begin{equation}
\nonumber
\Big[s-\frac{1}{Re}\big(D^2 - 1\big)\Big]\big(D^2-1\big)\bar{w}(s,z=0) = (D^2-1)\hat{w}(z=0,t=0)
\end{equation}
Here, $Bo^{-1}$ is the inverse of Bond number defined by $Bo = \rho_{ref} g/\tau k^2$. In the stream-function and vorticity formulation, these equations can be modified as
\begin{equation}
(D^2+1)\bar{\psi}(s,0) = 0\label{34}
\end{equation}
\begin{align}
 s\Big((D^2 - 3) - s Re\Big)D\bar{\psi}(s,0) &- Re(1+Bo^{-1})\bar{\psi}(s,0)\label{35}\\ &= Re(1+Bo^{-1})\textbf{i}\hat{\eta}(t=0) - s Re D\hat{\psi}(t=0,z=0)\nonumber
\end{align}
\begin{equation}
\Big[s-\frac{1}{Re}\big(D^2 - 1\big)\Big]\big(D^2-1\big)\bar{\psi}(s,0) = -\hat{\omega}(z=0,t=0)\label{36}
\end{equation}
Similarly, at the wall ($z = -H$), applying no-slip and no-penetration boundary conditions:
\begin{equation}
\begin{aligned}
    \bar{w} = 0,\hspace{1cm}\frac{\partial \bar{w}}{\partial z} = 0,\hspace{1cm} \bar{\rho} = 0
    \nonumber
\end{aligned}
\end{equation}
The Neumann boundary condition on $\bar{w}$ above is from continuity equation. The z-momentum equation has a $\bar{\rho}$ term. So substituting $\bar{\rho} = 0$  in z-momentum equation will give a boundary condition on $\bar{w}$. Therefore, doing that, applying a laplace transform and converting into stream-function and vorticity formulation will give
\begin{equation}
\begin{aligned}
    \bar{\psi}(s,-H) = 0,\hspace{0.3cm}\frac{\partial \bar{\psi}}{\partial z}(s,-H) = 0, \hspace{0.3cm}  \Big[s-\frac{1}{Re}\big(D^2 - 1\big)\Big]\big(D^2 - 1 \big)\bar{\psi}(s,-H) = -\hat{\omega}(z = -H,t=0)\label{37}
\end{aligned}
\end{equation}
For infinite-depth case, we choose conditions \ref{34}, \ref{35} and \ref{36} at $z = 0$ and $\bar{\psi} \rightarrow \textrm{finite}$ condition as $z\rightarrow-\infty$. For no-stratification case, we choose conditions \ref{34} and \ref{35} at $z=0$ and \ref{37}a and \ref{37}b at $z=-H$.
\section{Viscous - Finite Depth - Stratified Scenario}
We substitute the solution form given in equation \ref{31}, into boundary conditions \ref{34}-\ref{37}. The tangential stress condition condition is given by:
\begin{equation}
\sum_{i=1}^6 (\lambda_i^2 + 1)C_i(s) = -\sum_{i=1}^6 \mathscr{E}_i(s)\big((D+\lambda_i)^2 + 1\big)I_i(s,z=0)\label{2_1}
\end{equation}
Similarly, the normal stress boundary condition is given by
\begin{align}
\nonumber
&\sum_{i=1}^6 \bigg[s\lambda_i^3 - s\lambda_i(sRe+3) - Re(1+Bo^{-1})\bigg]C_i(s) =\\ \nonumber & -\sum_{i=1}^6 \mathscr{E}_i(s)\bigg[s(D+\lambda_i)^3- s(sRe + 3)(D+\lambda_i) -  Re(1+Bo^{-1})\bigg]I_i(s,z=0)\\ & + Re(1+Bo^{-1})\textbf{i}\hat{\eta}(t=0) - s Re D\hat{\psi}(t=0,z=0)\label{2_2}
\end{align}
The vanishing density fluctuation boundary condition is given by:
\begin{align}
\nonumber
\sum_{i=1}^6 \bigg[\lambda_i^4 &- \lambda_i^2(sRe+2) + (1+sRe)\bigg]C_i(s) = -\sum_{i=1}^6 \mathscr{E}_i(s)\bigg[(D+\lambda_i)^4\\ &- (sRe + 2)(D+\lambda_i)^2 +  (1+sRe)\bigg]I_i(s,z=0) + Re\hat{\omega}(z=0,t=0)\label{2_3}
\end{align}
The no-penetration boundary condition at the bottom surface is given by:
\begin{equation}
\sum_{i=1}^6e^{-\lambda_i H}C_i(s) = -\sum_{i=1}^6e^{-\lambda_i H}\mathscr{E}_i(s)I_i(s,z=-H)\label{2_4}
\end{equation}
The no-slip boundary condition is given by:
\begin{equation}
\sum_{i=1}^6\lambda_i e^{-\lambda_i H}C_i(s) = -\sum_{i=1}^6\mathscr{E}_i(s)e^{-\lambda_i H}(D+\lambda_i)I_i(s,z=-H)\label{2_5}
\end{equation}
The density fluctuation boundary condition is given by:
\begin{align}
\nonumber
\sum_{i=1}^6 e^{-\lambda_i H}\bigg[\lambda_i^4 &- (sRe+2) \lambda_i^2 \bigg]C_i(s) = -\sum_{i=1}^6 e^{-\lambda_i H} \mathscr{E}_i(s)\bigg[(D+\lambda_i)^4\\ &- (sRe + 2)(D+\lambda_i)^2 \bigg]I_i(s,z=-H) + Re\hat{\omega}(z=-H,t=0)\label{2_6}
\end{align}
The equations \ref{2_1}-\ref{2_6} can be written as
\begin{equation}
\begin{bmatrix}
(\lambda_1^2 + 1) &(\lambda_2^2 + 1) &(\lambda_3^2 + 1) &(\lambda_4^2 + 1) &(\lambda_5^2 + 1) &(\lambda_6^2 + 1) \\
L_1 &L_2 &L_3 &L_4 &L_5 &L_6\\
M_1 &M_2 &M_3 &M_4 &M_5 &M_6\\
e^{-\lambda_1 H} &e^{-\lambda_2 H} &e^{-\lambda_3 H} &e^{-\lambda_4 H} &e^{-\lambda_5 H} &e^{-\lambda_6 H}\\
\lambda_1e^{-\lambda_1 H} &\lambda_2e^{-\lambda_2 H} &\lambda_3e^{-\lambda_3 H} &\lambda_4e^{-\lambda_4 H} &\lambda_5e^{-\lambda_5 H} &\lambda_6e^{-\lambda_6 H}\\
O_1 &O_2 &O_3 &O_4 &O_5 &O_6 
\end{bmatrix}
\begin{bmatrix}
C_1 \\ C_2 \\ C_3 \\ C_4 \\ C_5 \\ C_6
\end{bmatrix}
=
\begin{bmatrix}
\mathscr{R}_1 \\ \mathscr{R}_2 \\ \mathscr{R}_3 \\ \mathscr{R}_4 \\ \mathscr{R}_5 \\ \mathscr{R}_6
\end{bmatrix}\label{2_7}
\end{equation}
where, $L_i(s)$, $M_i(s)$ and $O_i(s)$ are coefficients of $C_i(s)$ in equations \ref{2_2}, \ref{2_3} and \ref{2_6} respectively and $\mathscr{R}_i(s)$ are the RHS terms of the six equations. We observe that the matrix multiplying the $C_i(s)$ is the matrix used in the dispersion relation in the normal mode analysis and denote this by $\mathscr{D}(s)$. Therefore, we can write the coefficients as:
\begin{equation}
C_i(s) =  \frac{1}{|\mathscr{D}(s)|}\sum_{j=1}^6Adj(\mathscr{D}(s))_{ij}\mathscr{R}_j(s)\label{2_8}
\end{equation}
where, $Adj(\mathscr{D}(s))$ implies the adjugate matrix of a matrix $\mathscr{D}(s)$. The inverse laplace transform of the total solution, then, can be given by:
\begin{align}
\hat{\psi}(t,z) = \frac{1}{2\pi\textbf{i}}\int_{\gamma-\textbf{i}\infty}^{\gamma+\textbf{i}\infty}ds\exp{(st)}\sum_{i=1}^6\exp{(\lambda_i(s)z)}\Bigg[\frac{1}{|\mathscr{D}(s)|}\sum_{j=1}^{6}Adj(\mathscr{D}(s))_{ij}\mathscr{R}_j(s) + \mathscr{E}_{i}(s)I_i(s,z)\Bigg]\label{2_9}
\end{align}
Note that the initial conditions $\hat{\rho}(z,t=0)$ and $\hat{\omega}(z,t=0)$ are contained in $\mathscr{R}_j(s)$ and $I_i(s,z)$ in expression 2.9. 
\subsection{\textbf{Vorticity perturbation}}
Assuming all perturbations other than vorticity to be zero, we solve the Poisson equation to find initial stream-function from initial vorticity. The stream-function Poisson equation, after a Fourier transform in x-direction, is given by:
\begin{equation}
\big(D^2-1\big)\hat{\psi}(z,t=0) = -\hat{\omega}(z,t=0)\label{2_10}
\end{equation}
We solve this using variation of parameters:
\begin{align}
\nonumber \hat{\psi}(z,t=0) &= B_1 e^{z} + B_2 e^{-z} + e^z\int\frac{e^{-z'}\hat{\omega}(z',0)}{|\mathscr{W}|}dz' - e^{-z}\int \frac{e^{z'}\hat{\omega}(z',0)}{|\mathscr{W}|}dz'\\ & = \sum_{i=1}^2 e^{(-1)^{i+1} z}\bigg[B_i + \frac{(-1)^i}{2}J_i(z)\bigg]\label{2_11}
\end{align}
Here, the Wronskian $|\mathscr{W}| = -2$ and $J_i(z) \equiv \int^zdz'\omega(z',0)\exp\big[(-1)^iz\big]$. To find the constants $B_1$ and $B_2$, we use the no-penetration boundary condition at the bottom plane and the tangential stress boundary condition at the free surface. They are given as,
\begin{equation}
   \hat{\psi}(z=-H,t=0) = 0; \hspace{1cm} (D^2+1)\hat{\psi}(z=0,t=0) = 0. \nonumber
\end{equation}
The constants $B_1$ and $B_2$ are obtained as:
\begin{equation}
    B_1 = \dfrac{e^H\textrm{csch}H}{8}\bigg(2J_1(0)-2J_2(0) + 2J_2(-H)-2J_1(-H)e^{-2H}+2J'_1(0)+2J'_2(0)+J''_1(0)-J''_2(0)\bigg)\nonumber
\end{equation}
\begin{equation}
    B_2 = \dfrac{e^{-H}\textrm{csch}H}{8}\bigg(-2J_1(0)+2J_2(0) - 2J_2(-H)e^{2H}+2J_1(-H)-2J'_1(0)-2J'_2(0)-J''_1(0)+J''_2(0)\bigg)\nonumber
\end{equation}
For the case of $\hat{\omega}(z,t=0) = \Omega  \exp\left(-\frac{(z-z_d)^2}{d^2}\right)$, where $z_d \in ( -H, 0)$ and in order to obtain the stream function expression in 2.11, we require 
\begin{equation*}
J_1(z) = \frac{\Omega}{2}d\sqrt{\pi}\exp{\bigg(\frac{d^2}{4} - z_d\bigg)}\erf{\bigg(\frac{d}{2}+\frac{z-z_d}{d}\bigg)}
\end{equation*}
\begin{equation*}
J_2(z) = \frac{\Omega}{2}d\sqrt{\pi}\exp{\bigg(\frac{d^2}{4} + z_d\bigg)}\erf{\bigg(-\frac{d}{2}+\frac{z-z_d}{d}\bigg)}
\end{equation*}
For the case of an infinite domain, the no-penetration condition at $z = -H$ should be replaced by the finiteness condition at $z\to-\infty$. In equation 2.11, we neglect the term diverging as $z\to-\infty$ and only those terms which are bounded are considered. We require the following integrals: 
\begin{equation}
K_i(s,z) = \int \exp \bigg(\displaystyle \frac{-(z-z_d)^2}{d^2}\bigg) \exp (-\lambda_i z) dz =  \frac{1}{2}d\sqrt{\pi}\exp\big(d^2\lambda_i^2/4 -\lambda_i z_d \big)\erf{\bigg(\frac{d\lambda_i}{2}+\frac{z-z_d}{d}\bigg)}\label{2.12}
\end{equation}
where, $\erf{(\cdot)}$ is the error function. Finally, the initial condition:
\begin{equation}
D\hat{\psi}(z=0,t=0)  =-\frac{1}{2}\big(J_1(0)+J_2(0)\big)\label{2_13}
\end{equation}
Assuming no initial density perturbations, the RHS term is given by:
\begin{equation*}
\mathscr{R}(s,z) = Re\big[(D^2 - 1) - s RePr\big]\hat{\omega}(z,t=0)
\end{equation*}
The main integral is:
\begin{equation}
I_i(s,z) = \frac{Re\Omega}{d^2}\bigg[\big(d^2\lambda_i -  2(z-z_d)\big)\exp\bigg(-\lambda_i z - \frac{(z-z_d)^2}{d^2}\bigg)-(1-\lambda^2+sRePr)d^2K_i(s,z)\bigg]\label{2_14}
\end{equation}
The derivatives of which are given by:
\begin{equation*}
DI_i(s,z) = -\frac{Re\Omega}{d^4} \bigg(d^4(1+ s Re Pr) + 2d^2  - 4(z-z_d)^2\bigg)\exp\bigg(-\lambda_i z - \frac{(z-z_d)^2}{d^2}\bigg) 
\end{equation*}
\begin{align*}
D^2I_i(s,z) =& \frac{Re\Omega}{d^6}\bigg( d^6\lambda_i(1 + s Re Pr) + 2d^4\big((1+sRePr)(z-z_d) + \lambda_i\big)\\ &- 4d^2(z-z_d)(\lambda_i(z-z_d)-3)-8(z-z_d)^3\bigg)\exp\bigg(-\lambda_i z - \frac{(z-z_d)^2}{d^2}\bigg) 
\end{align*}
\begin{align*}
D^3I_i(s,z) =& \frac{Re\Omega}{d^8}\Bigg(- d^8(1+sRe Pr)\lambda_i^2+ d^6\Big((1+sRePr)\big(2-4(z-z_d)\lambda_i\big)-2\lambda_i^2\Big) \\ &+ 4d^4\bigg(3+(z-z_d)\big[(z-z_d)\lambda_i^2 - 6\lambda_i-(1+sRePr)(z-z_d)\big]\bigg) \\ &+ 16 d^2 (z-z_d)^2\big(\lambda_i(z-z_d)-3\big) +  16(z-z_d)^4 \Bigg)\exp\bigg(-\lambda_i z - \frac{(z-z_d)^2}{d^2}\bigg) 
\end{align*}
\begin{align*}
D^4I_i(s,z) =& \frac{Re\Omega}{d^{10}}\Bigg( d^{10}(1+sRePr)\lambda_i^3 + 2 d^8\lambda_i\bigg(\lambda_i^2 + 3\lambda_i(1+sRePr)(z-z_d) - 3(1+sRePr)\bigg) \\ &+ 4d^6\bigg(-9\lambda_i - 3(z-z_d)(1+sRePr-3\lambda_i^2) + (z-z_d)^2\lambda_i\big(3(1+sRePr)-\lambda_i^2\big)\bigg)\\ &+ 8d^4(z-z_d)\bigg(-15+18\lambda_i(z-z_d)+(z-z_d)^2\big(1+sRePr-3\lambda_i^2\big)\bigg) \\ &+ 16 d^2 (z-z_d)^3\big(10-3\lambda_i(z-z_d)\big)-32(z-z_d)^5\Bigg)\exp\bigg(-\lambda_i z - \frac{(z-z_d)^2}{d^2}\bigg) 
\end{align*}
Using these derivatives, we may find the RHS of matrix $\mathscr{R}_i(s)$ by substitution. Because we have these derivatives, we can easily find the RHS matrix $\mathscr{R}_i(s)$ by direct substitution.

\subsection{\textbf{Free surface perturbation}}
Assuming initial desnity $\hat{\rho}(z,0)$ and vorticity $\hat{\omega}(z,0)$ perturbations to be zero and a non-zero initial free-surface deformation $\hat{\eta}(t=0)$, the solution in 1.31 simplifies to 
\begin{equation}
\bar{\psi}(s,z) = \frac{1}{|\mathscr{D}(s)|}\sum_{i=1}^6\exp{(\lambda_iz)}\sum_{j=1}^6Adj(\mathscr{D}(s))_{ij}\mathscr{R}_j(s) \label{2_17}
\end{equation}
For this initial condition, the RHS matrix $\mathscr{R}_j(s)$ becomes independent of $s$ i.e. $\mathscr{R}_j$
\begin{equation}
\begin{bmatrix}
0 &Re(1+Bo^{-1})\textbf{i}\hat{\eta}(t=0) & 0 & 0 & 0 & 0
\end{bmatrix}^\intercal\label{2_18}
\end{equation}
Then the inverse laplace transform can be written as:
\begin{align}
\hat{\psi}(t,z) = \dfrac{1}{2\pi\mathbf{i}}\int_{\gamma-i\infty}^{\gamma+i\infty}\exp{(st)}ds\sum_{i=1}^6\exp{(\lambda_i(s)z)}\Bigg[\sum_{j=1}^6\frac{1}{|\mathscr{D}(s)|}Adj(\mathscr{D}(s))_{ij}\mathscr{R}_j(s)\Bigg]\label{2_19}
\end{align}
Assuming that the only poles in 2.17 arise from the denominator $\mathscr{D}(s)$, and these are simple poles ($n = 1,2,...$), we may write using Cauchy residue theorem
\begin{equation}
\hat{\psi}(t,z) = \sum_{n = 1}^\infty\frac{1}{|\mathscr{D}(s_n)|'}\exp{(s_nt)} \sum_{i=1}^6\exp{(\lambda_i(s_n)z)}\Bigg[\sum_{j=1}^6Adj(\mathscr{D}(s_n))_{ij}\mathscr{R}_j(s_n)\Bigg]\label{2_20}
\end{equation}
where prime indicates differentiation with respect to $s$. Then, we may write the surface evolution using the kinematic boundary condition written in Fourier-Laplace domain i.e. $s\bar{\eta}(s)-\hat{\eta}(0)=-\textbf{i}\bar{\psi}(z=0,s)$ and expression 2.18 to obtain
\begin{align}
\nonumber
\frac{\hat{\eta}(t)}{\hat{\eta}(0)} = 1&-\frac{\textbf{i}}{\hat{\eta}(0)}\sum_{n = 1}^\infty \exp{(s_nt)}\sum_{i=1}^6\Bigg[\frac{1}{s_n|\mathscr{D}(s_n)|'}\sum_{j=1}^6Adj(\mathscr{D}(s_n))_{ij}\mathscr{R}_j(s_n)\Bigg] \\ &- \frac{\textbf{i}}{\hat{\eta}(0)}\sum_{i=1}^6\Bigg[\frac{1}{|\mathscr{D}(0)|}\sum_{j=1}^6Adj(\mathscr{D}(0))_{ij}\mathscr{R}_j(0)\Bigg]\label{2_21}
\end{align}
\subsection{\textbf{Density perturbation}}
The initial density perturbation profile should satisfy the Dirichlet boundary conditions at both boundaries. Assuming no other perturbations, we observe that,
\begin{equation}
\mathscr{R}(z) = \textbf{i}Re^2Pr\hat{\rho}(z,t=0)\label{2_22}
\end{equation}
For $\hat{\rho}(z,t=0) = R  \exp(-(z-z_d)^2/d^2)$, where $z_d \in ( -H, 0)$, we use the following integrations:
\begin{equation*}
K_i(s,z) = \int \exp \bigg(\displaystyle \frac{-(z-z_d)^2}{d^2}\bigg) \exp (-\lambda_i z) dz =  \frac{1}{2}d\sqrt{\pi}\exp\big(d^2\lambda_i^2/4 -\lambda_i z_d \big)\erf{\bigg(\frac{d\lambda_i}{2}+\frac{z-z_d}{d}\bigg)}
\end{equation*}
where, $\erf{(\cdot)}$ is the error function. Assuming no other initial conditions, the main integral, is:
\begin{equation*}
I_i(s,z) = \textbf{i}Re^2Pr R K_i(s,z)
\end{equation*}
The derivatives of which are given by:
\begin{equation*}
DI_i(s,z) = \textbf{i}Re^2Pr R \exp \bigg(\displaystyle \frac{-(z-z_d)^2}{d^2}\bigg) \exp (-\lambda_i z)
\end{equation*}
\begin{align*}
D^2I_i(s,z) = -\textbf{i}Re^2Pr R \bigg(\frac{2(z-z_d)}{d^2} + \lambda_i\bigg) \exp \bigg(\displaystyle \frac{-(z-z_d)^2}{d^2}\bigg) \exp (-\lambda_i z) 
\end{align*}
\begin{align*}
D^3I_i(s,z) = \textbf{i}Re^2Pr R \Bigg[\bigg(\frac{2(z-z_d)}{d^2} + \lambda_i\bigg)^2 - \frac{2}{d^2}\Bigg] \exp \bigg(\displaystyle \frac{-(z-z_d)^2}{d^2}\bigg) \exp (-\lambda_i z)
\end{align*}
\begin{align*}
D^4I_i(s,z) = -\textbf{i}Re^2Pr R \Bigg[\bigg(\frac{2(z-z_d)}{d^2} + \lambda_i\bigg)^3 - \frac{6}{d^2}\bigg(\frac{2(z-z_d)}{d^2} + \lambda_i\bigg)\Bigg] \exp \bigg(\displaystyle \frac{-(z-z_d)^2}{d^2}\bigg) \exp (-\lambda_i z)
\end{align*}
The RHS matrix $\mathscr{R}_i(s)$ can now be found by direct substitution of the above derivatives.
\section{Infinite Depth - Stratified}
The governing equation \ref{31} and the free surface boundary conditions \ref{34}, \ref{35} and \ref{36} are same for both infinite and finite depth case. The boundedness condition at $z \rightarrow -\infty$ should be applied for the infinite-depth domain, whereas wall-boundary conditions \ref{37} are used for the finite-depth one. \subsection{\textbf{Vorticity perturbation}}
We assume $\hat{\omega}(z,t=0) = \Omega  \exp(-(z-z_d)^2/d^2)$, where $z_d \in ( -\infty, 0)$ and neglect all the other initial conditions. If the real part of the roots $\lambda_i > 0$, the solution terms in equation \ref{31}, for three of the roots, approaches zero as $z\rightarrow -\infty$. It will be seen that for the infinite depth case, there are three $\lambda_i$ with negative real parts i.e. $\mathscr{Re}(\lambda_i)<0$. For these roots, the solution diverges exponentially as $z\to-\infty$ because the integral $I_i(z)$ for the Gaussian initial conditions studied here, are bounded at $z\to-\infty$, irrespective of the sign of $\mathscr{Re}(\lambda_i)$. Consequently, to prevent this divergence we set for these three $\lambda_i$,
\begin{equation*}
C_i(s) = -\mathscr{E}_i(s)I_i(s,z=-\infty)
\end{equation*}
thus providing us three equations for these $C_i$ (say $i = 4,5,6$). The other three constants (i.e. $C_i$ with $i=1,2,3$) may be determined using the three boundary conditions \ref{34}, \ref{35} and \ref{36}. The final solution then can be given by:
\begin{align}
\nonumber
\bar{\psi}(s,z) = \sum_{i=1}^6\exp{(\lambda_iz)}\Bigg[C_i(s) + \mathscr{E}_{i}(s)I_i(s,z)\Bigg],
\end{align}
where for $i = 1,2$ and $3$
\begin{equation}
C_i(s) = \sum_{j=1}^3 \frac{1}{|\mathscr{D}(s)|}Adj(\mathscr{D}(s))_{ij}\bigg[\mathscr{R}_j(s)-\sum_{m=4}^6C_m(s)\mathscr{C}_{j}^{(m)}(s)\bigg]\label{3_2}
\end{equation}
and for $i = 4$ to $6$ (say)
\begin{equation}
C_i(s) = -\mathscr{E}_i(s)I_i(s,z=-\infty)\label{3_1}
\end{equation}
Here $\mathscr{D}$(s) is the dispersion matrix generated from the coefficients of the LHS terms in equations \ref{34}, \ref{35} and \ref{36} with $C_4$, $C_5$ and $C_6$ moved to the RHS. $\mathscr{R}_j(s)$ is the column matrix containing RHS terms of equations \ref{34}, \ref{35} and \ref{36}. The coefficient matrix $\mathscr{C}_{j}^{(m)}(s)$ is given by:
\begin{equation*}
\mathscr{C}_{j}^{(m)}(s) = [(\lambda_m^2 + 1)\hspace{0.5cm} L_m\hspace{0.5cm} M_m]^\intercal
\end{equation*}
where $L_m(s)$ and $M_m(s)$ are defined in 2.7.
\subsection{\textbf{Density perturbation}}
The procedure is similar to the previous section. For the case of $\hat{\rho}(z,t=0) = R\exp(-(z-z_d)^2/d^2)$, where $z_d\in(-\infty,0)$, the integral, its derivatives and the $\mathscr{R}_j(s)$ from the three boundary conditions at the interface, can be written directly from subsection 2.3.1. The constants can be written from subsection 3.1.1.
\subsection{\textbf{Free surface perturbation}}
The solution is quite similar to subsection 2.2, but the changes are: a) Only three roots that have non-negative real part are chosen because of boundedness conditions at $z \rightarrow -\infty$. b) The dispersion matrix $\mathscr{D}(s)$ is a $3\times3$ matrix generated from the LHS coefficients of boundary conditions \ref{34}, \ref{35} and \ref{36}. c) There will be a continuous spectrum along with a discrete spectrum.
\section{Finite Depth - Unstratified}
We substitute the solution form given in equation \ref{33}, into boundary conditions \ref{34}, \ref{35} and \ref{35}a,b. The tangential stress condition condition is given by:
\begin{equation}
\sum_{i=1}^4 (\lambda_i^2 + 1)C_i(s) = -\sum_{i=1}^4 \mathscr{F}_i(s)\big((D+\lambda_i)^2 + 1\big)I_i(s,z=0)\label{4_1}
\end{equation}
Similarly, the normal stress boundary condition is given by
\begin{align}
\nonumber
\sum_{i=1}^4 &\bigg[s\lambda_i^3 - s\lambda_i(sRe+3) - Re(1+Bo^{-1})\bigg]C_i(s) = \\ \nonumber &-\sum_{i=1}^4 \mathscr{F}_i(s)\bigg[s(D+\lambda_i)^3- s(sRe + 3)(D+\lambda_i) -  Re(1+Bo^{-1})\bigg]I_i(s,z=0) \\ &+ Re(1+Bo^{-1})\textbf{i}\hat{\eta}(t=0) - s Re D\hat{\psi}(z=0,t=0)\label{4_2}
\end{align}
The no-penetration boundary condition at the bottom surface is given by:
\begin{equation}
\sum_{i=1}^4e^{-\lambda_i H}C_i(s) = -\sum_{i=1}^4e^{-\lambda_i H}\mathscr{F}_i(s)I_i(s,z=-H)\label{4_3}
\end{equation}
The no-slip boundary condition is given by:
\begin{equation}
\sum_{i=1}^4\lambda_i e^{-\lambda_i H}C_i(s) = -\sum_{i=1}^4\mathscr{F}_i(s)e^{-\lambda_i H}(D+\lambda_i)I_i(s,z=-H)\label{4_4}
\end{equation}
The equations \ref{4_1}-\ref{4_4} can be written as
\begin{equation}
\begin{bmatrix}
(\lambda_1^2 + 1) &(\lambda_2^2 + 1) &(\lambda_3^2 + 1) &(\lambda_4^2 + 1)  \\
L_1 &L_2 &L_3 &L_4\\
e^{-\lambda_1 H} &e^{-\lambda_2 H} &e^{-\lambda_3 H} &e^{-\lambda_4 H} \\
\lambda_1e^{-\lambda_1 H} &\lambda_2e^{-\lambda_2 H} &\lambda_3e^{-\lambda_3 H} &\lambda_4e^{-\lambda_4 H}
\end{bmatrix}
\begin{bmatrix}
C_1 \\ C_2 \\ C_3 \\ C_4
\end{bmatrix}
=
\begin{bmatrix}
\mathscr{R}_1 \\ \mathscr{R}_2 \\ \mathscr{R}_3 \\ \mathscr{R}_4
\end{bmatrix}\label{4_5}
\end{equation}
where, $L_i(s)$ are coefficients of $C_i(s)$ in equation \ref{4_5} and $\mathscr{R}_i$ are the RHS terms of the four equations. We observe that the matrix multiplying the $C_i(s)$ is the matrix used in the dispersion relation in the normal mode analysis and denote this by $\mathscr{D}(s)$ Therefore, we can write the coefficients as:
\begin{equation}
C_i(s) = \sum_{j=1}^4 \frac{1}{|\mathscr{D}(s)|}Adj(\mathscr{D}(s))_{ij}\mathscr{R}_j(s)\label{4_6}
\end{equation}
where, $Adj(\mathscr{D}(s))$ implies the adjugate matrix of a matrix $\mathscr{D}(s)$. The inverse laplace transform of the total solution, then, can be given by:
\begin{align}
\hat{\psi}(z) = \dfrac{1}{2\pi\mathbf{i}}\int_{\gamma-\textbf{i}\infty}^{\gamma+\textbf{i}\infty}\exp{(st)}ds\sum_{i=1}^4\exp{(\lambda_i(s)z)}\Bigg[\sum_{j=1}^4\frac{1}{|\mathscr{D}(s)|}Adj(\mathscr{D}(s))_{ij}\mathscr{R}_j(s) + \mathscr{F}_{i}(s)I_i(s,z)\Bigg]\label{4_7}
\end{align}
\subsection{Vorticity perturbation}
For the case of $\hat{\omega}(z,t=0) = \Omega  \exp(-(z-z_d)^2/d^2)$, where $z_d \in ( -H, 0)$, we have
\begin{equation*}
I(s,z) = Re\Omega K_i(s,z)
\end{equation*}
where $K_i(s,z)$ has been defined earlier in equation 2.12. The derivatives of which are given by:
\begin{equation*}
DI_i(s,z) = Re\Omega \exp \bigg(\displaystyle \frac{-(z-z_d)^2}{d^2}\bigg) \exp (-\lambda_i z)
\end{equation*}
\begin{align*}
D^2I_i(s,z) = -Re\Omega \bigg(\frac{2(z-z_d)}{d^2} + \lambda_i\bigg) \exp \bigg(\displaystyle \frac{-(z-z_d)^2}{d^2}\bigg) \exp (-\lambda_i z) 
\end{align*}
\begin{align*}
D^3I_i(s,z) = Re\Omega \Bigg[\bigg(\frac{2(z-z_d)}{d^2} + \lambda_i\bigg)^2 - \frac{2}{d^2}\Bigg] \exp \bigg(\displaystyle \frac{-(z-z_d)^2}{d^2}\bigg) \exp (-\lambda_i z)
\end{align*}
\begin{align*}
D^4I_i(s,z) = -Re\Omega \Bigg[\bigg(\frac{2(z-z_d)}{d^2} + \lambda_i\bigg)^3 - \frac{6}{d^2}\bigg(\frac{2(z-z_d)}{d^2} + \lambda_i\bigg)\Bigg] \exp \bigg(\displaystyle \frac{-(z-z_d)^2}{d^2}\bigg) \exp (-\lambda_i z)
\end{align*}
The RHS matrix $\mathscr{R}_j$(s) can be written directly by substituting the above derivatives of the integral into equations \ref{4_1}-\ref{4_4}.
\subsection{Free-surface perturbation}
Assuming all the initial perturbations other than the amplitude to be zero, the solution of the homogeneous equation \ref{32} is given by:
\begin{equation}
\bar{\psi}(s,z) = \sum_{i=1}^4\exp{(\lambda_iz)}\frac{1}{|\mathscr{D}(s)|}\sum_{j=1}^4Adj(\mathscr{D}(s))_{ij}\mathscr{R}_j(s) \label{4_8}
\end{equation}
The RHS column vector $\mathscr{R}_j(s)$ is given by:
\begin{equation}
\begin{bmatrix}
0 &Re(1+Bo^{-1})\textbf{i}\hat{\eta}(t=0) & 0 & 0
\end{bmatrix}^\intercal\label{4_9}
\end{equation}
Then the inverse laplace transform can be written as:
\begin{align}
\hat{\psi}(z) = \dfrac{1}{2\pi\mathbf{i}}\int_{\gamma-\textbf{i}\infty}^{\gamma+\textbf{i}\infty}\exp{(st)}ds\sum_{i=1}^4\exp{(\lambda_i(s)z)}\Bigg[\frac{1}{|\mathscr{D}(s)|}\sum_{j=1}^4Adj(\mathscr{D}(s))_{ij}\mathscr{R}_j(s)\Bigg]\label{4_10}
\end{align}
If the numerators of equation \ref{4_10} don't have any poles, and if the poles from the denominator are simple (and countably infinite), we can write:
\begin{equation}
\hat{\psi}(z) = \sum_{n = 1}^\infty \sum_{i=1}^4\exp{(\lambda_i(s_n)z)}\Bigg[\frac{1}{|\mathscr{D}(s_n)|'}\sum_{j=1}^4Adj(\mathscr{D}(s_n))_{ij}\mathscr{R}_j(s_n)\Bigg]\exp{(s_nt)}\label{4_11}
\end{equation}
Then, we can write the amplitude evolution, from the kinematic boundary condition, as:
\begin{align}
\nonumber
\frac{\hat{\eta}(t)}{\hat{\eta}(0)} = 1-&\frac{\textbf{i}}{\hat{\eta}(0)}\sum_{n = 1}^\infty \sum_{i=1}^4\Bigg[\frac{1}{s_n|\mathscr{D}(s_n)|'}\sum_{j=1}^4Adj(\mathscr{D}(s_n))_{ij}\mathscr{R}_j(s_n)\Bigg]\exp{(s_nt)} \\ &- \frac{\textbf{i}}{\hat{\eta}(0)}\sum_{i=1}^4\Bigg[\frac{1}{|\mathscr{D}(0)|}\sum_{j=1}^4Adj(\mathscr{D}(0))_{ij}\mathscr{R}_j(0)\Bigg]\label{4_12}
\end{align}
\section{Infinite Depth - unstratified}
\subsection{Vorticity initial condition}
After the integral derivations of sub-subsection 1.1, we see that $I_i(s,z=-\infty)$ is bounded. So, we get $C_3(s) = -\mathscr{F}_3(s) I_3(s,z=-\infty)$ and $C_4(s) = -\mathscr{F}_4(s) I_4(s,z=-\infty)$ by applying finiteness boundary conditions. Therefore, we write:
\begin{equation}
C_i(s) = \sum_{j=1}^2 \frac{1}{|\mathscr{D}(s)|}Adj(\mathscr{D}(s))_{ij}\bigg[\mathscr{R}_j(s)-\sum_{m = 3}^4 C_m(s)\mathscr{C}_{j}^{(m)}(s)\bigg]\label{5_1}
\end{equation}
where, $\mathscr{D}(s)$ is the dispersion matrix from boundary conditions \ref{4_1} and \ref{4_2}, we can write $\mathscr{R}_j(s)$ by substituting derivatives of the integral in subsection 4.1 in the RHS of boundary condition \ref{4_1} and \ref{4_2}. And $\mathscr{C}_{j}^{(m)}(s)$ are the coefficients in the LHS of the same boundary condition viz.
\begin{equation*}
    \mathscr{C}_j^{(m)}(s) = [(\lambda_m^2+1)\hspace{0.3cm}s\lambda_m^3-s\lambda_m(s\textrm{Re}+3)-\textrm{Re}(1+\textrm{Bo}^{-1})]^\intercal
\end{equation*}
\subsection{Free surface perturbation}
Assuming all the initial perturbations other than the amplitude to be zero, the solution of the homogeneous equation \ref{32} is given by:
\begin{equation}
\bar{\psi}(s,z) = \sum_{i=1}^2\exp{(\lambda_iz)}\frac{1}{|\mathscr{D}(s)|}\sum_{j=1}^2Adj(\mathscr{D}(s))_{ij}\mathscr{R}_j(s)\label{5_3} 
\end{equation}
The matrix $\mathscr{D}(s)$ can be written from boundary conditions \ref{4_1} and \ref{4_2}. The RHS matrix $\mathscr{R}_j(s)$ is given by:
\begin{equation}
\begin{bmatrix}
0 &Re(1+Bo^{-1})\textbf{i}\hat{\eta}(t=0)
\end{bmatrix}^\intercal\label{5_4}
\end{equation}
Then the inverse laplace transform can be written as:
\begin{align}
\hat{\psi}(z) = \dfrac{1}{2\pi\mathbf{i}}\int_{\gamma-\textbf{i}\infty}^{\gamma+\textbf{i}\infty}\exp{(st)}ds\sum_{i=1}^2\exp{(\lambda_i(s)z)}\Bigg[\frac{1}{|\mathscr{D}(s)|}\sum_{j=1}^2Adj(\mathscr{D}(s))_{ij}\mathscr{R}_j(s)\Bigg]\label{5_5}
\end{align}
We expect a continuous spectrum and a discrete spectrum in this case as the integrand of this inverse laplace integral will have poles and a branch-point.

\section{Roots of the polynomial from ODE \citep{abramowitz1988handbook}}
\subsection{For stratified case}
After substituting the solution form in he homogeneous form of equation \ref{29}
\begin{equation}
\Big[(\lambda_i^2 - 1)-sRePr\Big]\Big[(\lambda_i^2-1)-sRe\Big](\lambda_i^2-1)- RiFr^2Re^2Pr = 0\label{6_1}
\end{equation}
Assuming $\lambda_i^2 - 1 = l$, we can reduce this sixth order polynomial in $\lambda_i$, into a third order polynomial in $l$. Therefore,
\begin{equation}
\Big[l-sRePr\Big]\Big[l-sRe\Big]l- RiFr^2Re^2Pr = 0\label{6_2}
\end{equation}
\begin{equation*}
l^3 - sRe(Pr+1)l^2 + s^2 Re^2 Pr l - RiFr^2 Re^2 Pr = 0
\end{equation*}
we write,
\begin{equation*}
q = \frac{1}{3}s^2Re^2Pr-\frac{1}{9}s^2Re^2(Pr+1)^2 
\end{equation*}
\begin{equation*}
r = \frac{1}{6}\bigg[3RiFr^2Re^2Pr - s^3Re^3Pr(Pr+1)\bigg] + \frac{1}{27}\big[s^3Re^3(Pr+1)^3\big]
\end{equation*}
and
\begin{equation*}
p_1 = \big(r + \sqrt{q^3+r^2}\big)^{1/3}, \hspace{1cm} p_2 = (r - \sqrt{q^3+r^2})^{1/3}
\end{equation*}
then, the roots of the cubic polynomial are
\begin{equation*}
l_1 = (p_1 + p_2) +  \frac{1}{3} s Re(Pr+1)
\end{equation*}
\begin{equation*}
l_2 = -\frac{1}{2}(p_1 + p_2) + \frac{-\textbf{i}\sqrt{3}}{2}(p_1-p_2) +  \frac{1}{3} s Re(Pr+1)
\end{equation*}
\begin{equation*}
l_3 =-\frac{1}{2}(p_1 + p_2) - \frac{-\textbf{i}\sqrt{3}}{2}(p_1-p_2) +  \frac{1}{3} s Re(Pr+1)
\end{equation*}
Therefore, we get
\begin{align}
\lambda_1 = +\sqrt{1+l_1}, \hspace{0.5cm} &\lambda_2 = -\sqrt{1+l_1}, \hspace{0.5cm} \lambda_3 = +\sqrt{1+l_2}, \hspace{0.5cm} \lambda_4 = -\sqrt{1+l_2}, \hspace{0.5cm}\label{6_3}\\ \nonumber & \lambda_5 = +\sqrt{1+l_3}, \hspace{0.5cm} \lambda_6 = -\sqrt{1+l_3}
\end{align}
\subsection{Without stratification}
Substituting the solution form in the homogeneous form of equation \ref{32}
\begin{equation}
\Big[l-sRe\Big]l = 0\label{6_4}
\end{equation}
Therefore, the solutions are given by:
\begin{equation*}
l = 0,\hspace{1cm} l = sRe
\end{equation*}
And the roots of the fourth order polynomial:
\begin{equation}
\lambda_1 = +1, \hspace{0.5cm} \lambda_2 = -1, \hspace{0.5cm} \lambda_3 = +\sqrt{1+sRe},\hspace{0.5cm}\lambda_4 = -\sqrt{1+sRe}\label{6_5}
\end{equation}
\bibliographystyle{jfm}
\bibliography{apssamp}